\pdfoutput=1
\documentclass[12pt,a4paper]{article}

\usepackage{ifthen} 
\newboolean{pdflatex}
\setboolean{pdflatex}{true} 

\newboolean{articletitles}
\setboolean{articletitles}{true} 

\newboolean{uprightparticles}
\setboolean{uprightparticles}{false} 

\newcommand{\ptjet}{\ensuremath{\pt^{\text{jet}}}\xspace}
\newcommand{\z}{\ensuremath{z}\xspace}
\newcommand{\jt}{\ensuremath{j_{\text{T}}}\xspace}
\newcommand{\rhad}{\ensuremath{r}\xspace}

\newcommand{\bdtbc}{BDT\textsubscript{$b|c$}\xspace}
\newcommand{\dNdx}{dN/d\ensuremath{x}\xspace}
\newcommand{\Njets}{\ensuremath{\text{N}_{\text{jets}}}\xspace}
\newcommand{\Nbjets}{\ensuremath{\text{N}_{b\text{-jets}}}\xspace}
\newcommand{\Ncjets}{\ensuremath{\text{N}_{c\text{-jets}}}\xspace}
\newcommand{\NormdNdz}{\ensuremath{\frac{1}{\text{N}_{\text{jets}}}\frac{\text{dN}}{\text{d}z}}\xspace}
\newcommand{\NormdNdjt}{\ensuremath{\frac{1}{\text{N}_{\text{jets}}}\frac{\text{dN}}{\text{d}j_{\text{T}}}}\xspace}
\newcommand{\NormdNdr}{\ensuremath{\frac{1}{\text{N}_{\text{jets}}}\frac{\text{dN}}{\text{d}r}}\xspace}

\def\paperauthors{LHCb collaboration} 
\def\paperasciititle{Measurement of charged-hadron distributions in heavy-flavor jets in proton-proton collisions at \sqrt{s} = 13 TeV} 
\def\papertitle{Measurement of charged-hadron distributions in heavy-flavor jets in proton-proton collisions at $\sqs = 13\tev$} 
\def\paperkeywords{{High Energy Physics}, {LHCb}} 
\def\papercopyright{\the\year\ CERN for the benefit of the LHCb collaboration} 
\def\paperlicence{CC BY 4.0 licence}
\def\paperlicenceurl{https://creativecommons.org/licenses/by/4.0/}

\newif\ifEnableSectionTOCLinks
\EnableSectionTOCLinksfalse 

\usepackage[top=1in, bottom=1.25in, left=1in, right=1in]{geometry}

\usepackage{microtype}
\usepackage{lineno}  
\usepackage{xspace} 
\usepackage{caption} 

\usepackage{graphicx}  
\usepackage{color}
\usepackage{colortbl}
\graphicspath{{./figs/}} 

\usepackage{amsmath} 
\usepackage{amssymb}
\usepackage{amsfonts}
\usepackage{upgreek} 

\newcommand*\patchAmsMathEnvironmentForLineno[1]{%
\expandafter\let\csname old#1\expandafter\endcsname\csname #1\endcsname
\expandafter\let\csname oldend#1\expandafter\endcsname\csname
end#1\endcsname
 \renewenvironment{#1}%
   {\linenomath\csname old#1\endcsname}%
   {\csname oldend#1\endcsname\endlinenomath}%
}
\newcommand*\patchBothAmsMathEnvironmentsForLineno[1]{%
  \patchAmsMathEnvironmentForLineno{#1}%
  \patchAmsMathEnvironmentForLineno{#1*}%
}
\AtBeginDocument{%
\patchBothAmsMathEnvironmentsForLineno{equation}%
\patchBothAmsMathEnvironmentsForLineno{align}%
\patchBothAmsMathEnvironmentsForLineno{flalign}%
\patchBothAmsMathEnvironmentsForLineno{alignat}%
\patchBothAmsMathEnvironmentsForLineno{gather}%
\patchBothAmsMathEnvironmentsForLineno{multline}%
\patchBothAmsMathEnvironmentsForLineno{eqnarray}%
}

\usepackage[pdftex,
            pdfauthor={\paperauthors},
            pdftitle={\paperasciititle},
            pdfkeywords={\paperkeywords}]{hyperref}
\usepackage{hyperxmp}
\hypersetup{
    pdfcopyright={Copyright (C) \papercopyright},
    pdflicenseurl={\paperlicenceurl}
}

\usepackage[colorinlistoftodos,textsize=scriptsize]{todonotes}

\usepackage[bottom,flushmargin,hang,multiple]{footmisc}

\usepackage[all]{hypcap} 

\usepackage{xspace}
\usepackage{upgreek}

\def\lhcb   {\mbox{LHCb}\xspace}
\def\atlas  {\mbox{ATLAS}\xspace}
\def\cms    {\mbox{CMS}\xspace}
\def\alice  {\mbox{ALICE}\xspace}

\def\MagUp {\mbox{\em Mag\kern -0.05em Up}\xspace}

\ifthenelse{\boolean{uprightparticles}}%
{

 \def\Peta        {\ensuremath{\upeta}\xspace}

 \def\PDelta      {\ensuremath{\Delta}\xspace}
 \def\PXi         {\ensuremath{\Xi}\xspace}
 \def\PLambda     {\ensuremath{\Lambda}\xspace}
 \def\PSigma      {\ensuremath{\Sigma}\xspace}
 \def\POmega      {\ensuremath{\Omega}\xspace}
 \def\PUpsilon    {\ensuremath{\Upsilon}\xspace}
 \let\oldPi\Pi
 \def\PPi         {\ensuremath{\oldPi}\xspace}

 \def\PB      {\ensuremath{\mathrm{B}}\xspace}
 \def\PD      {\ensuremath{\mathrm{D}}\xspace}

 \def\PK      {\ensuremath{\mathrm{K}}\xspace}
 \def\PZ      {\ensuremath{\mathrm{Z}}\xspace}
 \def\Pb      {\ensuremath{\mathrm{b}}\xspace}
 \def\Pc      {\ensuremath{\mathrm{c}}\xspace}
 
 \def\Pe      {\ensuremath{\mathrm{e}}\xspace}
 \def\Ps      {\ensuremath{\mathrm{s}}\xspace}

 \def\thebaroffset{0.0em}
}
{

 \def\Peta        {\ensuremath{\eta}\xspace}

 \mathchardef\PDelta="7101
 \mathchardef\PXi="7104
 \mathchardef\PLambda="7103
 \mathchardef\PSigma="7106
 \mathchardef\POmega="710A
 \mathchardef\PUpsilon="7107
 \mathchardef\PPi="7105
 \def\PB      {\ensuremath{B}\xspace}
 \def\PD      {\ensuremath{D}\xspace}

 \def\PK      {\ensuremath{K}\xspace}
 \def\PZ      {\ensuremath{Z}\xspace}
 \def\Pb      {\ensuremath{b}\xspace}
 \def\Pc      {\ensuremath{c}\xspace}
 
 \def\Pe      {\ensuremath{e}\xspace}
 \def\Ps      {\ensuremath{s}\xspace}

 \def\thebaroffset{0.18em}
}
\newcommand{\offsetoverline}[2][\thebaroffset]{\kern #1\overline{\kern -#1 #2}}%

\makeatletter
\ifcase \@ptsize \relax
  \newcommand{\miniscule}{\@setfontsize\miniscule{4}{5}}
\or
  \newcommand{\miniscule}{\@setfontsize\miniscule{5}{6}}
\or
  \newcommand{\miniscule}{\@setfontsize\miniscule{5}{6}}
\fi
\makeatother

\DeclareRobustCommand{\optbar}[1]{\shortstack{{\miniscule (\rule[.5ex]{1.25em}{.18mm})}
  \\ [-.7ex] $#1$}}

\def\en         {{\ensuremath{\Pe^-}}\xspace}   
\def\ep         {{\ensuremath{\Pe^+}}\xspace}

\def\Z      {{\ensuremath{\PZ}}\xspace}

\def\squark    {{\ensuremath{\Ps}}\xspace}

\def\cquark    {{\ensuremath{\Pc}}\xspace}
\def\cquarkbar {{\ensuremath{\overline \cquark}}\xspace}
\def\ccbar     {{\ensuremath{\cquark\cquarkbar}}\xspace}
\def\bquark    {{\ensuremath{\Pb}}\xspace}
\def\bquarkbar {{\ensuremath{\overline \bquark}}\xspace}
\def\bbbar     {{\ensuremath{\bquark\bquarkbar}}\xspace}

\def\KorKbar {\kern \thebaroffset\optbar{\kern -\thebaroffset \PK}{}\xspace}

\newcommand{\etaz}{\ensuremath{\Peta}\xspace}

\def\D       {{\ensuremath{\PD}}\xspace}

\def\DorDbar {\kern \thebaroffset\optbar{\kern -\thebaroffset \PD}\xspace}
\def\Dz      {{\ensuremath{\D^0}}\xspace}

\def\Dp      {{\ensuremath{\D^+}}\xspace}
\def\Dm      {{\ensuremath{\D^-}}\xspace}

\def\DpDm    {\ensuremath{\Dp {\kern -0.16em \Dm}}\xspace}

\def\Dstarpm {{\ensuremath{\D^{*\pm}}}\xspace}

\def\B       {{\ensuremath{\PB}}\xspace}

\def\BorBbar {\kern \thebaroffset\optbar{\kern -\thebaroffset \PB}\xspace}

\def\Bd      {{\ensuremath{\B^0}}\xspace}

\def\BdorBdbar {\kern \thebaroffset\optbar{\kern -\thebaroffset \Bd}\xspace}

\def\Bpm     {{\ensuremath{\B^\pm}}\xspace}

\def\Bs      {{\ensuremath{\B^0_\squark}}\xspace}

\def\BsorBsbar {\kern \thebaroffset\optbar{\kern -\thebaroffset \Bs}\xspace}

\def\Y#1S{\ensuremath{\PUpsilon{(#1S)}}\xspace}

\def\Lz          {{\ensuremath{\PLambda}}\xspace}

\def\LorLbar     {\kern \thebaroffset\optbar{\kern -\thebaroffset \PLambda}\xspace}

\def\Lc          {{\ensuremath{\Lz^+_\cquark}}\xspace}

\def\AT#1     {\ensuremath{A_{\mathrm{T}}^{#1}}\xspace}           

\def\C#1      {\ensuremath{\mathcal{C}_{#1}}\xspace}                       
\def\Cp#1     {\ensuremath{\mathcal{C}_{#1}^{'}}\xspace}                    
\def\Ceff#1   {\ensuremath{\mathcal{C}_{#1}^{\mathrm{(eff)}}}\xspace}        
\def\Cpeff#1  {\ensuremath{\mathcal{C}_{#1}^{'\mathrm{(eff)}}}\xspace}       
\def\Ope#1    {\ensuremath{\mathcal{O}_{#1}}\xspace}                       
\def\Opep#1   {\ensuremath{\mathcal{O}_{#1}^{'}}\xspace}                    

\newcommand{\nospaceunit}[1]{\ensuremath{\text{#1}}}
\newcommand{\aunit}[1]{\ensuremath{\text{\,#1}}}

\newcommand{\tev}{\aunit{Te\kern -0.1em V}\xspace}
\newcommand{\gev}{\aunit{Ge\kern -0.1em V}\xspace}
\newcommand{\mev}{\aunit{Me\kern -0.1em V}\xspace}
\newcommand{\kev}{\aunit{ke\kern -0.1em V}\xspace}
\newcommand{\ev}{\aunit{e\kern -0.1em V}\xspace}

\newcommand{\mevc}{\ensuremath{\aunit{Me\kern -0.1em V\!/}c}\xspace}
\newcommand{\gevc}{\ensuremath{\aunit{Ge\kern -0.1em V\!/}c}\xspace}
\newcommand{\mevcc}{\ensuremath{\aunit{Me\kern -0.1em V\!/}c^2}\xspace}
\newcommand{\gevcc}{\ensuremath{\aunit{Ge\kern -0.1em V\!/}c^2}\xspace}

\def\mum  {\ensuremath{\,\upmu\nospaceunit{m}}\xspace}

\def\fb   {\ensuremath{\aunit{fb}}\xspace}
\def\invfb   {\ensuremath{\fb^{-1}}\xspace}

\def\gsim{{~\raise.15em\hbox{$>$}\kern-.85em
          \lower.35em\hbox{$\sim$}~}\xspace}
\def\lsim{{~\raise.15em\hbox{$<$}\kern-.85em
          \lower.35em\hbox{$\sim$}~}\xspace}

\def\sqs   {\ensuremath{\protect\sqrt{s}}\xspace}

\def\pt         {\ensuremath{p_{\mathrm{T}}}\xspace}

\def\ptot       {\ensuremath{p}\xspace}

\def\evtgen     {\mbox{\textsc{EvtGen}}\xspace}

\def\geant      {\mbox{\textsc{Geant4}}\xspace}

\def\photos     {\mbox{\textsc{Photos}}\xspace}

\def\pythia     {\mbox{\textsc{Pythia}}\xspace}

\def\tell1  {TELL1\xspace}
\def\ukl1   {UKL1\xspace}

\newcommand{\ie}{\mbox{\itshape i.e.}\xspace}

\newcommand{\lhcborcid}[1]{\href{https://orcid.org/#1}{\hspace*{0.1em}\raisebox{-0.45ex}{\includegraphics[width=1em]{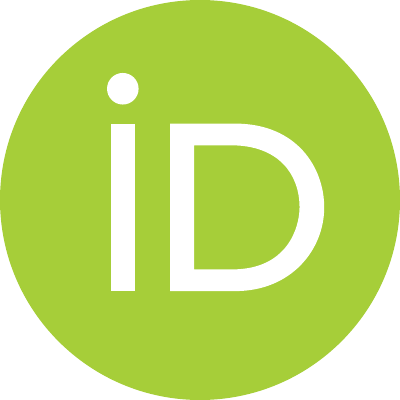}}}}

\hypersetup{
  colorlinks   = true, 
  urlcolor     = blue, 
  linkcolor    = blue, 
  citecolor    = red   
}

\ifEnableSectionTOCLinks
    \usepackage[explicit]{titlesec} 
    
    \let\oldcontentsline\contentsline
    \renewcommand

    \titleformat{\section}{\normalfont\Large\bf}{\hyperlink{tocsection.\thesection}{{\thesection} \parbox[t]{\dimexpr\textwidth-1pc}{#1}}}{1pc}{}

    \titleformat{\subsection}{\normalfont\bf}{\hyperlink{tocsubsection.\thesubsection}{{\thesubsection} \parbox[t]{\dimexpr\textwidth-1pc}{#1}}}{1pc}{}

    \titleformat{name=\section,numberless}[display]{}{}{0pt}{\normalfont\Huge\bfseries #1}
\fi

\usepackage{cite} 
\usepackage{mciteplus}

\usepackage{longtable}
\usepackage{booktabs}

\begin{document}

\renewcommand{\thefootnote}{\fnsymbol{footnote}}
\setcounter{footnote}{1}


\begin{titlepage}
\pagenumbering{roman}

\vspace*{-1.5cm}
\centerline{\large EUROPEAN ORGANIZATION FOR NUCLEAR RESEARCH (CERN)}
\vspace*{1.5cm}
\noindent
\begin{tabular*}{\linewidth}{lc@{\extracolsep{\fill}}r@{\extracolsep{0pt}}}
\ifthenelse{\boolean{pdflatex}}
{\vspace*{-1.5cm}\mbox{\!\!\!\includegraphics[width=.14\textwidth]{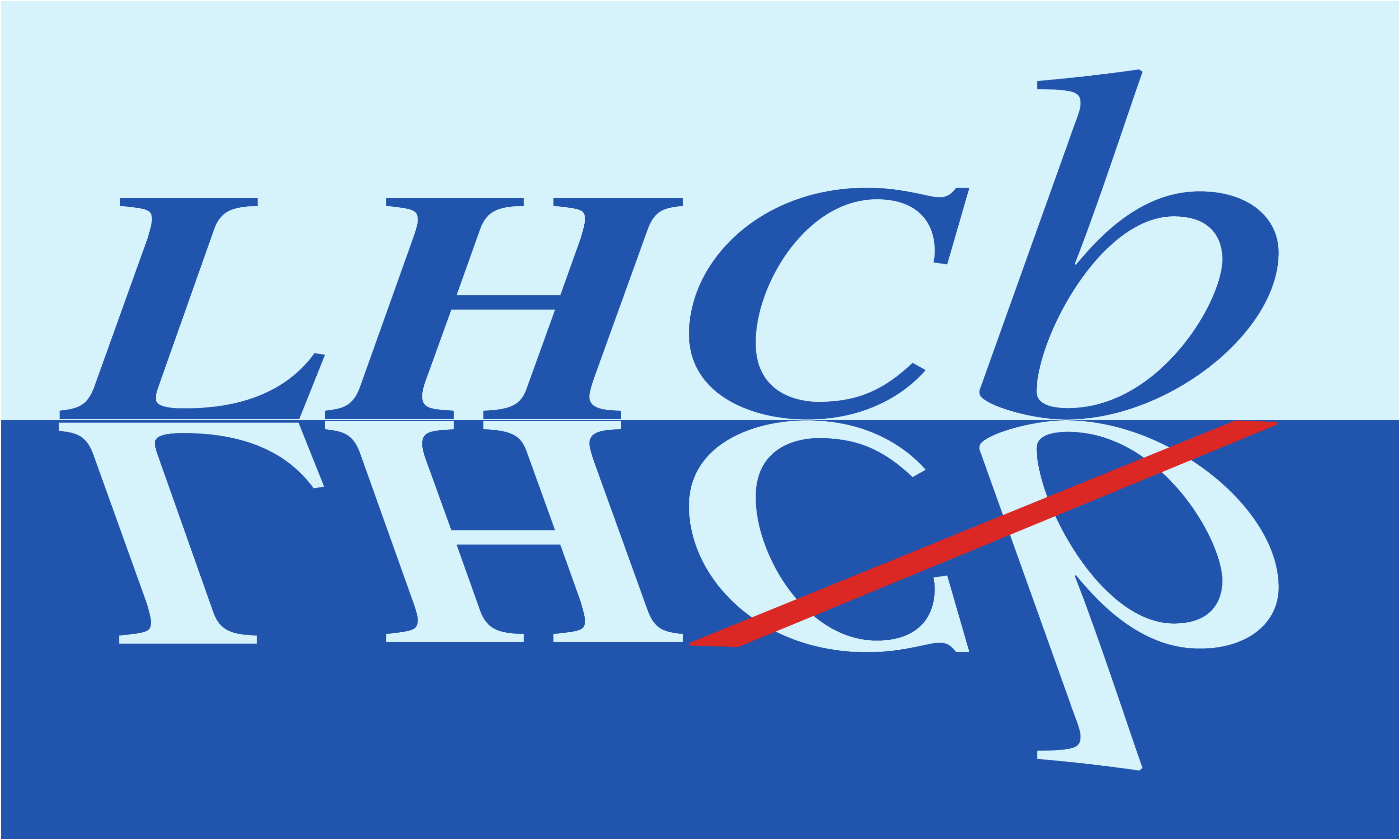}} & &}%
{\vspace*{-1.2cm}\mbox{\!\!\!\includegraphics[width=.12\textwidth]{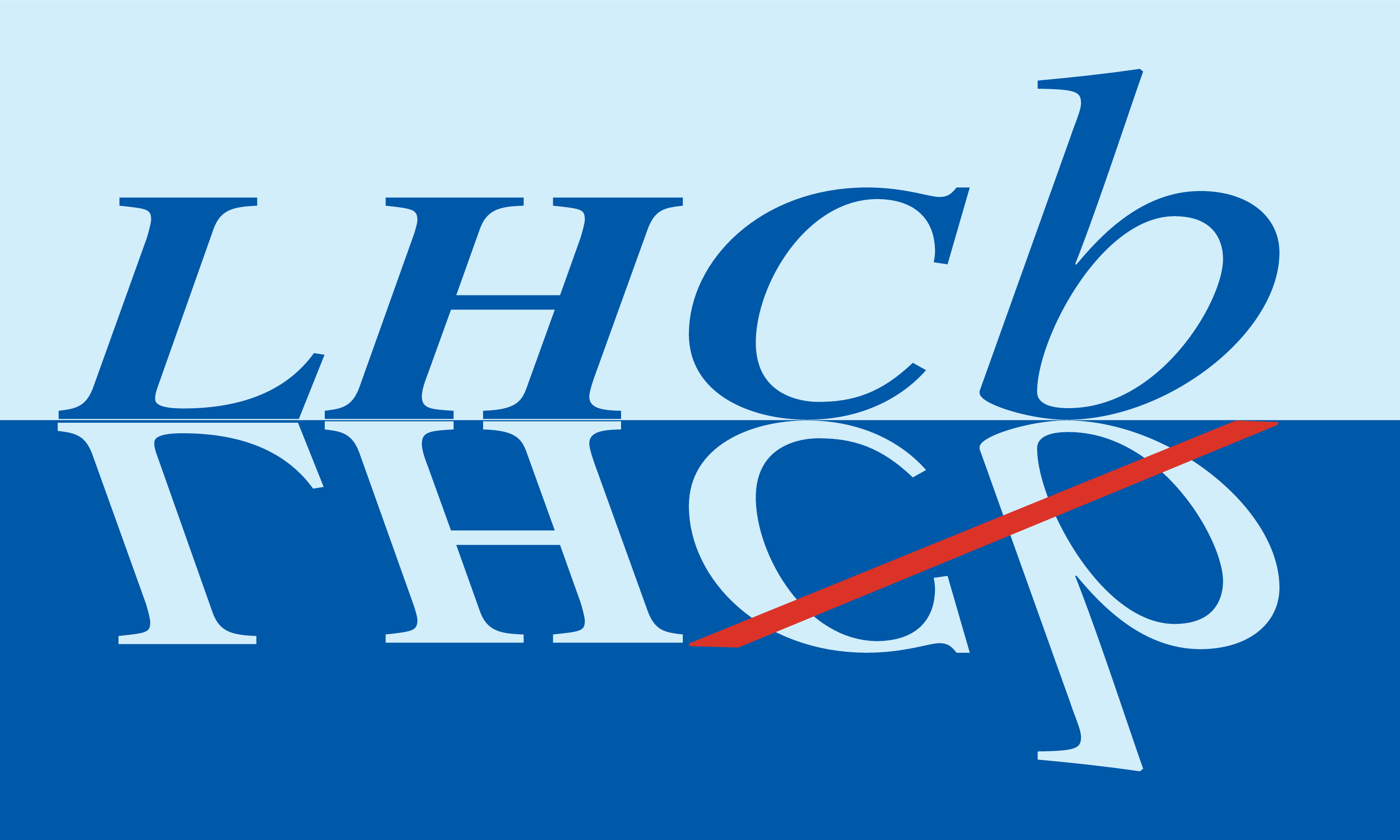}} & &}%
\\
 & & CERN-EP-2025-230 \\  
 & & LHCb-PAPER-2025-038 \\  
 & & 8 April 2026 \\ 
 & & \\
\end{tabular*}

\vspace*{2cm}

{\normalfont\bfseries\boldmath\huge
\begin{center}
  \papertitle 
\end{center}
}

\vspace*{0.4cm}

\begin{center}
\paperauthors\footnote{Authors are listed at the end of this paper.}
\end{center}

This paper is dedicated to the memory of our friend and colleague Jordan D. Roth. 
\vspace{0.4cm}

\begin{abstract}
  \noindent
 Charged-hadron distributions in heavy-flavor jets are measured in proton-proton collisions at a center-of-mass energy of \mbox{$\sqs = 13 \tev$} collected by the LHCb experiment. Distributions of the longitudinal momentum fraction, transverse momentum, and radial profile of charged hadrons are measured separately in beauty and charm jets. The distributions are compared to those previously measured by the LHCb collaboration in jets produced back-to-back with a \Z boson, which in the forward region are primarily light-quark-initiated, to compare the hadronization mechanisms of heavy and light quarks. The observed differences between the heavy- and light-jet distributions are consistent with the heavy-quark dynamics expected to arise from the dead-cone effect, as well as with a hard fragmentation of the heavy-flavor hadron as previously measured in single-hadron fragmentation functions. This measurement provides additional constraints for the extraction of collinear and transverse-momentum-dependent heavy-flavor fragmentation functions and offers another approach to probing the mechanisms that govern heavy-flavor hadronization.  
  
\end{abstract}

\vspace*{1.0cm}

\begin{center}
  Published in 
  JHEP 04 (2026) 029 
\end{center}

\vspace{\fill}

{\footnotesize 
\centerline{\copyright~\papercopyright. \href{\paperlicenceurl}{\paperlicence}.}}
\vspace*{2mm}

\end{titlepage}


\newpage
\setcounter{page}{2}
\mbox{~}
%
%
%
%


\renewcommand{\thefootnote}{\arabic{footnote}}
\setcounter{footnote}{0}


\cleardoublepage


\pagestyle{plain} 
\setcounter{page}{1}
\pagenumbering{arabic}


\section{Introduction}
\label{sec:Introduction}

Hadronization, the process by which a colored quark or gluon becomes confined within a color-neutral hadron, is one of the most fundamental processes in quantum chromodynamics (QCD), yet its underlying mechanisms remain poorly understood. The nonperturbative energy scales associated with bound-state formation and color neutralization prevent a theoretical description based on perturbative QCD. Instead, the hadronization process is parameterized with the use of fragmentation functions (FFs), which describe the probability density for a particular flavor of outgoing parton to produce a specific hadron~\cite{CTEQ:1993hwr}. These functions are assumed to be universal, meaning that they are independent of the collision system in which they are measured, and since they are nonperturbative they must be constrained from fits to data~\cite{Metz:2016swz, Bertone:2018ecm, Soleymaninia:2018uiv, Soleymaninia:2019sjo, Soleymaninia:2020bsq, Gao:2024nkz, Gao:2024dbv}. Measurements sensitive to FFs in a variety of collision systems and for different hadron species are therefore necessary to improve the current understanding of hadronization.

Different types of fragmentation functions reveal varying levels of detail about the hadronization process. Collinear FFs describe the probability for a parton to form a hadron carrying a fraction \z of the parton's initial momentum. Transverse-momentum-dependent (TMD) FFs~\cite{Metz:2016swz} additionally describe the probability for a parton to form a hadron carrying a certain transverse momentum, expanding the one-dimensional description of hadronization to three dimensions. In recent years, the FF formalism has been extended to jets, collimated sprays of hadrons produced during the hadronization of a high-energy quark or gluon. Measurements of hadrons within jets have been shown to be sensitive to the collinear and TMD FFs~\cite{Procura:2009vm, Jain:2011xz, Procura:2011aq, Kaufmann:2015hma, Kang:2016ehg, Kang:2017glf}. Jets are therefore powerful tools for studying hadronization, not only because they provide additional constraints on FFs but also because they enable a way to access both the initial and final states of the hadronization process. The jet kinematics serve as a proxy for the initial-state parton, while the particles within the jet constitute the final-state hadrons. 

Heavy-flavor jets, which originate from the hadronization of a beauty or charm quark, are ideal systems for studying hadronization as they provide access from an experimental point of view to the flavor content of both the initial fragmenting parton and final-state hadrons. Measurements of hadron production in heavy-flavor jets in proton-proton ($pp$) collisions complement previous measurements of inclusive collinear heavy-flavor FFs measured in \ep\en collisions~\cite{DELPHI:2011aa, OPAL:2002plk, ALEPH:2001pfo, SLD:2002poq, ALEPH:1999syy, OPAL:1994cct, ARGUS:1991vjh}, provide additional access to TMD FFs, and allow to test the FFs' universality. Measurements in jets of the longitudinal momentum fraction \z carried by an open heavy-flavor hadron, \ie~a bound state of at least one charm or beauty quark and one or more lighter quarks, have been performed by the ATLAS and ALICE collaborations using fully reconstructed \Bpm~\cite{ATLAS:2021agf}, \Dstarpm~\cite{ATLAS:2011chi}, \Dz~\cite{ALICE:2019cbr, ALICE:2022mur}, and \Lc~\cite{ALICE:2023jgm} decays. These measurements are sensitive to the collinear FFs. Similar measurements using partially reconstructed inclusive heavy-flavor decays have been performed in beauty jets by the ATLAS~\cite{ATLAS:2022miz} and CMS~\cite{CMS:2024gds} collaborations. Fewer measurements exist of the hadrons produced along with the heavy-flavor hadron in the jet, which also originate from the fragmentation of the initial heavy quark. These measurements, which are typically based on charged hadrons, also benefit from the fact that they can be directly compared with previous measurements in light-parton-initiated jets. The charged-particle multiplicity and jet shape~\cite{Ellis:1992qq}, an observable that describes the energy distribution of particles within the jet, have been compared between beauty and inclusive jets by the \cms collaboration, with the former category found to have larger particle multiplicities and a broader energy distribution~\cite{CMS:2020geg}. The energy-energy correlator, a measure of the correlation function between particles in a jet, is sensitive to the transition between partons and hadrons and was recently studied in charm and inclusive jets by the \alice collaboration, with observed differences in the correlation functions attributed to quark mass effects~\cite{ALICE:2025igw}. Measurements of charged-hadron production within heavy-flavor jets complement measurements of the heavy-flavor hadron itself and are necessary to understand the full picture of heavy-quark hadronization. 

Fragmentation studies within heavy-flavor jets are complemented by jet substructure measurements, which enhance understanding of the parton shower stage leading up to hadronization. Recent measurements~\cite{ALICE:2021aqk, LHCb-PAPER-2025-010} have confirmed the presence of the dead-cone effect~\cite{Dokshitzer:1991fd}, in which quark radiation is suppressed for emission angles smaller than the quark mass divided by its energy. Heavy-flavor quarks correspondingly have larger dead-cone angles, which influence their radiation patterns and the distribution of hadrons produced during their fragmentation.

In this paper, measurements of charged-hadron fragmentation distributions in beauty and charm jets are presented. The measured observables are the longitudinal momentum fraction \z, the transverse momentum relative to the jet axis \jt, and the radial position within the jet \rhad, defined as
\begin{equation} \label{eqn:z}
    z \equiv \frac{\mathbf{p}_{\rm had} \cdot \mathbf{p}_{\rm jet}}{|\mathbf{p}_{\rm jet}|^{2}},   
\end{equation}

\begin{equation} \label{eqn:jT}
    j_{\text{T}} \equiv \frac{|\mathbf{p}_{\rm had} \times \mathbf{p}_{\rm jet}|}{|\mathbf{p_{\rm jet}}|},  
\end{equation}
and
\begin{equation} \label{eqn:r}
    r \equiv \sqrt{(\phi_{\rm had} - \phi_{\rm jet})^{2} + (\eta_{\rm had} - \eta_{\rm jet})^{2}}.
\end{equation}
Here, \textbf{p} is the three-momentum vector, $\phi$ the azimuthal angle and $\eta$ the pseudorapidity, and the subscripts refer to the jet or a single hadron within the jet. Each distribution is normalized by the total number of beauty or charm jets. These observables have previously been measured in $pp$ collisions for inclusive jets by the \atlas~\cite{ATLAS:2011myc, ATLAS:2019rqw} collaboration and in jets recoiling against a \Z boson (``\Z-tagged'' jets) by the \lhcb collaboration~\cite{LHCb-PAPER-2019-012, LHCb-PAPER-2022-013}. The $j_\text{T}$ and similar \z observables have also been studied by the \alice collaboration using $pp$ and $p$Pb collisions~\cite{ALICE:2020pga, ALICE:2023oww, ALICE:2018ype}. 

The data sample used in this measurement is the same as that used in the \bbbar and \ccbar cross-section measurement~\cite{LHCb-PAPER-2020-018}, consisting of $pp$ collision data collected by the \lhcb experiment during its Run 2 in 2016 at a center-of-mass energy of \mbox{$\sqs = 13\tev$} and corresponding to an integrated luminosity of 1.6\invfb. The fragmentation distributions of beauty and charm quarks are measured as a function of the jet transverse momentum, \ptjet, in the intervals 20--30\gevc, 30--50\gevc, and 50--100\gevc. The distributions are compared to those previously measured by the \lhcb collaboration in \Z-tagged jets~\cite{LHCb-PAPER-2019-012, LHCb-PAPER-2022-013}. In the forward region accessible to the \lhcb detector, the \Z-tagged jets are primarily light-quark-initiated, therefore comparing the \z, \jt, and \rhad distributions in beauty, charm, and \Z-tagged jet samples allows for comparison of beauty, charm, and light-quark hadronization. 

\section{Detector and simulation}
\label{sec:Detector}

The \lhcb detector~\cite{LHCb-DP-2008-001,LHCb-DP-2014-002} is a single-arm forward spectrometer covering the \mbox{pseudorapidity} range $2 < \eta < 5$, designed for the study of particles containing \bquark or \cquark quarks. The detector used to collect the data analysed in this paper includes a high-precision tracking system consisting of a silicon-strip vertex detector surrounding the $pp$ interaction region~\cite{LHCb-DP-2014-001}, a large-area silicon-strip detector located upstream of a dipole magnet with a bending power of approximately $4{\mathrm{\,T\,m}}$, and three stations of silicon-strip detectors and straw drift tubes placed downstream of the magnet~\cite{LHCb-DP-2017-001}. The tracking system provides a measurement of the momentum, \ptot, of charged particles with a relative uncertainty that varies from 0.5\% at low momentum to 1.0\% at 200\gevc~\cite{LHCb-DP-2014-002}. The minimum distance of a track to a primary $pp$ collision vertex (PV), the impact parameter, is measured with a resolution of $(15+29/\pt)\mum$, where \pt is the component of the momentum transverse to the beam, in\,\gevc. Different types of charged hadrons are distinguished using information from two ring-imaging Cherenkov detectors~\cite{LHCb-DP-2012-003}. Photons, electrons and hadrons are identified by a calorimeter system consisting of scintillating-pad and preshower detectors, an electromagnetic calorimeter and a hadronic calorimeter. Muons are identified by a system composed of alternating layers of iron and multiwire proportional chambers~\cite{LHCB-DP-2012-002}. The online event selection is performed by a trigger~\cite{LHCb-DP-2012-004}, which consists of a hardware stage, based on information from the calorimeter and muon systems, followed by a software stage, which applies a full event reconstruction. Triggered data further undergo a centralized, offline processing step~\cite{Stripping}.

The same trigger selection used in the \bbbar and \ccbar cross-section measurements~\cite{LHCb-PAPER-2020-018} is also applied for this analysis. At the hardware trigger stage, events are required to have a muon with high \pt or a hadron, photon or electron with high transverse energy in the calorimeters. A global event cut is applied on the number of hits in the scintillating-pad detector. The software trigger consists of two stages. The first stage requires a displaced secondary vertex (SV) or a high-\pt track inconsistent with originating from the PV. The second stage requires two jets with $\ptjet >$ 17\gevc each and a SV consistent with a heavy-flavor hadron decay. More details about the jet reconstruction and heavy-flavor tagging are provided in Sec.~\ref{sec:jetrecoandeventsel}.

Simulation is required to model the effects of the detector acceptance and the imposed selection requirements on the jets and charged hadrons. In the simulation, $pp$ collisions are generated using
\pythia~\cite{Sjostrand:2007gs,*Sjostrand:2006za} with a specific \lhcb configuration~\cite{LHCb-PROC-2010-056}. Decays of unstable particles are described by \evtgen~\cite{Lange:2001uf}, in which final-state radiation is generated using \photos~\cite{davidson2015photos}. The interaction of the generated particles with the detector, and its response, are implemented using the \geant toolkit~\cite{Allison:2006ve, *Agostinelli:2002hh} as described in Ref.~\cite{LHCb-PROC-2011-006}. Simulated samples of \bbbar and \ccbar dijets are used in this analysis to compute the efficiencies and purities required to correct the charged-hadron and jet yields, and to construct the response matrices used in the unfolding procedure, which corrects for bin migrations introduced by the smearing of reconstructed observables at the detector level. Several data-driven corrections are also applied to the simulation in order to improve its agreement with data. These corrections are described in more detail in Sec.~\ref{sec:analysisstrategy}.

\section{Jet reconstruction and event selection} \label{sec:jetrecoandeventsel}
Charged and neutral particles are selected as inputs to the jet reconstruction by a particle-flow algorithm~\cite{LHCb-PAPER-2013-058}. The anti-$k_\text{T}$ algorithm~\cite{Cacciari:2008gp} implemented in the \mbox{\textsc{Fastjet}}\xspace package~\cite{Cacciari:2011ma} is used to reconstruct jets with a radius parameter of $R = 0.5$. Selections are applied to remove fake jets arising from either combinatorial background or high-energy leptons mistakenly reconstructed as jets. The fraction of charged particles in the jet must be at least 6\%, at least one particle in the jet must have $\pt > 1.4\gevc$, and no particle is allowed to carry more than 75\% of \ptjet.

Jets are identified as originating from heavy-flavor quarks with the use of a dedicated flavor-tagging algorithm. The SV-tagging algorithm, described in detail in Ref.~\cite{LHCb-PAPER-2015-016}, reconstructs SVs within the jet that have properties consistent with originating from the decay of a heavy-flavor hadron. The algorithm selects tracks displaced from the PV, without requiring that the selected tracks are in a jet. All possible two-body SVs are reconstructed using the displaced tracks, then two-body SVs that share tracks are merged to form an $n$-body SV. Selected $n$-body SVs must have a significant \pt sum of the constituent tracks and displacement from the PV. Additional selection criteria are applied to reject light-hadron decays. If an $n$-body SV passing the selection requirements is found within the jet cone, the jet is ``SV-tagged'' and is very likely to have originated from a beauty or charm quark. The SV-tagging efficiency in Run 2 simulation is about 60\% for beauty jets and 20\% for charm jets, with a misidentification rate of light jets around 1\%~\cite{LHCb-PAPER-2020-018}. Due to the higher particle multiplicity in \mbox{$\sqs = 13\tev$} $pp$ collisions compared to \mbox{$\sqs = 7$} and 8\tev collisions in Run 1, the beauty and charm jet efficiencies in 2016 data are slightly lower and the light jet misidentification rate is slightly higher than the corresponding efficiencies measured in Run 1 data of 65\%, 25\%, and 0.3\%, respectively \cite{LHCb-PAPER-2015-016}. 

Once a jet is SV-tagged, two boosted decision tree (BDT) classifiers~\cite{Breiman,AdaBoost}, also described in detail in Ref.~\cite{LHCb-PAPER-2015-016}, are used to further discriminate among light, beauty, and charm jets. The BDTs are trained on simulated beauty, charm, and light-parton jets, and take as input variables related to the SV in the jet, some of which also depend on the jet kinematics. One BDT is trained to separate heavy jets from light jets, while the other one is trained to separate beauty from charm jets (\bdtbc). Since the data sample used in this analysis was already measured to have a negligible light dijet contribution~\cite{LHCb-PAPER-2020-018}, only the \bdtbc classifier is used in this measurement.

Events passing the trigger selection already contain two reconstructed and SV-tagged jets. Further criteria are applied offline to select jets and charged hadrons for the measurement of the fragmentation observables. Both jets are required to originate from the same PV, and their difference in azimuthal angle, $\Delta\phi = |\phi_{\text{jet 1}} - \phi_{\text{jet 2}}|$, is required to be at least 1.5 radians. The jet pseudorapidity is required to satisfy $2.5 < \etaz_{\text{jet}} < 4.0$ to ensure that the full jet cone lies within the fiducial acceptance of the \lhcb detector. Each jet is required to have 17 $<$ \ptjet  $< 100\gevc$. In a small fraction of events, multiple dijets pass the selection criteria and the dijet with the highest \pt is chosen. 

Jets passing the kinematic selections are divided into beauty- and charm-enhanced samples by requiring a response of the \bdtbc classifier as larger or smaller than 0.1, respectively. The selection requirement is chosen based on previous studies~\cite{LHCb-PAPER-2015-016} of the \bdtbc performance on data, which demonstrated that beauty jets typically have high \bdtbc scores, while charm jets are typically concentrated around zero or negative values. Only jets in the beauty- or charm-enhanced samples are used for the measurement of the charged-hadron distributions. If only one of the jets in a dijet passes the \bdtbc selection requirement, only that jet is assigned to the beauty- or charm-enhanced sample. The purity of the samples is estimated using the previously measured flavor-tagged dijet yields in this data sample from Ref.~\cite{LHCb-PAPER-2020-018}, corrected by the misidentification rates obtained in simulation. The resulting purities are 95\% for the beauty-enhanced sample and 70--75\% for the charm-enhanced sample, depending on the \ptjet. Charged hadrons in the jet are identified with the particle-flow algorithm~\cite{LHCb-PAPER-2013-058}, which uses information from the particle identification (PID) detectors to discriminate between leptons and charged hadrons. The selected hadrons are required to have a good track quality, a momentum between 4\gevc and 1\tev/$c$, \pt $>$ 250\mevc, and to lie within the jet cone ($\Delta R$(track, jet) $<$ 0.5). 

\section{Determination of the charged-hadron distributions} \label{sec:analysisstrategy}
The general strategy for measuring the charged-hadron distributions is the following. The number of charged hadrons in each \z, \jt, and \rhad interval is computed according to Eqs.~\ref{eqn:z}--\ref{eqn:r} using the candidates selected as described above. The \dNdx ($x =$ \z, \jt, or \rhad) distributions are purity corrected, then unfolded two-dimensionally in \ptjet and each fragmentation observable in order to correct for bin migrations, and finally efficiency corrected. The \dNdx distributions are normalized by the total number of jets, \Nbjets or \Ncjets, which is determined independently using the same procedure of purity corrections, unfolding, and then efficiency corrections as a function of the \ptjet distribution. Finally, the fully corrected \dNdx distributions are normalized by the fully corrected \Nbjets or \Ncjets.

Simulation is used to determine the purity and efficiency corrections, and the response matrices used in the unfolding procedure. Several data-driven corrections are applied to correct for known discrepancies between data and simulation. Weights are applied per-jet to correct the trigger and jet SV-tagging efficiencies in simulation, and per-event to correct the global-event-cut efficiency~\cite{LHCb-PAPER-2015-016, LHCb-PAPER-2020-018}. A second set of corrections is applied to correct the energy and momentum resolution of jets in simulation. Jet energy scale (JES) and jet energy resolution (JER) corrections are measured from $Z$+jet events in data and simulation, where the $Z$ boson provides a good approximation of \ptjet when produced approximately back-to-back with the jet~\cite{LHCb-PAPER-2016-011}. The JES and JER corrections are applied to the reconstructed jets in simulation to improve the description of the detector response to jets.

The jet purity correction accounts for reconstructed dijets in simulation that are matched to generator-level dijets not passing the kinematic selections discussed in Sec.~\ref{sec:jetrecoandeventsel}. The purity correction for the normalization factor is computed as a function of \ptjet. The correction is largest at low \ptjet, around 15\%, where the reconstructed jets are more likely to be matched to generator-level jets below the \ptjet requirement, and subsequently decreases to less than 5\% in the highest \ptjet interval. The purity correction for the charged-hadron distributions is computed two-dimensionally as a function of \ptjet and each fragmentation observable, and consists of two components. The jet purity is determined as the fraction of reconstructed charged hadrons in selected reconstructed dijets that are also matched to generator-level dijets passing the kinematic selections. The correction is independent of \z, \jt, and \rhad, with its size being consistent with that observed in the purity correction for the normalization factor. A separate purity correction is applied to the charged hadrons to account for extra hadrons included in the jet at the reconstructed level that are not present in the generator-level jet, and for charged leptons misidentified as charged hadrons in the reconstruction. The charged-hadron purity correction is largest at low \z and at large \rhad, where low-energy particles not correlated to the hard fragmentation are expected to be located. The purity is largely independent of \ptjet when computed as a function of \jt and \rhad, while some \ptjet dependence is observed when the correction is computed as a function of \z. The minimum momentum requirement on the charged hadrons results in higher \pt jets probing lower \z values, and the lowest \z values have the largest purity corrections. 

The jet efficiency correction includes the jet reconstruction, the SV-tagging, the \bdtbc selection requirement used to define the \bquark- and \cquark-enhanced samples, and the trigger efficiency. The efficiency correction for the normalization factor is determined as a function of \ptjet, and the efficiency correction for the charged-hadron distributions is computed two-dimensionally as a function of \ptjet and each fragmentation observable. The jet reconstruction efficiency increases from around 70\% in the lowest \ptjet interval to over 90\% in the highest interval, with similar performance for both beauty and charm jets. The SV-tagging efficiency is lower, as both jets in the dijet are required to be SV-tagged, and reaches up to 40\% for beauty jets, while it is around 3\% for charm jets. The efficiency for beauty jets to pass the \bdtbc selection requirement is larger than 85\% in all \ptjet intervals, while for charm jets it varies between 70 and 80\%. The jet trigger efficiency includes the selection requirements in both the hardware and the first stage of the software trigger, as the jet reconstruction and flavor tagging (SV-tagging and \bdtbc) requirements imposed in the second stage are already accounted for in the previous efficiencies. It is around 60\% for both beauty and charm jets. When computed as a function of \z, \jt, and \rhad, the jet reconstruction and trigger efficiencies are generally uniform, while the jet SV-tagging and \bdtbc efficiencies show a small but non-negligible dependence on the fragmentation observables. This arises from the fact that the SV-tagging algorithm and the \bdtbc use some information about tracks within the jet, which can slightly bias the charged-hadron distributions. A systematic uncertainty is assigned to cover the size of the observed bias, as discussed in more detail in Sec.~\ref{sec:sysunc}.

A separate efficiency correction is applied to the charged hadrons to account for the efficiency of their track reconstruction within the jet radius and their identification. For each charged hadron in a generator-level dijet that is matched to a reconstructed dijet, the efficiencies are determined as the product of the probability that the generator-level charged hadron is reconstructed as a track, that the track is reconstructed within the reconstructed jet, and that the track is identified as a charged hadron. The track reconstruction efficiency is around 80\%, consistent with the values previously measured in Ref.~\cite{LHCb-PAPER-2022-013}. The efficiency varies slightly depending on the \z, \jt, and \rhad interval and the \ptjet interval, as higher-\pt jets can have significantly larger multiplicities than lower-\pt jets for the same jet radius. The second correction accounts for the efficiency that a reconstructed track is actually selected by the anti-$k_\text{T}$ algorithm as belonging to the reconstructed jet. It varies depending on the specific \z, \jt, and \rhad interval, and is larger than 80\% in the majority of intervals. The charged-hadron PID efficiency is generally larger than 98\%. 

Response matrices for the iterative Bayesian unfolding procedure~\cite{DAgostini:1994fjx} are also constructed using simulation. For the unfolding of \Njets, a response matrix composed of the true and reconstructed \ptjet value is determined using the generator-level simulation and the purity-corrected reconstructed simulation. For the unfolding of the \dNdx distributions, two-dimensional response matrices are constructed as a function of \ptjet and each fragmentation observable. Convergence is obtained after two unfolding iterations. Some intervals at the edges of the \z and \jt distributions display large deviations from the generator-level distributions after the two unfolding iterations, and are dropped from the final measurement. Only jets with \ptjet $>$ 20\gevc are reported in the final measurement, as jets with 17 $<$ \ptjet $< 20\gevc$ are only used to improve the characterization of bin migrations at low \ptjet in the unfolding procedure.

The measurement procedure is verified with a closure test of the entire analysis chain using independent simulation samples generated with different LHCb magnet polarities. The charged-hadron distributions after passing the entire analysis chain, including purity correction, unfolding, efficiency correction, and normalization, are found to recover the distributions in the generator-level simulation within a few percent, excluding edge intervals. The residual nonclosure is assigned as a systematic uncertainty.

\section{Systematic uncertainties} \label{sec:sysunc}
Systematic uncertainties are considered for the jet and charged-hadron selection, determination of the purity corrections, response matrices, and efficiency corrections, bias induced by the jet SV-tagging algorithm on the charged-hadron distributions, unfolding procedure and analysis chain, and the difference in the detector response between data and simulation. Possible variations of the distributions due to imperfect knowledge of the \bquark- and \cquark-hadron branching fractions in simulation were not studied as a source of systematic uncertainty. The total systematic uncertainty is obtained as the sum in quadrature of the individual components. Table~\ref{tab:sysunc} summarizes the systematic uncertainties for the \z, \jt, and \rhad distributions in beauty and charm jets. 

\begin{table}
\caption{Summary of relative systematic uncertainties on the beauty- and charm-jet charged-hadron distributions. The given ranges include variations over \ptjet. The additional 2\% uncertainty due to the description of the jet identification in simulation, which only affects the normalization factor, is not included in the ranges below. }
\begin{tabular}{ l c c c c c c} 
&  \multicolumn{3}{c}{Beauty jets} & \multicolumn{3}{c}{Charm jets} \\
Source & \z (\%) & \jt (\%) & \rhad (\%) & \z (\%) & \jt (\%) & \rhad (\%)\\
 \hline
 Jet identification & 0--13 & 0--6 & 0--3 & 0--8 & 0--6 & 0--5\\
 Track purity & 2--7 & 1--7 & 2--4 & 1--10 & 1--10 & 3--7 \\
 Track-in-jet reconstruction efficiency & 4--8 & 4--5 & 4--42 & 4--10 & 4--6 & 4--54\\
 Particle misidentification & 1--3 & 1--2 & 0--2 & 0--2 & 0--2 & 0--1 \\
 JES & 0--20 & 1--10 & 1--3 & 0--19 & 0--13 & 0--8\\
 JER & 0--16 & 0--6 & 0--4 & 0--10 & 0--10 & 0--9\\
 SV+\bdtbc tag bias & 1--30 & 1--15 & 1--12 & 1--38 & 1--21 & 1--20\\
 Unfolding & 0--17 & 0--19 & 0--10 & 0--7 & 0--23 & 0--10\\
 Nonclosure & 0--25& 0--20 & 0--7 & 0--14 & 0--14 & 0--5\\
 \hline
 Total (excl.~2\% sys. on normalization) & 6--42 & 6--40 & 6--43 & 6--39 & 7--34 & 7--57\\
\end{tabular}
\label{tab:sysunc}
\end{table}

The choice of selection requirements used to reject fake jets, discussed in Sec.~\ref{sec:jetrecoandeventsel}, can affect the total number of selected jets, and could also bias the charged-hadron distributions as they use some information about the particles within the jet. A systematic uncertainty is assigned by tightening the selection requirements applied to the data, repeating the entire analysis, and taking the ratio of the charged-hadron distributions with the tightened jet selection cuts to the baseline ones. The difference from unity is assigned as a systematic uncertainty in each interval, and is less than 5\% in most intervals and slightly larger at the extremes of the \z and \jt distributions. A second source of systematic uncertainty accounts for the imperfect modeling of the jet selection efficiency in simulation. The tightened jet selection requirements are also applied to the simulation, and the resulting difference in efficiency with respect to the baseline requirements is computed. The same efficiency is computed with data, and the double ratio of the jet selection efficiency in data and simulation is obtained. The difference from unity reaches a maximum of around 2\% in the highest \ptjet interval, and is taken as a systematic uncertainty on the normalization factors \Nbjets and \Ncjets.

Sources of systematic uncertainty for charged-hadron selection include the track purity, the track reconstruction within the jet cone, and the description of the charged-hadron PID in the simulation. The purity of the charged-hadron selection is studied by applying a tighter selection requirement to remove fake tracks~\footnote{Tracks that do not correspond to a charged particle trajectory, but are instead reconstructed from combinations of uncorrelated hits in the tracking detectors.}, repeating the entire analysis, and computing the ratio with respect to the baseline distributions. The deviation from unity is assigned as a systematic uncertainty, and is within 5\% for most intervals and slightly larger at the extremes of the \z and \jt distributions and in the highest \ptjet interval. A systematic uncertainty is applied to the track-reconstruction efficiency in simulation to account for the uncertainty on the material budget of the LHCb detector~\cite{LHCb-DP-2013-002}. In the previous measurement of identified charged-hadron distributions in $Z$-tagged jets, this uncertainty was found to be 1.5\% for pions, 1.3\% for kaons, and 3.3\% for protons~\cite{LHCb-PAPER-2022-013}. Since the current analysis measures unidentified charged-hadron distributions, which are expected to be dominated by pions and kaons, a conservative uncertainty of 2\% is assigned. The angular resolution of the jet cone is also considered as a source of systematic uncertainty, as it can determine whether or not tracks are included within the jet. The jet angular resolution is studied by looking at the $\Delta R$ distribution between matched generator-level and reconstructed jets in simulation, which peaks at around 0.01--0.02 depending on the \ptjet interval. The charged-hadron efficiencies and purities are recomputed after shifting the jet radius requirement within the observed resolution, and the resulting efficiency difference is added in quadrature with the track-reconstruction uncertainty to obtain a total uncertainty on the charged-hadron efficiency and purity. The entire analysis is repeated with the charged-hadron efficiency shifted within its uncertainty, and the envelope of the resulting charged-hadron distributions is taken as a systematic uncertainty in each interval. The uncertainty is a few percent in most intervals, but is larger in the \rhad interval closest to the edge of the jet, which is expected as this is where it is most difficult to determine if a charged hadron is correlated with the jet fragmentation. A systematic uncertainty for the charged-hadron PID is studied by using calibration samples from data to estimate the misidentification probabilities for leptons to be mistakenly identified as charged hadrons and vice versa~\cite{LHCb-DP-2018-001}. The efficiencies and purities are recomputed taking into account the misidentification rates measured in data, and the ratio to the baseline distributions is computed. The difference from unity is assigned as a systematic uncertainty, and is less than 3\% in all intervals.

The JES and JER corrections are also considered as a source of systematic uncertainty, as they modify the reconstructed jet energy and momentum in simulation which in turn influence the \z, \jt, and \rhad distributions used to construct the purities, efficiencies, and response matrices. Each correction is shifted within its uncertainty, and the purities, efficiencies, and response matrices are computed for each shift and used to correct and unfold the data. An envelope of the charged-hadron distributions obtained with the variations is computed. The larger value between the size of the envelope and the maximum deviation from the baseline distribution is assigned as a systematic uncertainty. For most intervals, the resulting systematic uncertainty is a few percent; however, at the extremes of the \z distribution, at large \jt and in the highest \ptjet interval, the uncertainty approaches the 10--20\% level.

As previously mentioned in Sec.~\ref{sec:analysisstrategy}, a systematic uncertainty is included to account for the bias induced by the SV-tagging and \bdtbc algorithms on the \z, \jt, and \rhad distributions. The bias is studied in simulation for beauty and charm jets separately. Reconstructed dijets in simulation are selected with the same requirements as data, with the exception of the SV-tag and \bdtbc requirements, and matched to generator-level dijets. The \z, \jt, and \rhad distributions are computed in three samples of jets: all selected jets, those passing the SV-tagging requirement, and those satisfying the \bdtbc requirement. The generator-level information is used to compute the distributions in order to avoid including bin migration effects, which are corrected separately with the unfolding procedure. Ratios of the SV-tagged and \bdtbc-tagged jets to the total sample are made to identify where the SV-tagging and \bdtbc biases affect the distributions. Since the \bdtbc selection is only applied once a jet has been SV-tagged, the ratio of the \bdtbc- to the SV-tagged jets is also studied to isolate the effect of the \bdtbc classifier. Most of the observed bias is due to the SV-tagging algorithm, and is typically larger for charm than for beauty jets. In the beauty-jet \z distributions, the SV-tagging and \bdtbc requirements preferentially select charged hadrons located at intermediate \z, approximately between 0.04 and 0.3. For most of the intervals, the bias is within 10\%, except at very low \z where the bias reaches 30\% in the highest \ptjet interval. For charm jets, the low-\z behavior is similar to that in beauty jets, but at high \z the bias is localized towards \z values between 0.1 and 0.5, where it can be as large as 38\%, depending on the \ptjet. Beauty hadrons tend to have larger decay multiplicities than charm hadrons, which could explain why the bias in the beauty \z distribution is shifted to lower \z values, and the bias in the charm distributions is more localized and shifted to higher \z values. For the \jt distributions, the bias is generally less than 10\%. It is slightly higher in some intervals in the lowest \ptjet interval, and in the highest \jt interval for charm jets it is of the order of 20\% in some \ptjet intervals. The beauty \rhad distribution bias is less than 10\% in most of the intervals, with the peak of the bias located between 0.05 and 0.25 depending on the \ptjet interval. For charm jets the bias is larger, with the largest deviation of up to 20\% at low \rhad and around 10\% in the large-\rhad intervals. The behavior of the bias in the \rhad distributions is qualitatively consistent with the dead-cone effect. The beauty hadrons are located at larger \rhad compared to charm hadrons because of the larger beauty dead-cone, and the charged-hadron decay products are therefore also close to the position of the beauty hadron. The charm dead-cone is smaller, therefore the bias is concentrated in the small-\rhad intervals. For all distributions, either the size of the bias or the size of the error on the bias, whichever is larger, is taken as a systematic uncertainty.

A systematic uncertainty on the unfolding is determined by swapping the beauty and charm response matrices and unfolding the \bquark-enhanced data sample with the charm response matrix and vice versa. This allows the use of a different but still physically motivated prior for the response matrix, to determine how much the charged-hadron distributions depend on the choice of prior. The ratio of the distributions unfolded with the opposite-flavor response matrices to the baseline distributions is computed, and the deviation from unity is assigned as a systematic uncertainty. The uncertainties are generally of a few percent, but are larger at the extremes of the \z and \jt distributions. A final systematic uncertainty is assigned for the closure of the full analysis chain, as previously discussed in Sec.~\ref{sec:analysisstrategy}. This uncertainty is less than 5$\%$ for most intervals, and is larger in some edge intervals of the \z and \jt distributions that also have large statistical uncertainties. 

\begin{figure}[tb]
  \begin{center}
     \includegraphics[width=0.5\linewidth]{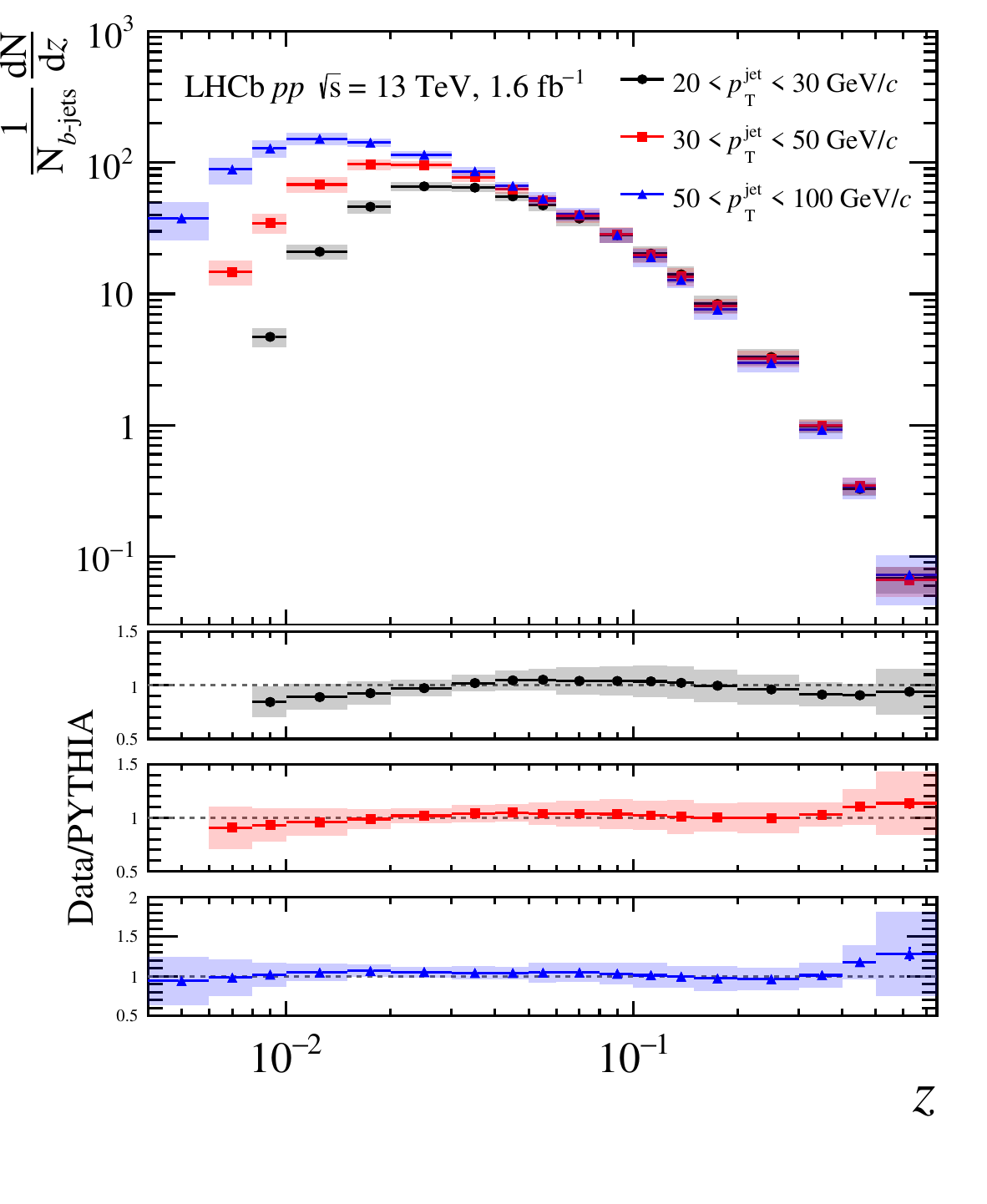}
     \includegraphics[width=0.5\linewidth]{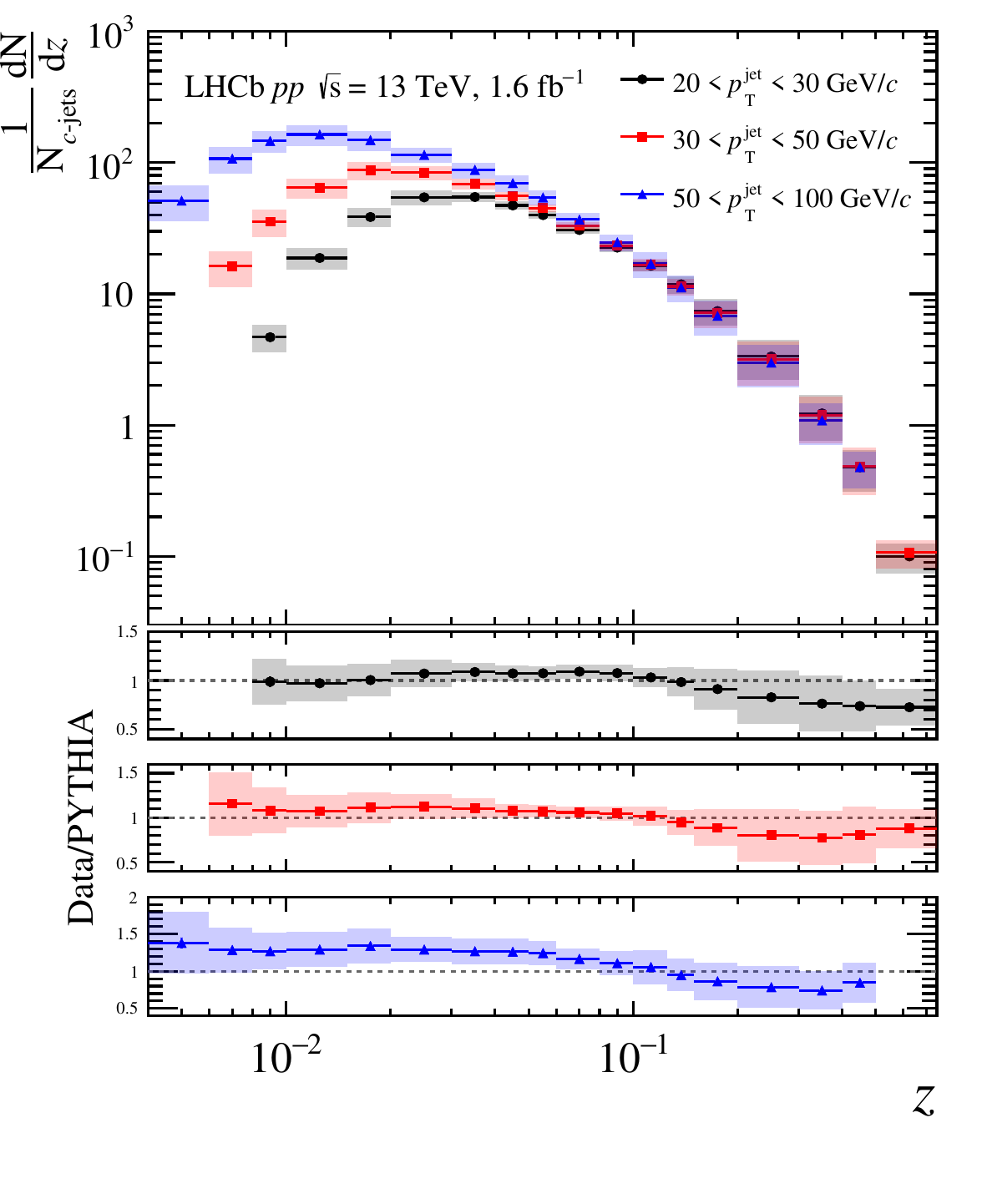}
     \vspace*{-0.5cm}
   \end{center}
   \caption{Charged-hadron distributions measured as a function of the longitudinal momentum fraction \z carried by the charged hadron within the jet for (left) beauty and (right) charm jets. Statistical uncertainties are drawn with error bars but are often too small to be observed, and systematic uncertainties are drawn with shaded boxes. The lower panels show the ratio of the distributions in data to predictions from the \pythia model.}
  \label{fig:zDataMC}
\end{figure}

\section{Results}
 The charged-hadron distributions measured as a function of the longitudinal momentum fraction \z carried by a charged hadron within the jet are shown in Fig.~\ref{fig:zDataMC}.  The distributions in heavy-flavor jets display features qualitatively similar to those measured in light-parton-initiated jets~\cite{LHCb-PAPER-2019-012, LHCb-PAPER-2022-013}. At high \z, the distributions are independent of \ptjet. At low \z, higher-\pt jets probe lower \z values due to the minimum track-momentum requirement being the same for all \ptjet intervals. Higher-\pt jets also have larger particle multiplicities, as shown by the \ptjet-ordering of the distributions. The panels below the distributions show the ratio of the data to the distributions from the \pythia generator~\cite{Sjostrand:2007gs,*Sjostrand:2006za} for each \ptjet interval. The \pythia package uses a string-fragmentation model to simulate hadronization and describes the data well within the uncertainties. The ratios of the charm distributions in data and simulation show a hint of fewer charged hadrons at high \z in data than in the simulation, as the central values are consistently lower than unity. However, when interpreting the charm ratio between data and simulation, it is important to consider that the charm-enhanced sample purity is 70--75\%, therefore some discrepancies may be attributed to beauty-jet contamination.

\begin{figure}[tb]
  \begin{center}
     \includegraphics[width=0.5\linewidth]{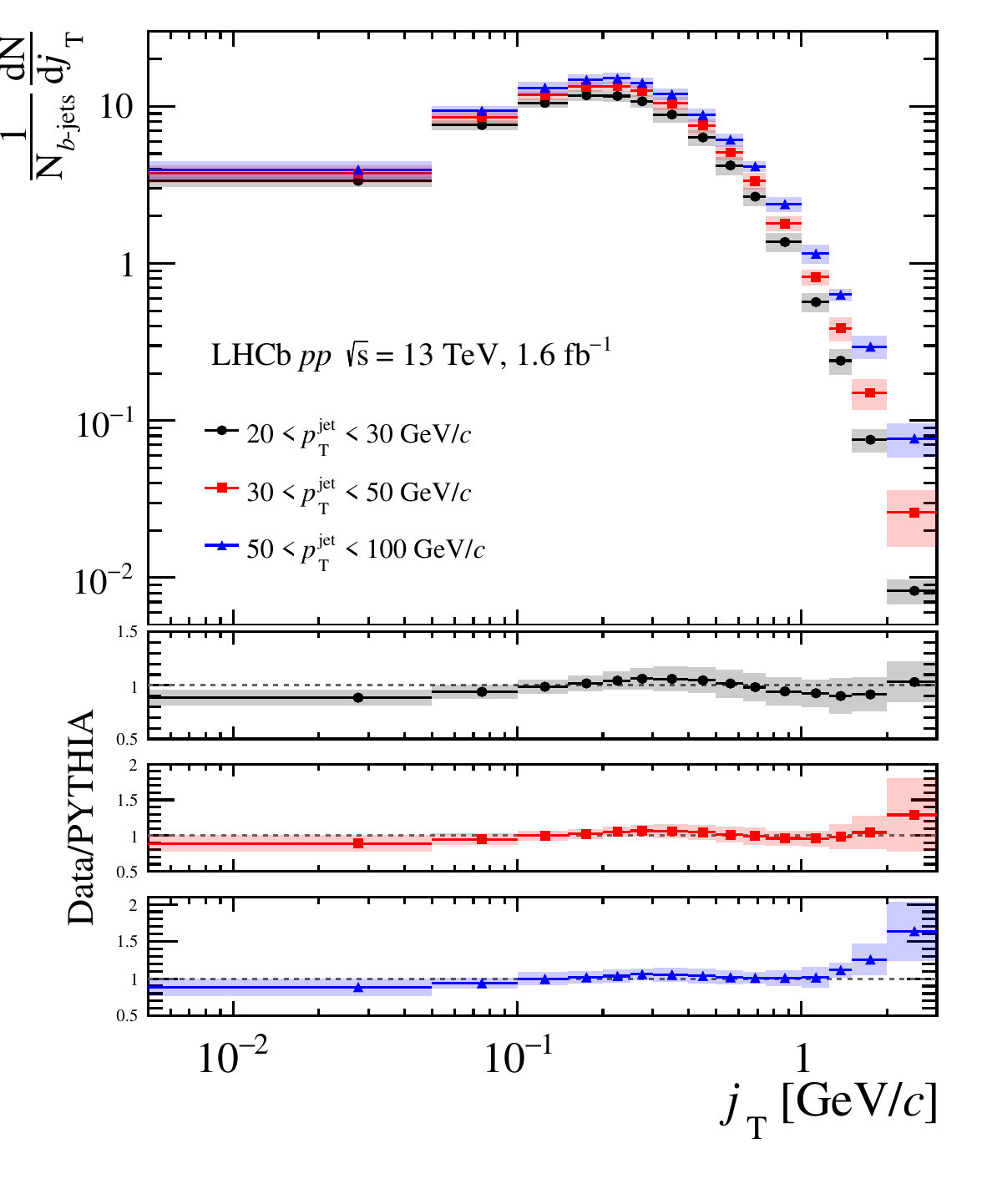}
     \includegraphics[width=0.5\linewidth]{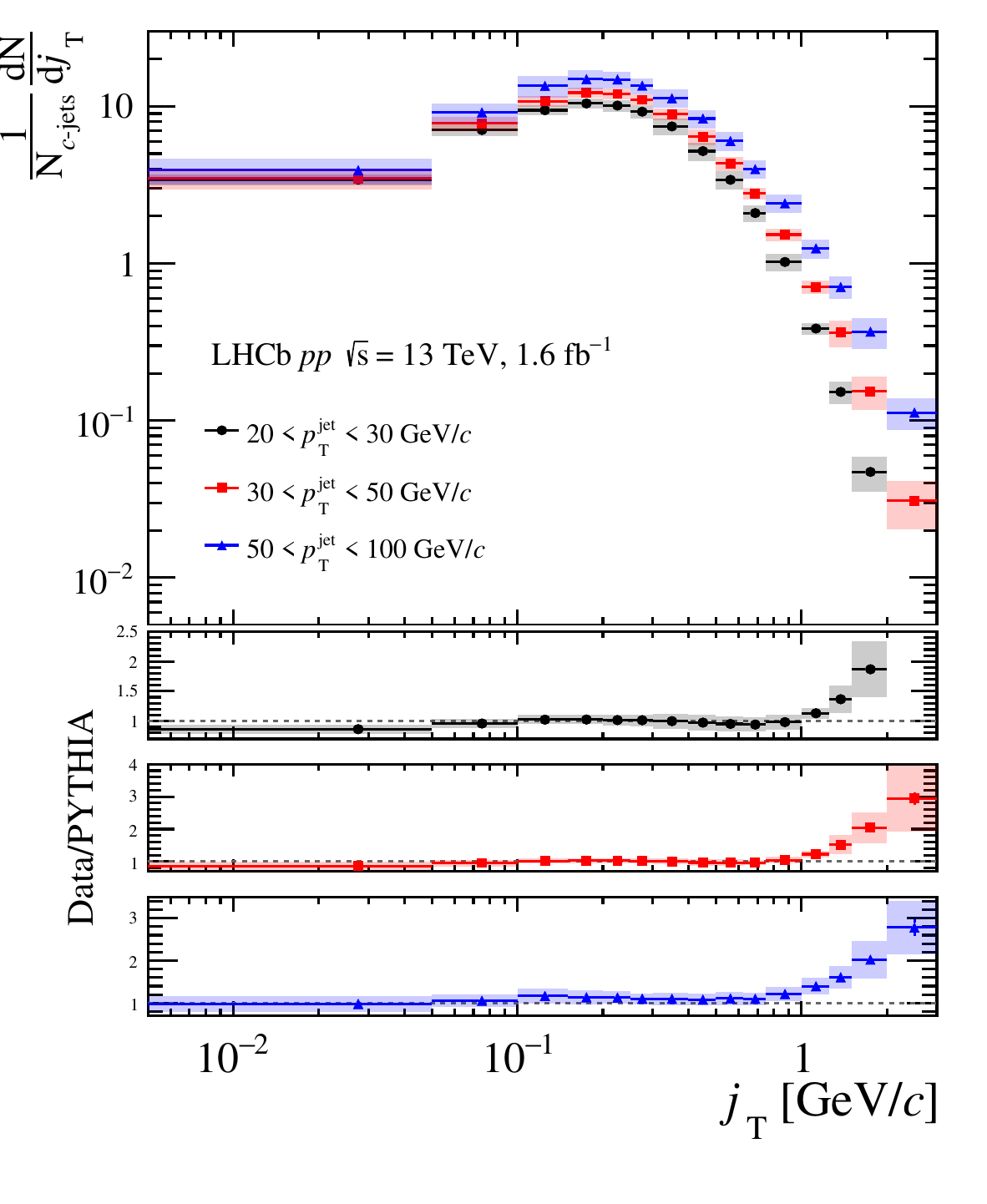}
     \vspace*{-0.5cm}
   \end{center}
   \caption{Charged-hadron distributions measured as a function of the transverse momentum of the charged hadron relative to the jet axis, \jt, for (left) beauty and (right) charm jets. Statistical uncertainties are drawn with error bars but are often too small to be observed, and systematic uncertainties are drawn with shaded boxes. The lower panels show the ratio of the distributions in data to predictions from the \pythia model.}
   \label{fig:jTDataMC}
\end{figure}

The charged-hadron distributions as a function of \jt are shown in Fig.~\ref{fig:jTDataMC}. A similar \ptjet-ordering is observed, with higher-\pt jets containing more charged hadrons overall, and in particular more charged hadrons with large \jt values. \pythia generally describes the data well across most of the \jt range, but appears to predict fewer charged hadrons at large \jt. 
 
\begin{figure}[tb]
  \begin{center}
     \includegraphics[width=0.5\linewidth]{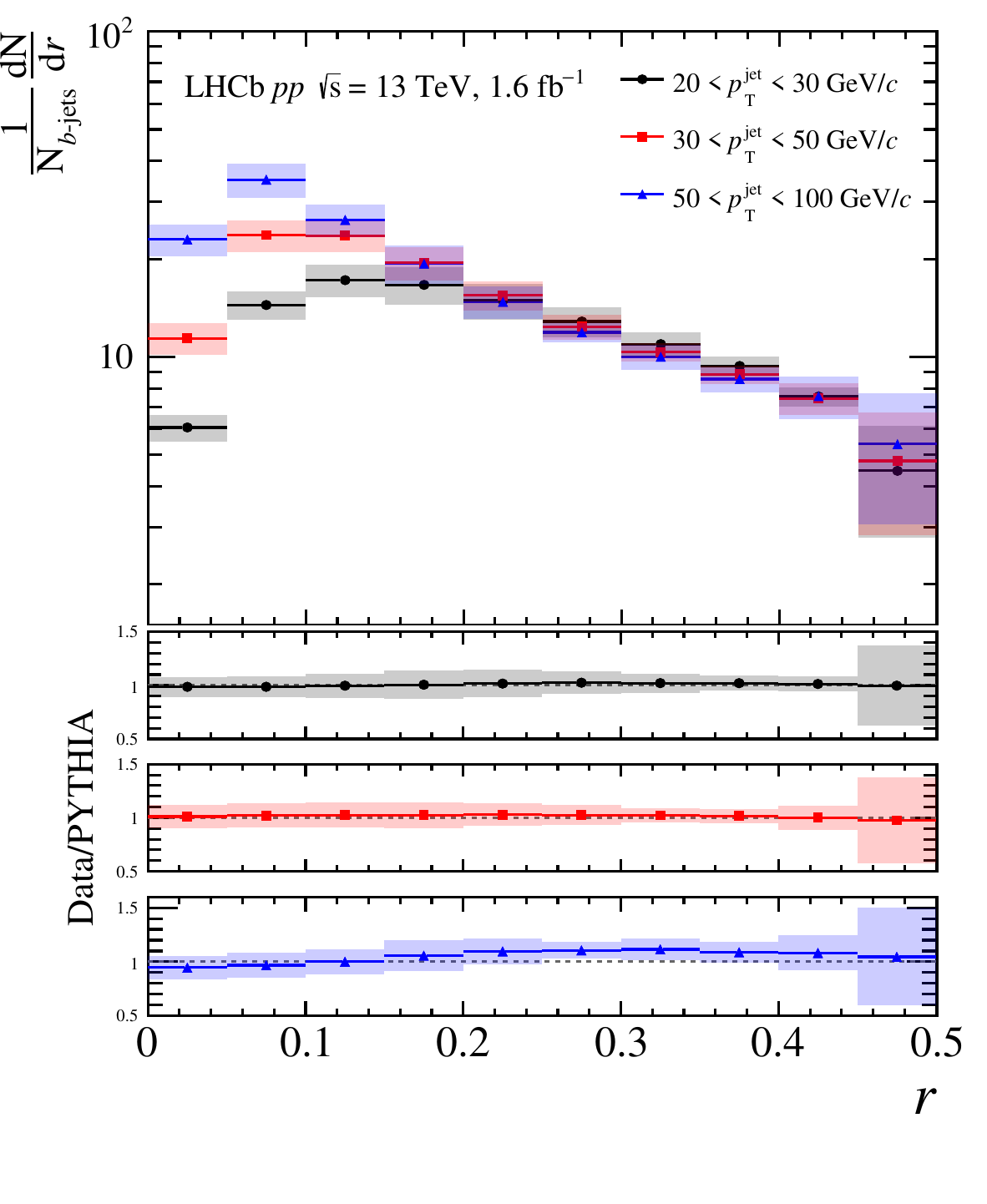}
     \includegraphics[width=0.5\linewidth]{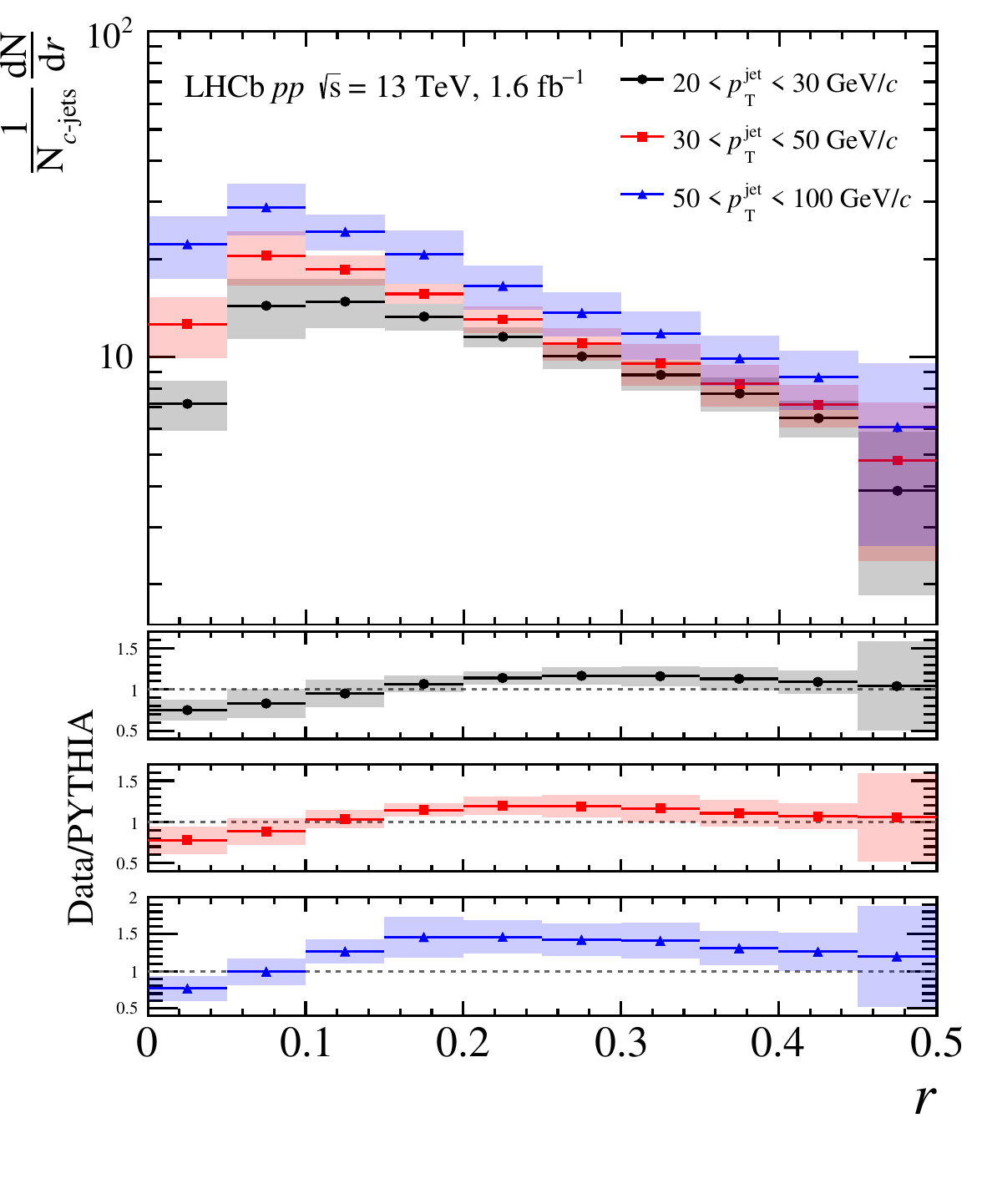}
     \vspace*{-0.5cm}
   \end{center}
   \caption{Charged-hadron distributions measured as a function of the radial position of the charged hadron within the jet cone, \rhad, for (left) beauty and (right) charm jets. Statistical uncertainties are drawn with error bars but are often too small to be observed, and systematic uncertainties are drawn with shaded boxes. The lower panels show the ratio of the distributions in data to predictions from the \pythia model.}
  \label{fig:rDataMC}
\end{figure}

The charged-hadron distributions as a function of \rhad are shown in Fig.~\ref{fig:rDataMC}. The difference between the number of charged hadrons in the lowest \rhad interval (0.00--0.05) and the next-to-lowest \rhad interval (0.05--0.10) decreases with increasing \ptjet for both beauty and charm jets. This is qualitatively consistent with the dynamics expected to arise from the heavy-flavor dead-cone effect, which is larger for beauty than for charm quarks. The resulting \rhad distributions demonstrate that mass-dependent effects in the parton shower impact the hadron distributions within the jets. \pythia nearly perfectly models the beauty-jet \rhad distributions, while it seems to predict more charged hadrons in charm jets at low \rhad and fewer charged hadrons at intermediate \rhad. The slight discrepancy between data and simulation at low \rhad for charm jets could be due to the beauty contamination in the charm-enhanced sample.

\begin{figure}[tb]
  \begin{center}
     \includegraphics[width=0.5\linewidth]{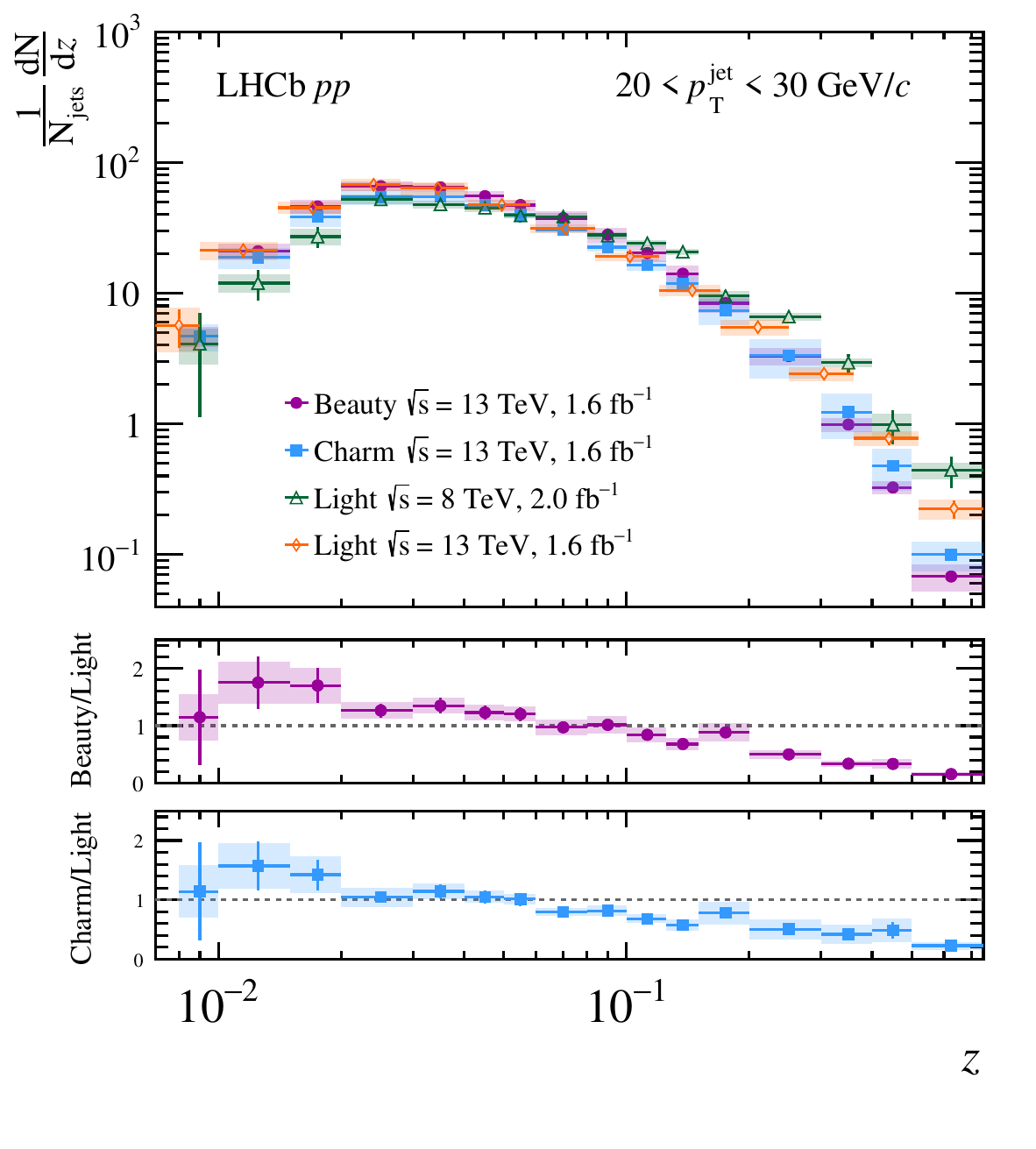}
     \includegraphics[width=0.5\linewidth]{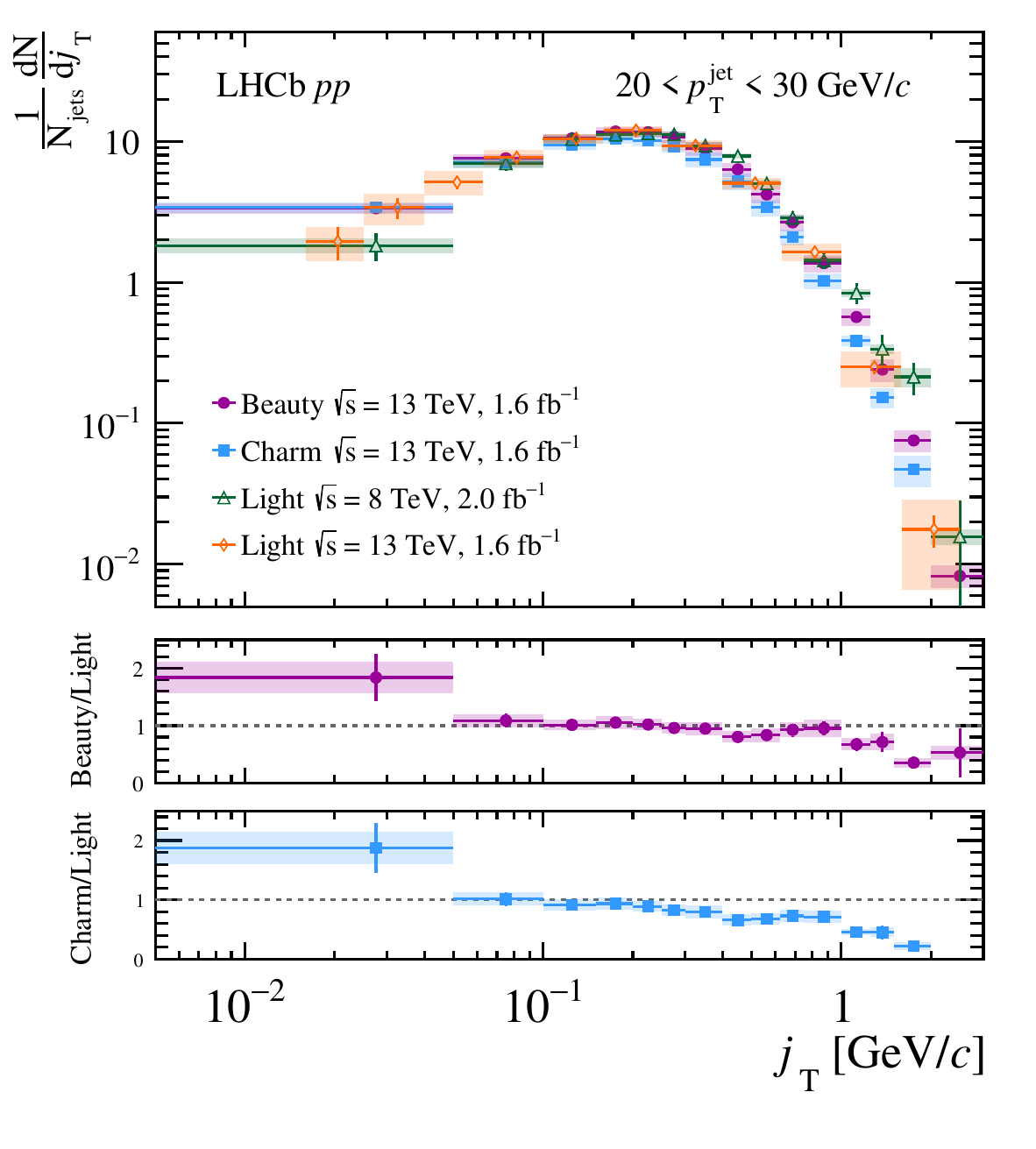}
     \vspace*{-0.5cm}
   \end{center}
   \caption{Charged-hadron distributions measured as a function of (left) \z and (right) \jt in beauty and charm jets, compared to the distributions measured in $Z$-tagged jets (denoted as ``Light'' in the figure) at \mbox{$\sqs = 8\tev$}~\cite{LHCb-PAPER-2019-012} and \mbox{$\sqs = 13\tev$}~\cite{LHCb-PAPER-2022-013}, in the lowest \ptjet interval. The lower panels show the ratios of the distributions in heavy-flavor to light-parton-initiated jets, using the \mbox{$\sqs = 8\tev$} $Z$-tagged jet distributions as they used the same binning as the present measurement.}
  \label{fig:compZjets}
\end{figure}

The charged-hadron distributions measured in beauty and charm jets are compared to previous LHCb measurements in $Z$-tagged jets~\cite{LHCb-PAPER-2019-012, LHCb-PAPER-2022-013}, which primarily probe light-quark-initiated jets, to study differences between light- and heavy-quark hadronization. In the forward acceptance of the LHCb detector, the dominant process to produce a \Z boson and a jet is quark-gluon scattering between a high-momentum quark and a low-momentum gluon, which results in an enhanced fraction of jets initiated by light quarks. Figure~\ref{fig:compZjets} shows the measured \z and \jt distributions in the lowest \ptjet interval in beauty and charm jets from the present analysis, along with the distributions measured in the same \ptjet interval in $Z$-tagged jets in $pp$ collisions at center-of-mass energies of 8 and 13\tev. The $Z$-tagged jet distributions at \mbox{$\sqs = 13\tev$} used a slightly different binning for \z and \jt, motivated by additional measurements of identified charged-hadron distributions. When accounting for the different binning, the dependence of the distributions on the center-of-mass energy is a small effect. Therefore, to obtain general observations about the differences between light- and heavy-quark hadronization, the ratios of the beauty and charm distributions at \mbox{$\sqs = 13\tev$} to the $Z$-tagged jet distributions at \mbox{$\sqs = 8\tev$}, which used the same binning, are computed and shown in the panels below the distributions. The \z ratios indicate that beauty and charm jets tend to have fewer charged hadrons at high \z compared to light-parton-initiated jets. Previous measurements of single heavy-flavor hadron fragmentation functions, in jets and in other collision systems, have shown that the heavy-flavor hadron has a large \z, \ie it carries a large fraction of the initial momentum of the heavy quark~\cite{DELPHI:2011aa, OPAL:2002plk, ALEPH:2001pfo, SLD:2002poq, ALEPH:1999syy, OPAL:1994cct, ARGUS:1991vjh, ATLAS:2021agf, ATLAS:2011chi, ALICE:2019cbr, ALICE:2022mur, ALICE:2023jgm}. In this measurement, although the heavy-flavor hadron is not reconstructed, its decay products are still present in the jet and would also be expected to be located at high \z values. Since the heavy-flavor hadron carries most of the momentum of the heavy-flavor quark, there is less momentum available for other hadrons within the jet and these are therefore expected to be located at lower \z. Light-parton-initiated jets do not have the same limitations, and the initial quark momentum can therefore be distributed differently among charged hadrons in the jets. The \jt ratios indicate that heavy-flavor jets also tend to have fewer charged hadrons at large \jt than light-parton-initiated jets, with the effect seeming to be slightly more pronounced for charm jets than for beauty jets. 

\begin{figure}[tb]
  \begin{center}
     \includegraphics[width=0.5\linewidth]{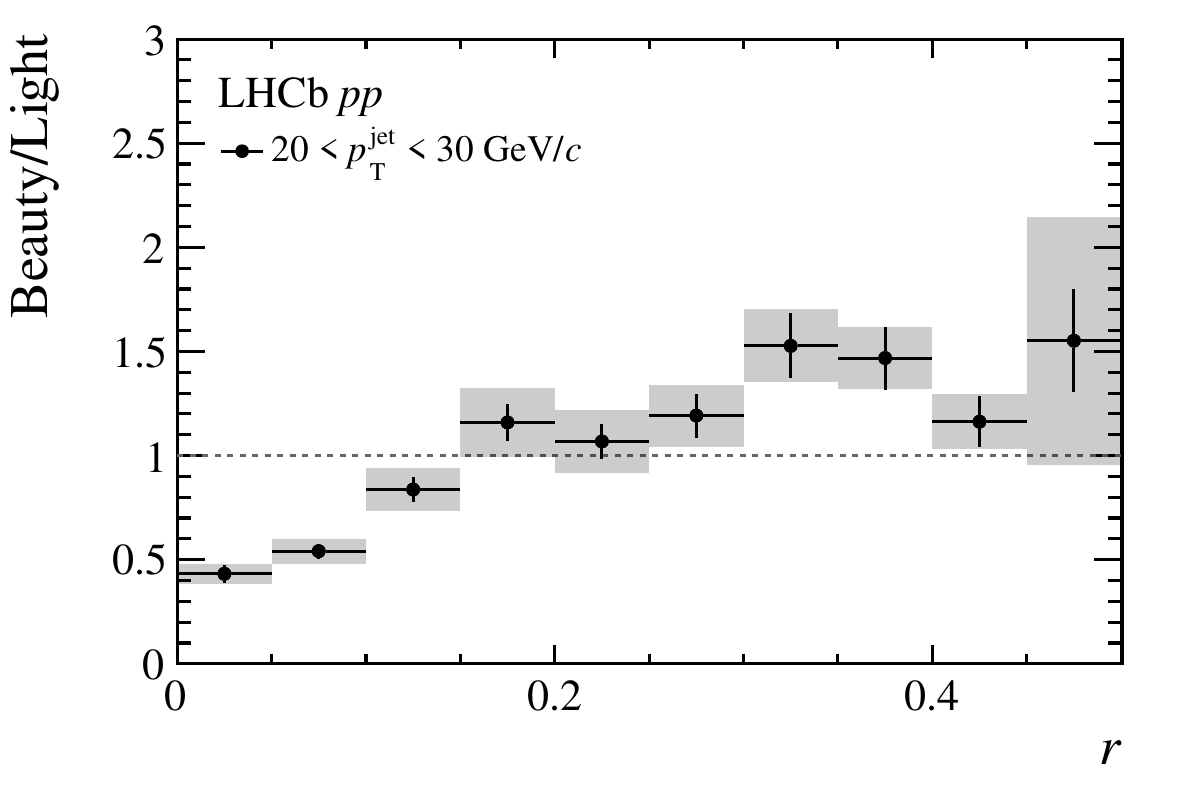}
     \vspace*{-0.5cm}
   \end{center}
   \caption{Ratio of the charged-hadron distributions as a function of \rhad in beauty jets at \mbox{$\sqs = 13\tev$} to $Z$-tagged jets at \mbox{$\sqs = 8\tev$}~\cite{LHCb-PAPER-2019-012}, in the lowest \ptjet interval.}
  \label{fig:compZjetsr}
\end{figure}

Figure~\ref{fig:compZjetsr} shows the ratio of the charged-hadron distributions in beauty and $Z$-tagged jets as a function of the radial position \rhad. A depletion of charged hadrons is observed at small \rhad in beauty jets compared to light-parton-initiated jets, which is qualitatively consistent with an indirect observation of the dead-cone effect~\cite{Battaglia:2004coa}.

Additional comparisons between beauty and charm jets as well as between heavy-flavor and light-quark-initiated jets can be found in Appendices~\ref{appendix:MoreBeautyCharm} and~\ref{appendix:MoreHFLight}, respectively.  Tables of \jt and \rhad distributions measured with $\z>0.01$ are provided in Appendix~\ref{appendix:withzcut}.

\section{Summary}
Charged-hadron distributions are measured in samples of beauty and charm jets using $pp$ collision data collected by the LHCb experiment in 2016. The measured observables, comprising the longitudinal momentum fraction \z carried by charged hadrons in the jets, the transverse momentum \jt of the charged hadrons relative to the jet axis, and the radial position \rhad of the charged hadrons in the jets, together provide an image of heavy-quark hadronization in momentum and position space. The distributions quantify the fragmentation of a single beauty or charm quark into multiple charged hadrons, complementary to similar measurements probing the fragmentation of a heavy quark into a single heavy-flavor hadron. The measurements also provide an additional constraint for the extraction of collinear and transverse-momentum-dependent heavy-flavor fragmentation functions. Comparisons are made to generator-level \pythia distributions, and \pythia is found to generally describe the beauty and charm data well within experimental uncertainties. The distributions are also compared to those previously measured in $Z$-tagged jets at LHCb, allowing for a direct comparison between beauty, charm, and light-quark hadronization. Differences between heavy- and light-quark hadronization are observed and are consistent with interpretations based on previous measurements of single-hadron fragmentation functions and the heavy-flavor dead-cone effect.

\newpage
\section*{Acknowledgements}
%
%
\noindent We express our gratitude to our colleagues in the CERN
accelerator departments for the excellent performance of the LHC. We
thank the technical and administrative staff at the LHCb
institutes.
We acknowledge support from CERN and from the national agencies:
ARC (Australia);
CAPES, CNPq, FAPERJ and FINEP (Brazil); 
MOST and NSFC (China); 
CNRS/IN2P3 (France); 
BMFTR, DFG and MPG (Germany);
INFN (Italy); 
NWO (Netherlands); 
MNiSW and NCN (Poland); 
MCID/IFA (Romania); 
MICIU and AEI (Spain);
SNSF and SER (Switzerland); 
NASU (Ukraine); 
STFC (United Kingdom); 
DOE NP and NSF (USA).
We acknowledge the computing resources that are provided by ARDC (Australia), 
CBPF (Brazil),
CERN, 
IHEP and LZU (China),
IN2P3 (France), 
KIT and DESY (Germany), 
INFN (Italy), 
SURF (Netherlands),
Polish WLCG (Poland),
IFIN-HH (Romania), 
PIC (Spain), CSCS (Switzerland), 
and GridPP (United Kingdom).
We are indebted to the communities behind the multiple open-source
software packages on which we depend.
Individual groups or members have received support from
Key Research Program of Frontier Sciences of CAS, CAS PIFI, CAS CCEPP, 
Fundamental Research Funds for the Central Universities,  and Sci.\ \& Tech.\ Program of Guangzhou (China);
Minciencias (Colombia);
EPLANET, Marie Sk\l{}odowska-Curie Actions, ERC and NextGenerationEU (European Union);
A*MIDEX, ANR, IPhU and Labex P2IO, and R\'{e}gion Auvergne-Rh\^{o}ne-Alpes (France);
Alexander-von-Humboldt Foundation (Germany);
ICSC (Italy); 
Severo Ochoa and Mar\'ia de Maeztu Units of Excellence, GVA, XuntaGal, GENCAT, InTalent-Inditex and Prog.~Atracci\'on Talento CM (Spain);
SRC (Sweden);
the Leverhulme Trust, the Royal Society and UKRI (United Kingdom).

\newpage

\appendix

\section{Ratios of distributions in beauty and charm jets}

\begin{figure}[h!]
    \begin{center}
        \includegraphics[width=0.45\linewidth]{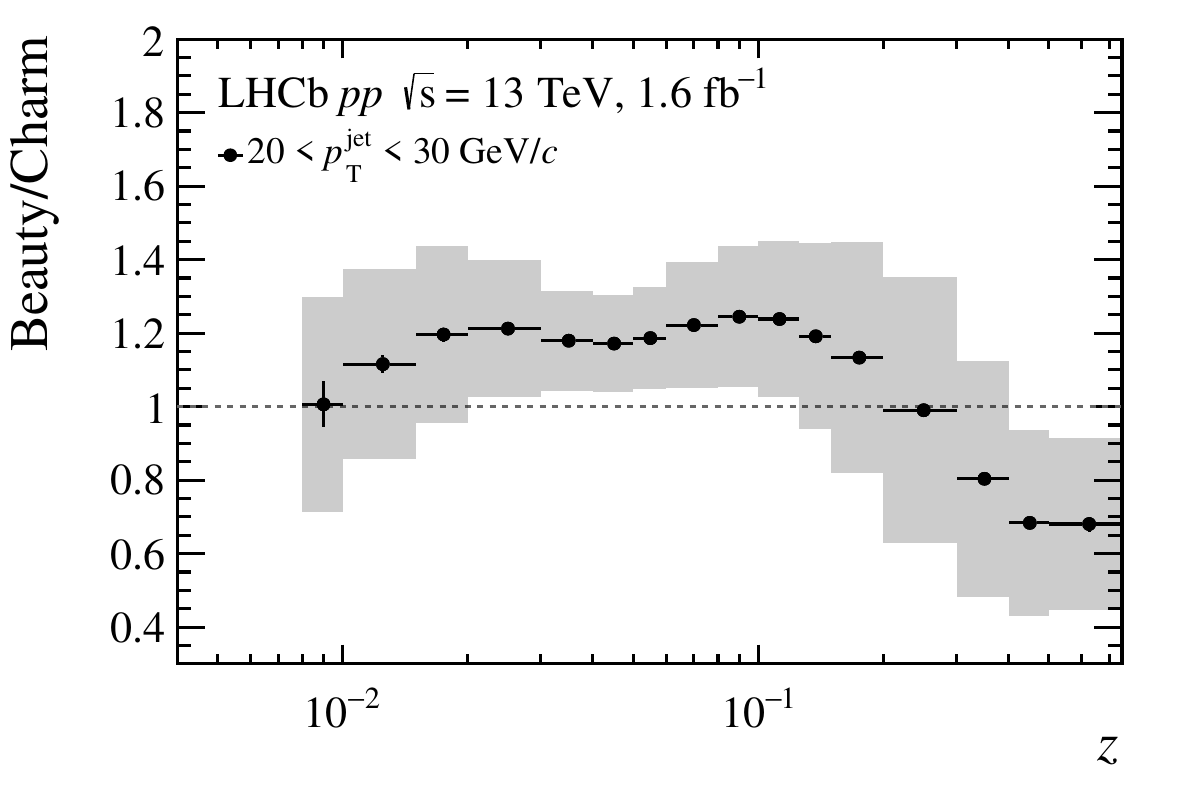}
        \includegraphics[width=0.45\linewidth]{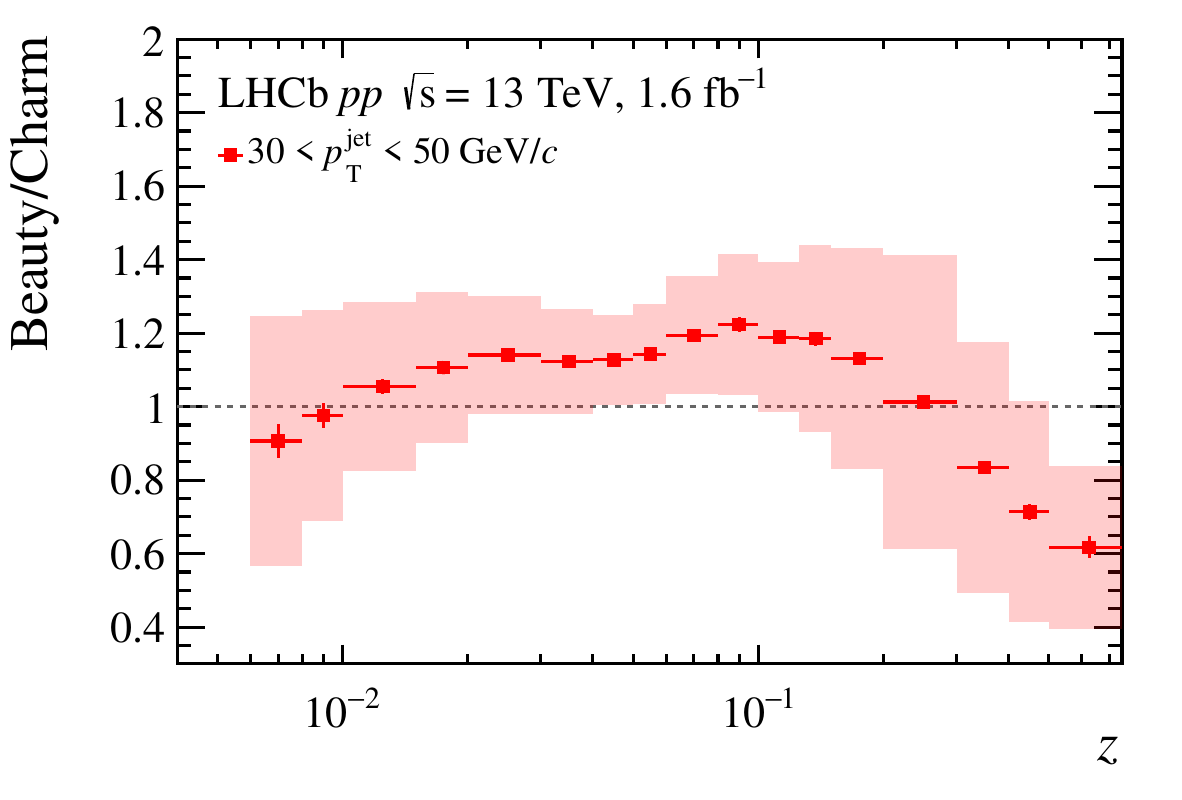}
        \includegraphics[width=0.45\linewidth]{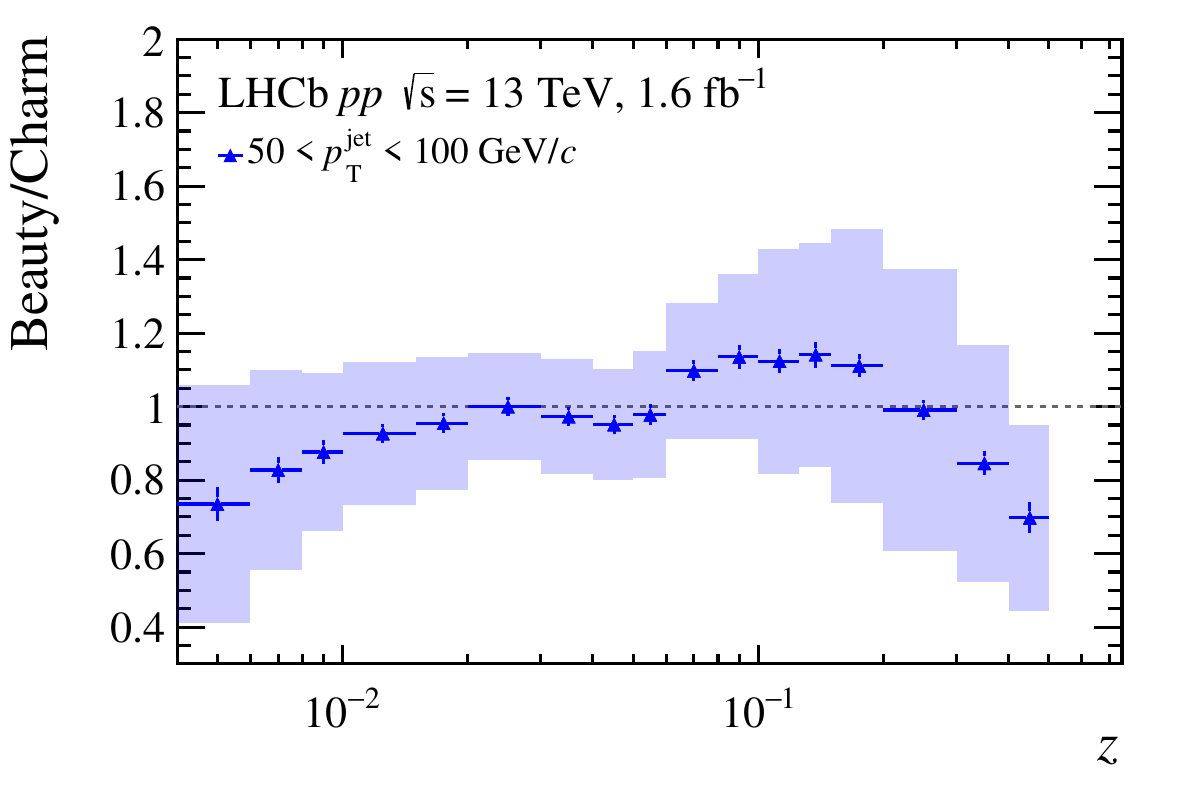}
        \vspace*{-0.5cm}
    \end{center}
    \caption{Ratio of the measured \bquark- to \cquark-jet \z distributions. Statistical uncertainties are indicated with bars and systematic uncertainties with shaded boxes.}
    \label{fig:BeautyCharmz}
\end{figure}

\begin{figure}[h!]
    \begin{center}
        \includegraphics[width=0.44\linewidth]{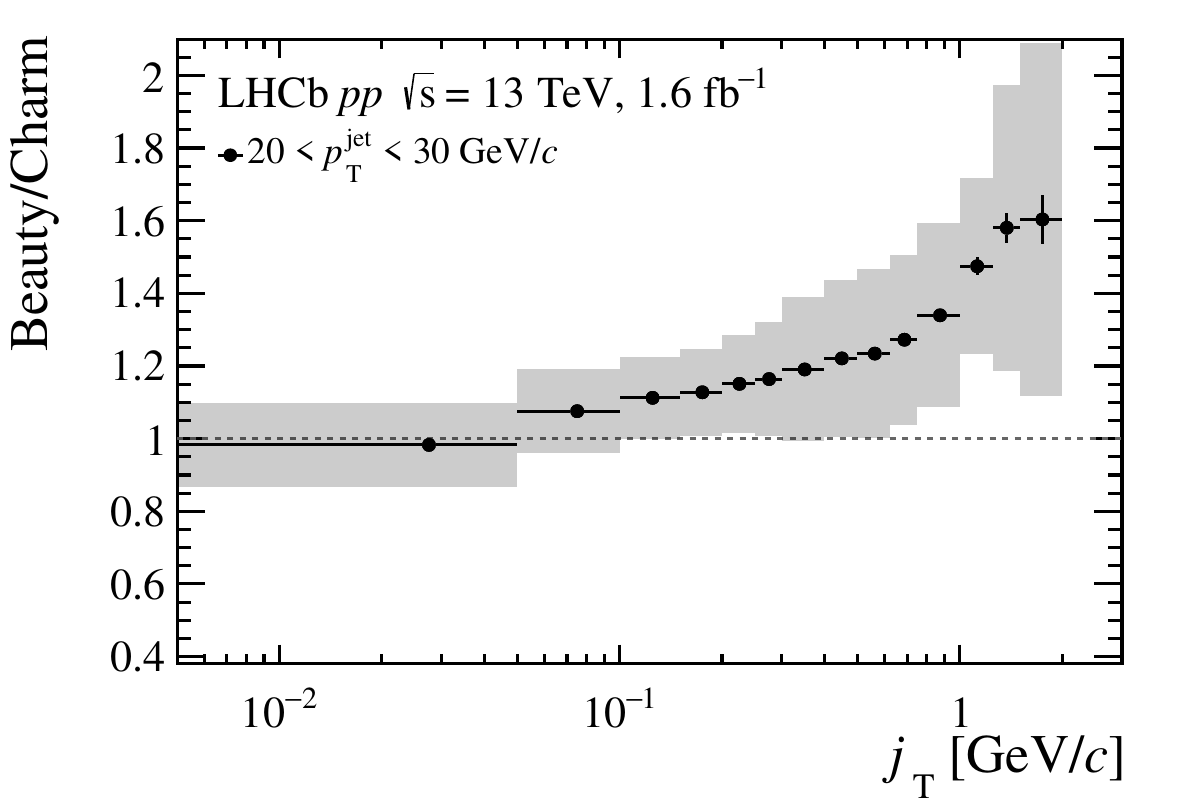}
        \includegraphics[width=0.44\linewidth]{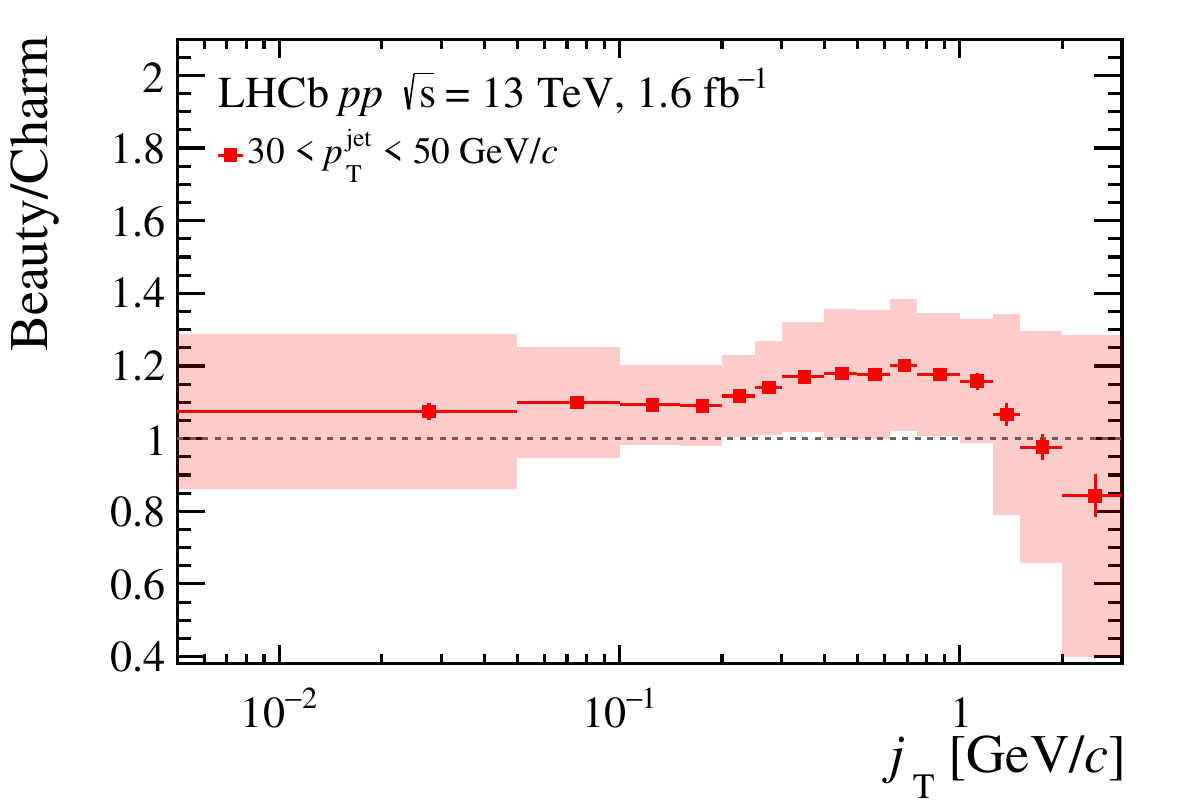}
        \includegraphics[width=0.44\linewidth]{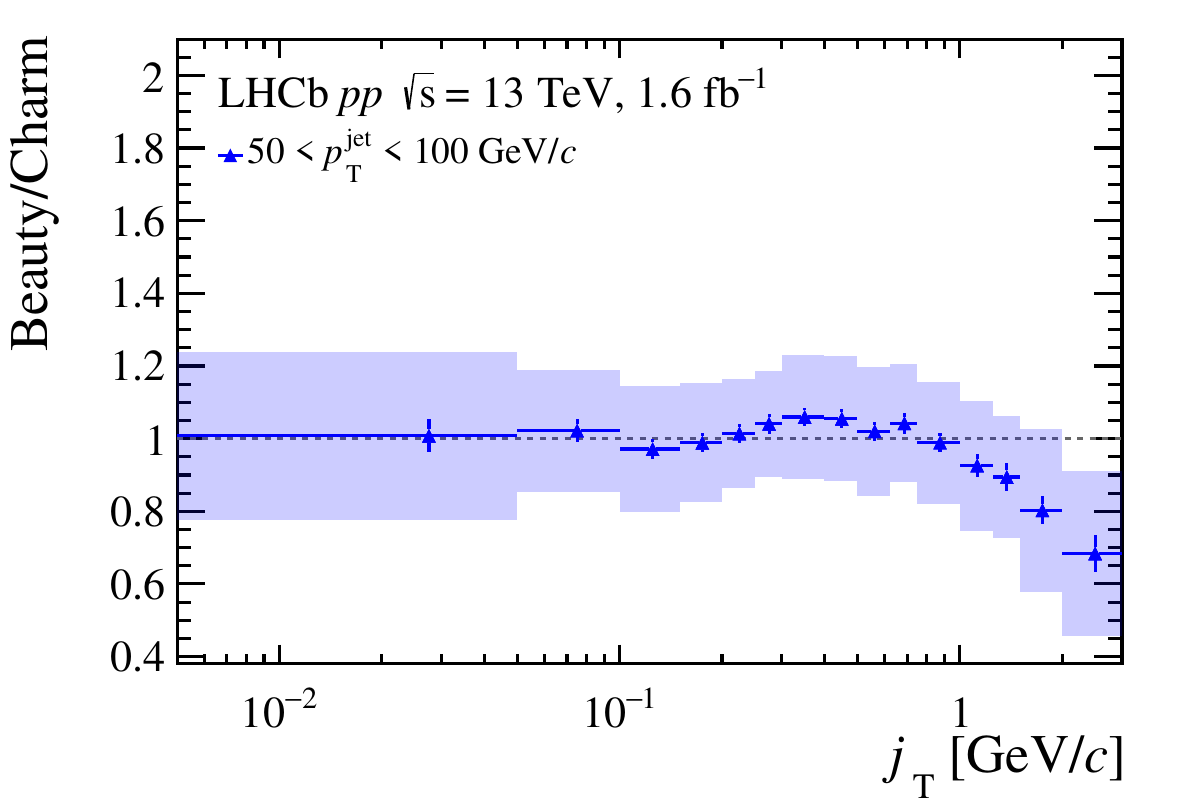}
        \vspace*{-0.5cm}
    \end{center}
    \caption{Ratio of the measured \bquark- to \cquark-jet \jt distributions. Statistical uncertainties are indicated with bars and systematic uncertainties with shaded boxes.}
    \label{fig:BeautyCharmjT}
\end{figure}

\begin{figure}[h!]
    \begin{center}
        \includegraphics[width=0.44\linewidth]{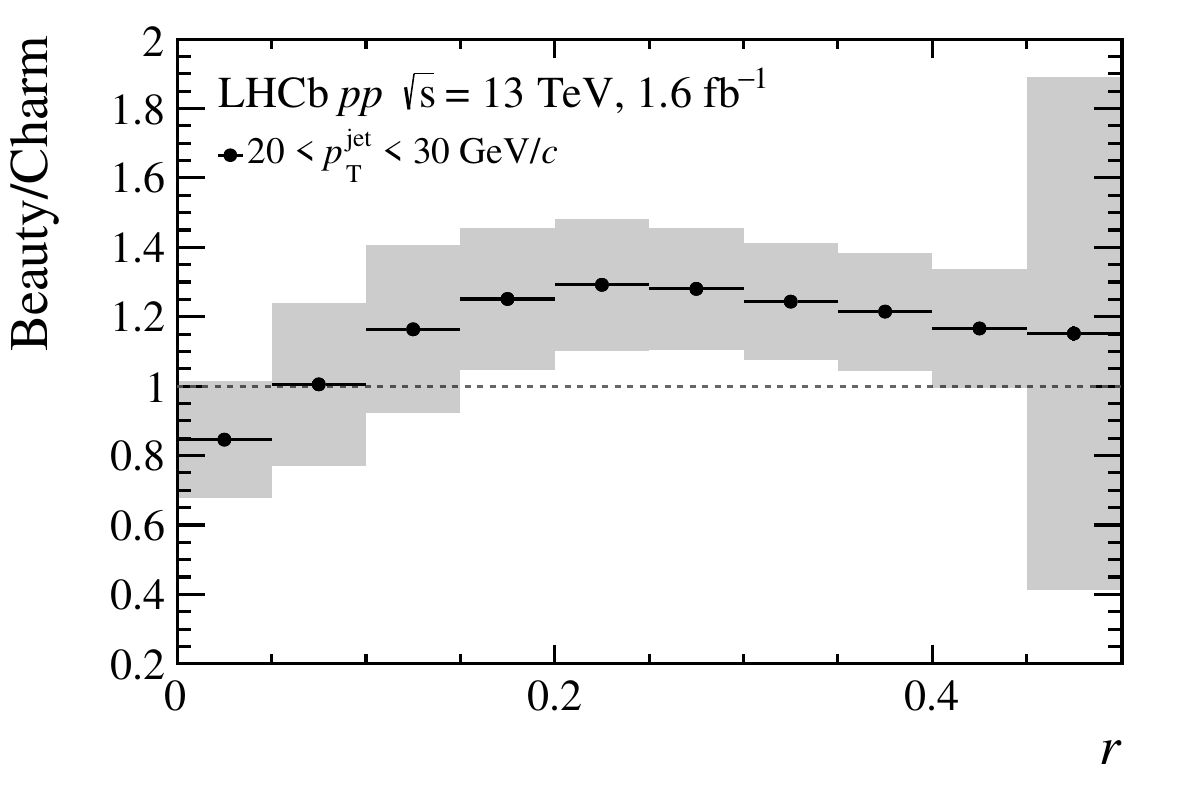}
        \includegraphics[width=0.44\linewidth]{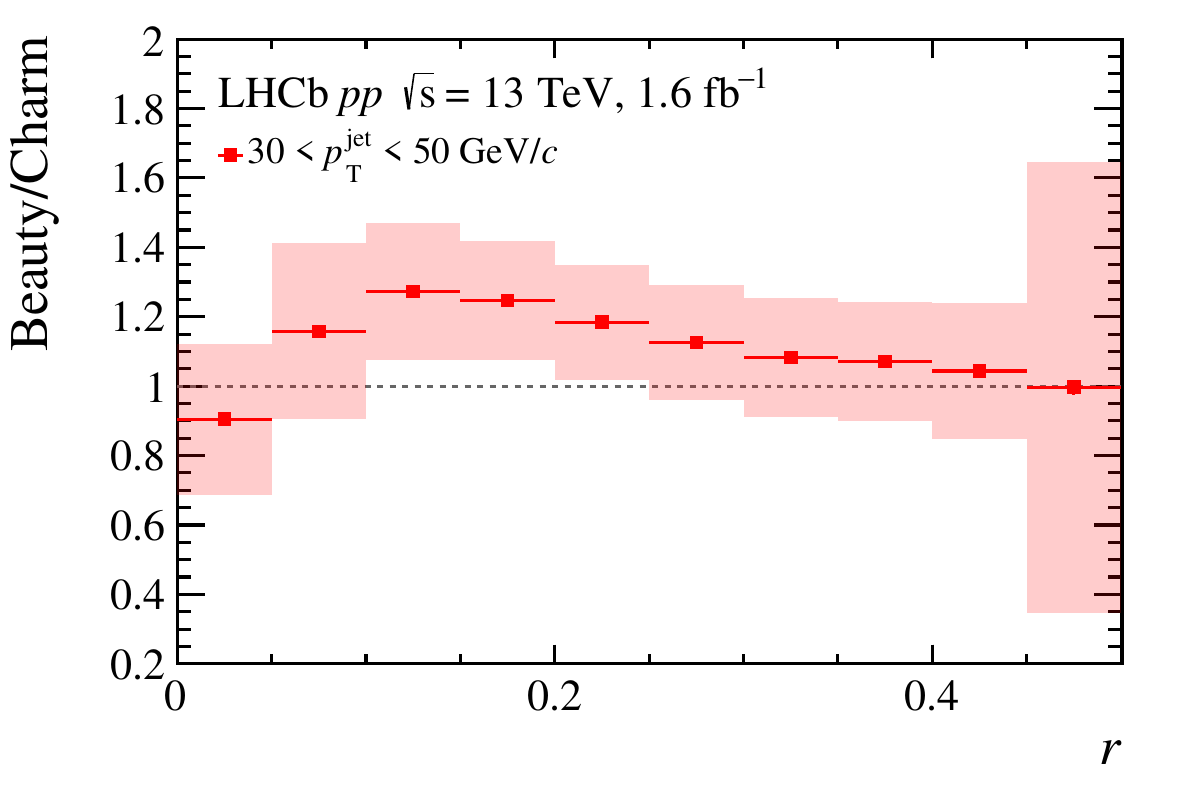}
        \includegraphics[width=0.44\linewidth]{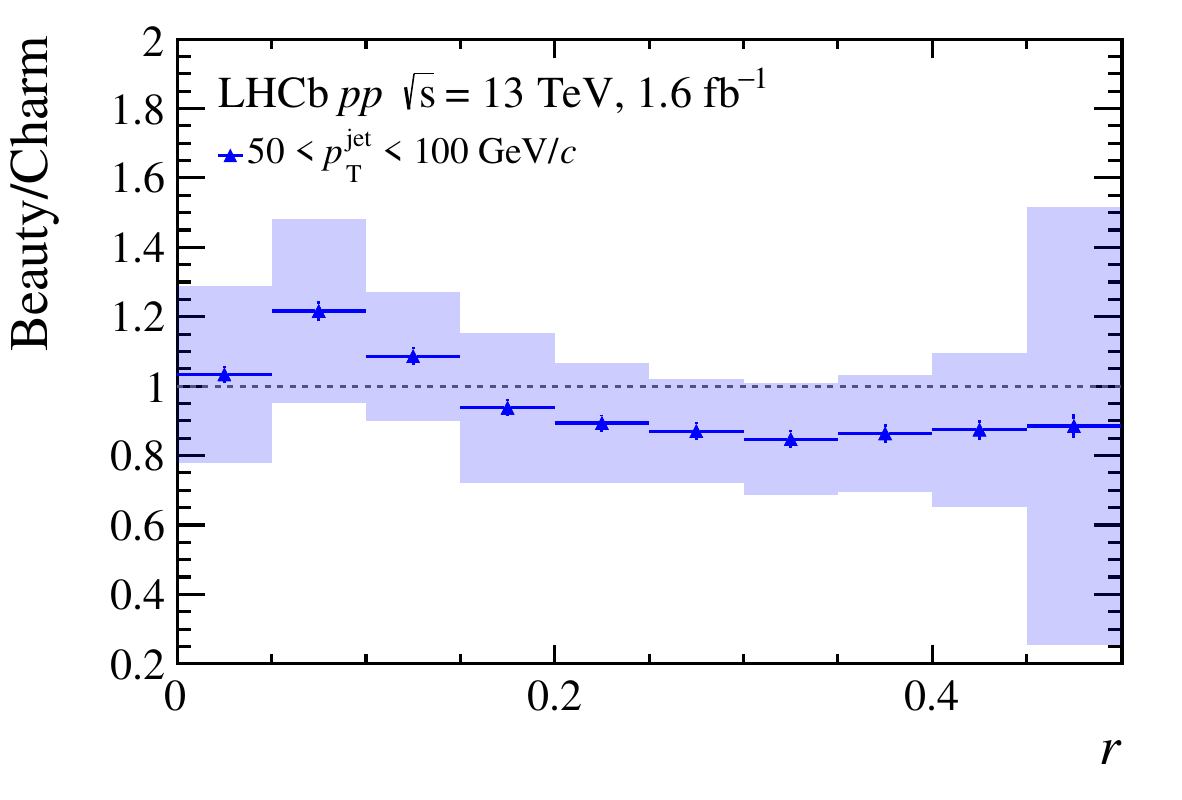}
        \vspace*{-0.5cm}
    \end{center}
    \caption{Ratio of the measured \bquark- to \cquark-jet \rhad distributions. Statistical uncertainties are indicated with bars and systematic uncertainties with shaded boxes.}
    \label{fig:BeautyCharmr}
\end{figure}

This section contains ratios of the charged-hadron distributions measured in beauty jets with respect to charm jets. Figure \ref{fig:BeautyCharmz} shows the ratio as a function of \z, Fig.~\ref{fig:BeautyCharmjT} as a function of \jt and Fig.~\ref{fig:BeautyCharmr} as a function of \rhad.

\label{appendix:MoreBeautyCharm}

\clearpage
\newpage
\section{Additional comparisons to \boldmath{$Z$}-tagged jets}
\label{appendix:MoreHFLight}

\begin{figure}[h!]
  \begin{center}
     \includegraphics[width=0.5\linewidth]{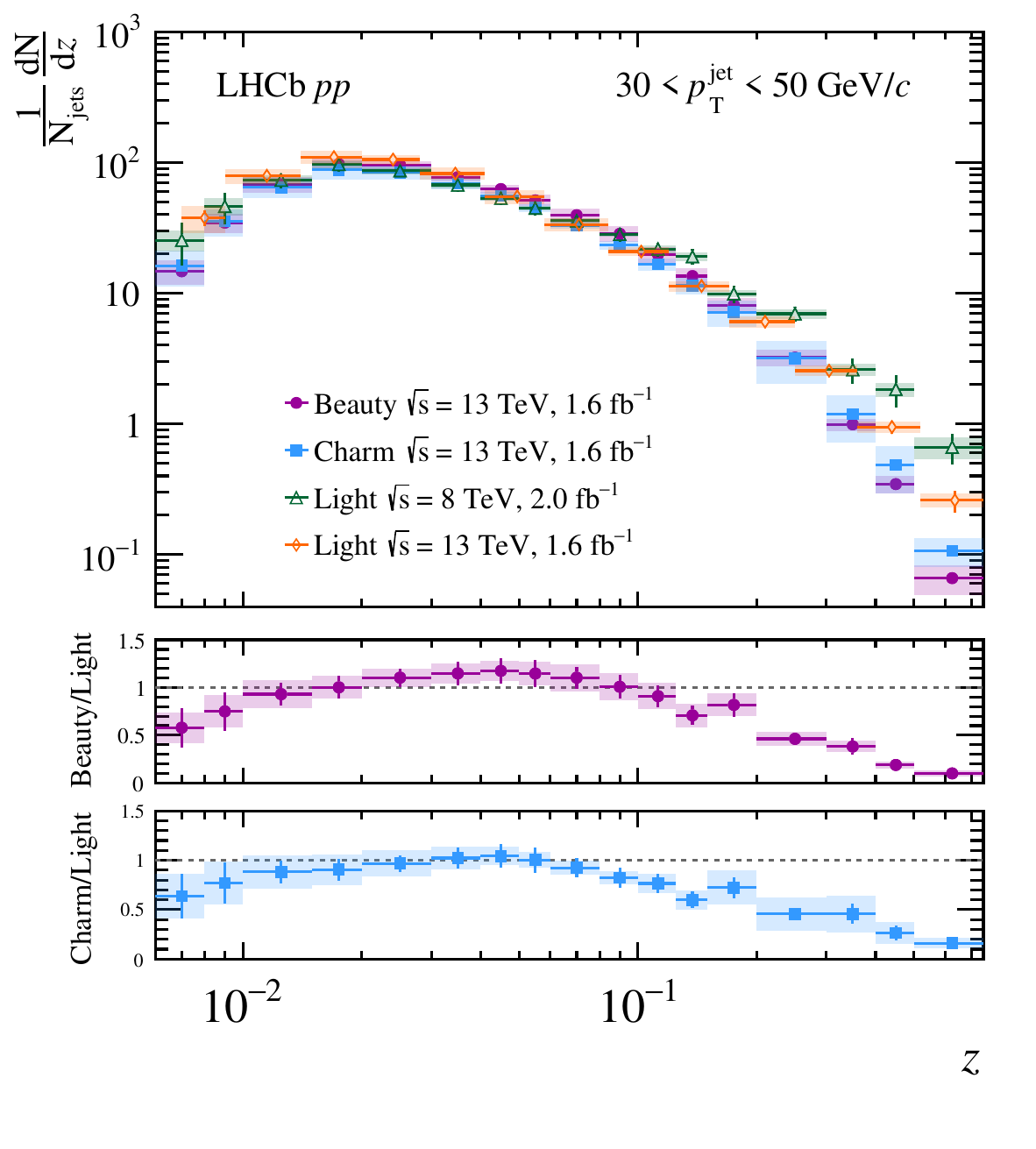}
     \includegraphics[width=0.5\linewidth]{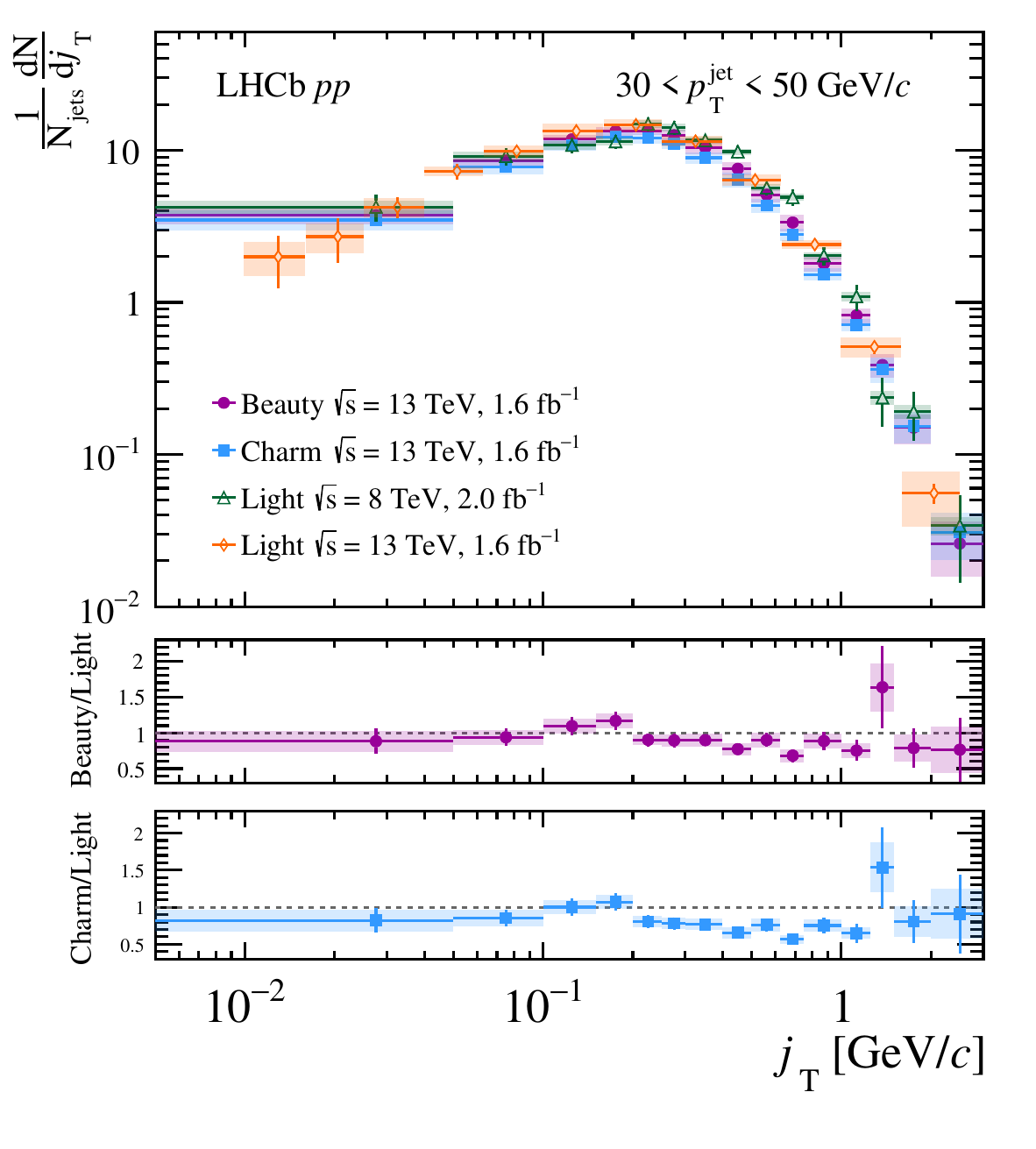}
     \vspace*{-0.5cm}
   \end{center}
   \caption{\small Charged-hadron distributions measured as a function of (left) \z and (right) \jt in beauty and charm jets, compared to those from $Z$-tagged jets (denoted as ``Light'' in the figure) at \mbox{$\sqs=8\tev$}~\cite{LHCb-PAPER-2019-012} and \mbox{$\sqs=13\tev$}~\cite{LHCb-PAPER-2022-013}, for 30 $<$ \ptjet $<$ 50\gevc. The lower panels show the ratios of the distributions in heavy-flavor to light-parton-initiated jets, using the \mbox{$\sqs=8\tev$} measurement that shares the same binning as this analysis. Statistical uncertainties are indicated with bars and systematic uncertainties with shaded boxes.}
  \label{fig:compZjets3050}
\end{figure}

\begin{figure}[h!]
  \begin{center}
     \includegraphics[width=0.5\linewidth]{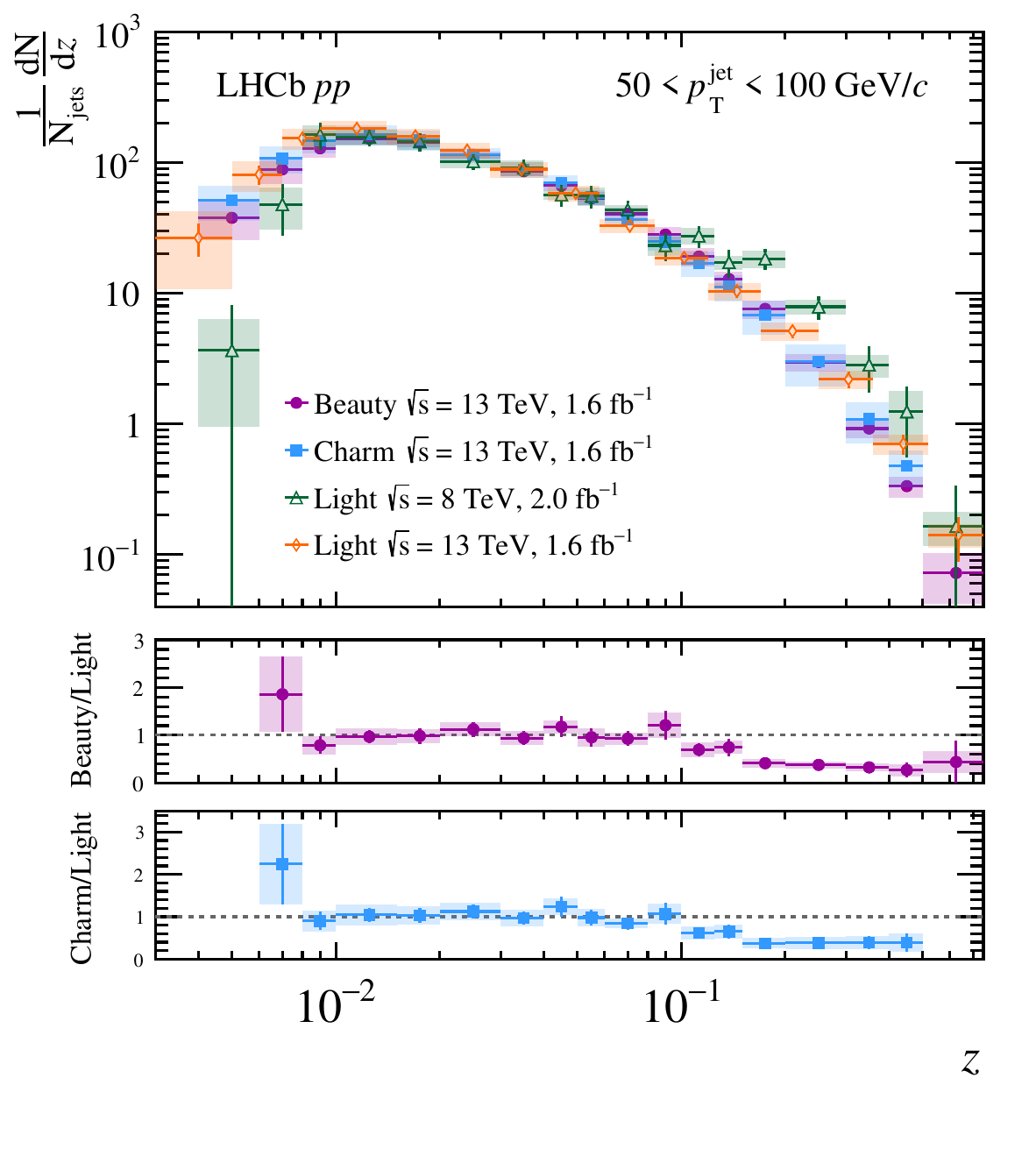}
     \includegraphics[width=0.5\linewidth]{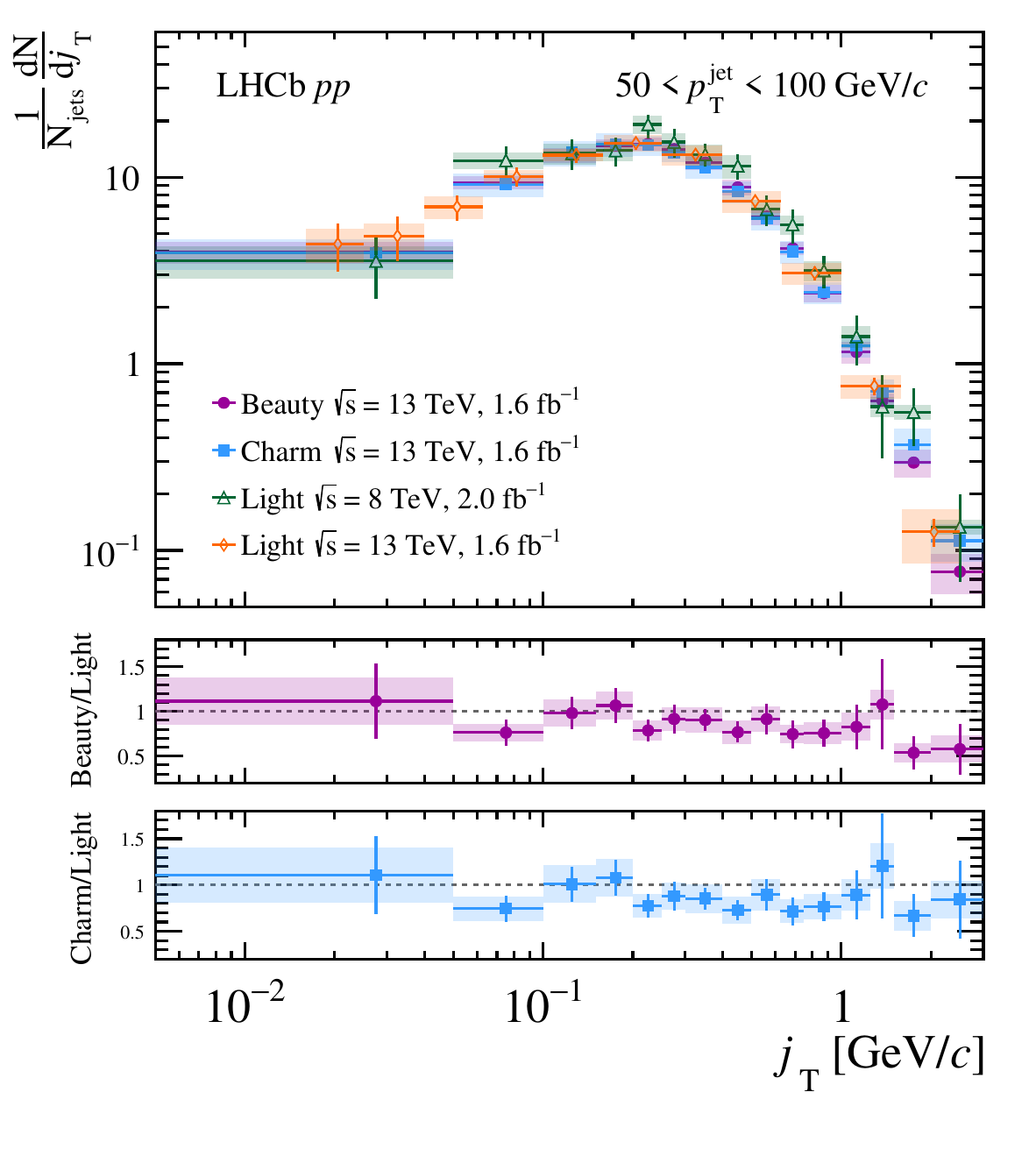}
     \vspace*{-0.5cm}
   \end{center}
   \caption{Charged-hadron distributions measured as a function of (left) \z and (right) \jt in beauty and charm jets, compared to those from $Z$-tagged jets (denoted as ``Light'' in the figure) at \mbox{$\sqs= 8\tev$}~\cite{LHCb-PAPER-2019-012} and \mbox{$\sqs = 13\tev$}~\cite{LHCb-PAPER-2022-013}, for 50 $<$ \ptjet $<$ 100\gevc. The lower panels show the ratios of the distributions in heavy-flavor to light-parton-initiated jets, using the \mbox{$\sqs = 8\tev$} measurement that shares the same binning as this analysis. Statistical uncertainties are indicated with bars and systematic uncertainties with shaded boxes.}
  \label{fig:compZjets50100}
\end{figure}

\begin{figure}
    \begin{center}
        \includegraphics[width=0.45\linewidth]{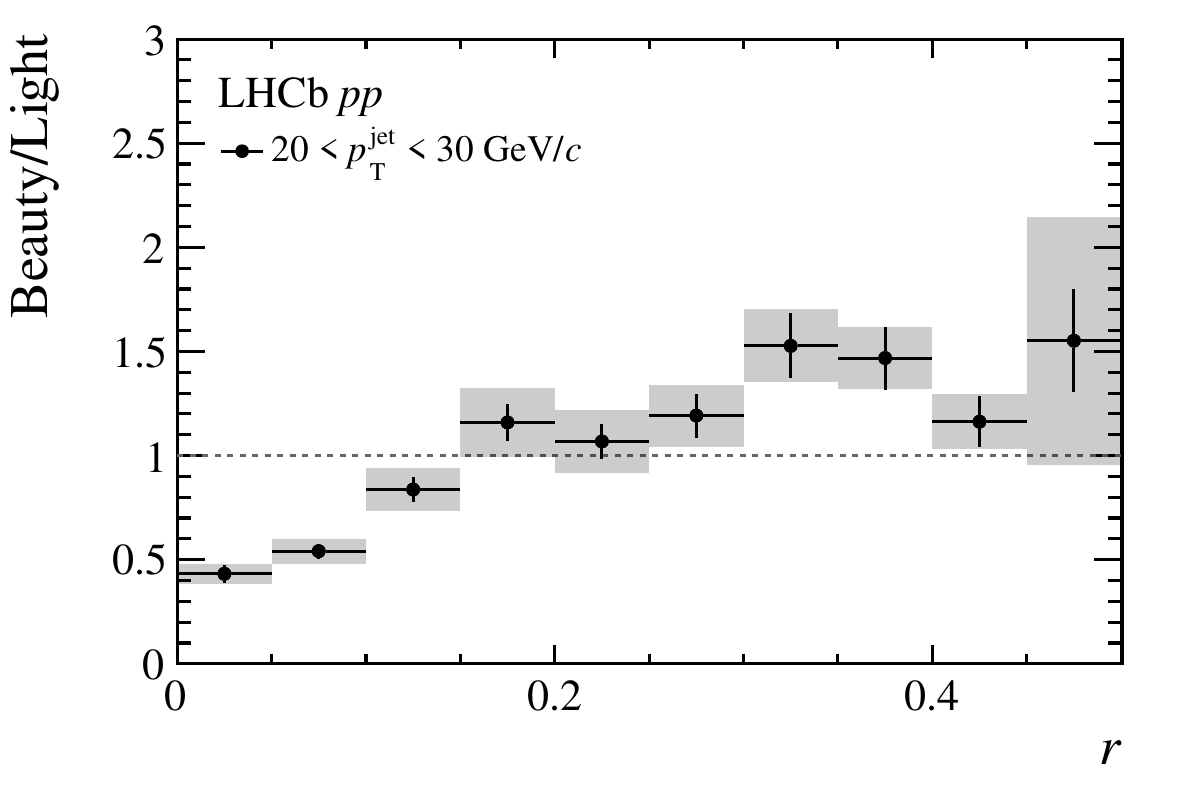}
        \includegraphics[width=0.45\linewidth]{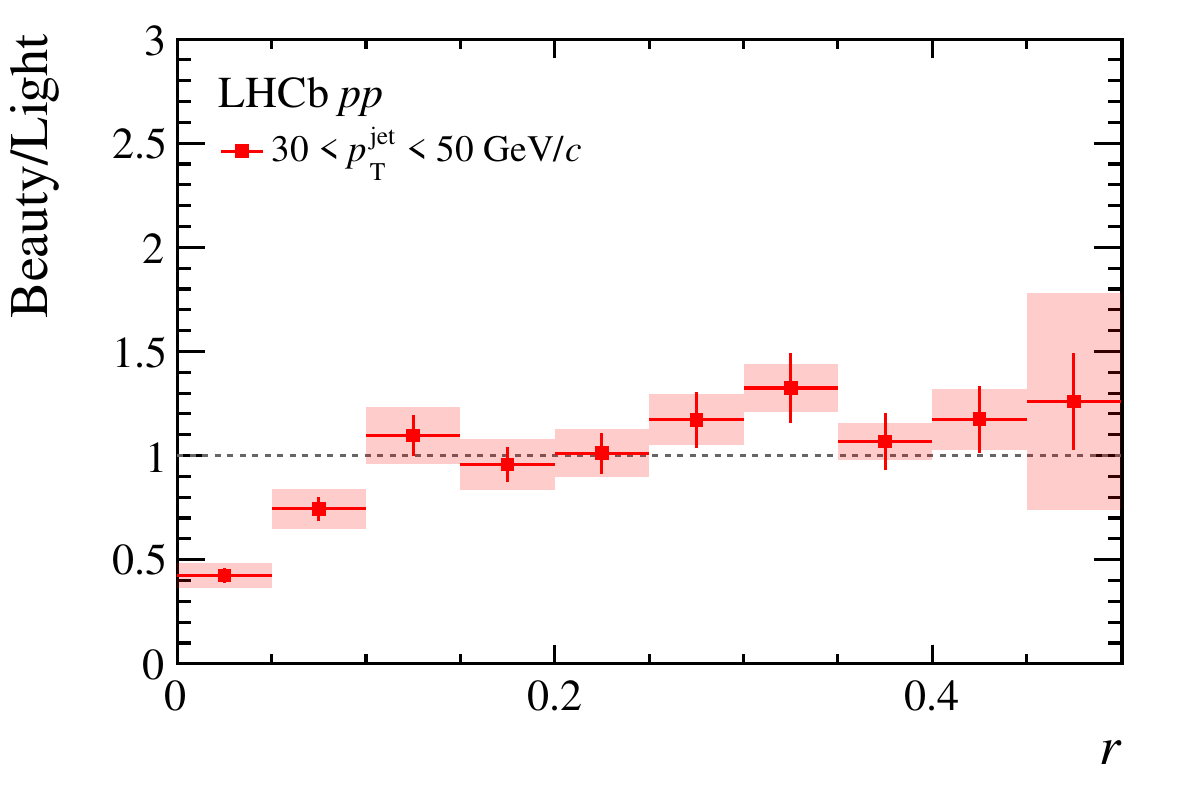}
        \includegraphics[width=0.45\linewidth]{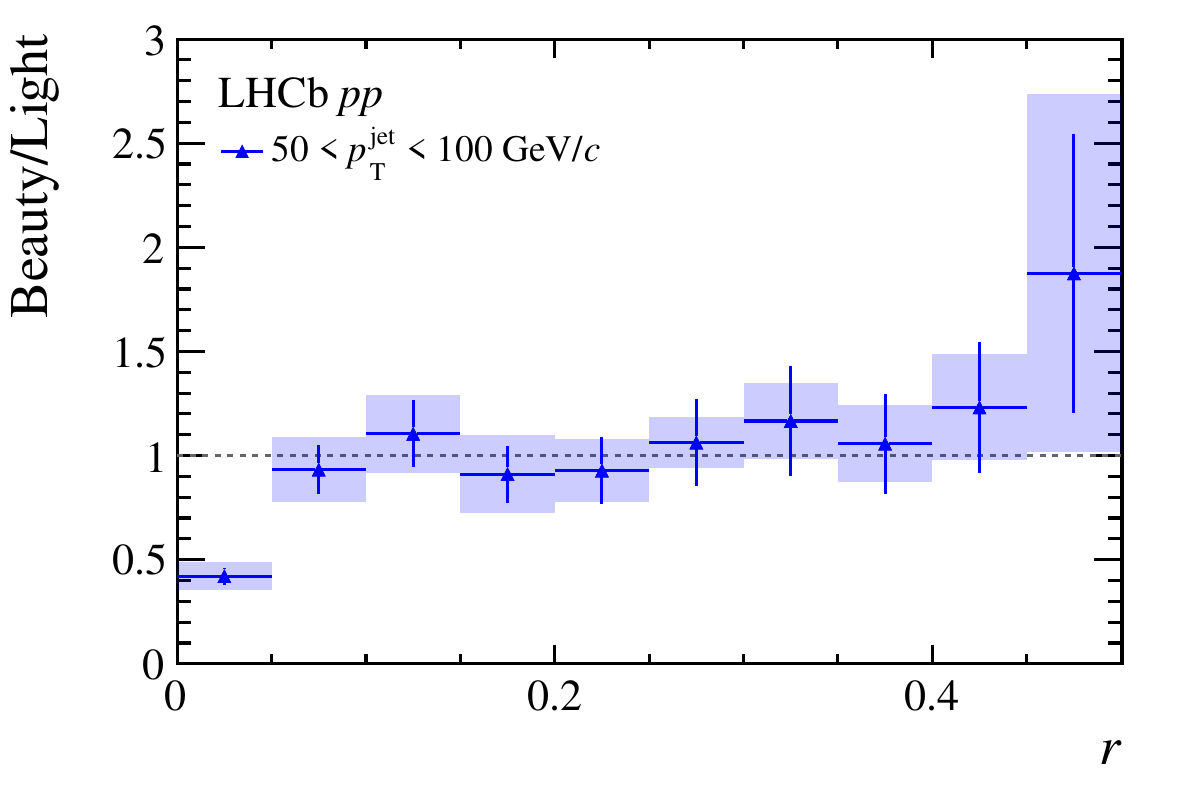}
    \end{center}
    \caption{Ratio of the measured \bquark-jet \rhad distributions at \mbox{$\sqs = 13\tev$} to the measured \rhad distributions in \Z-tagged jets at \mbox{$\sqs = 8\tev$}~\cite{LHCb-PAPER-2019-012}. Statistical uncertainties are indicated with bars and systematic uncertainties with shaded boxes.}
    \label{fig:BeautyLightr}
\end{figure}

\begin{figure}
    \begin{center}
        \includegraphics[width=0.42\linewidth]{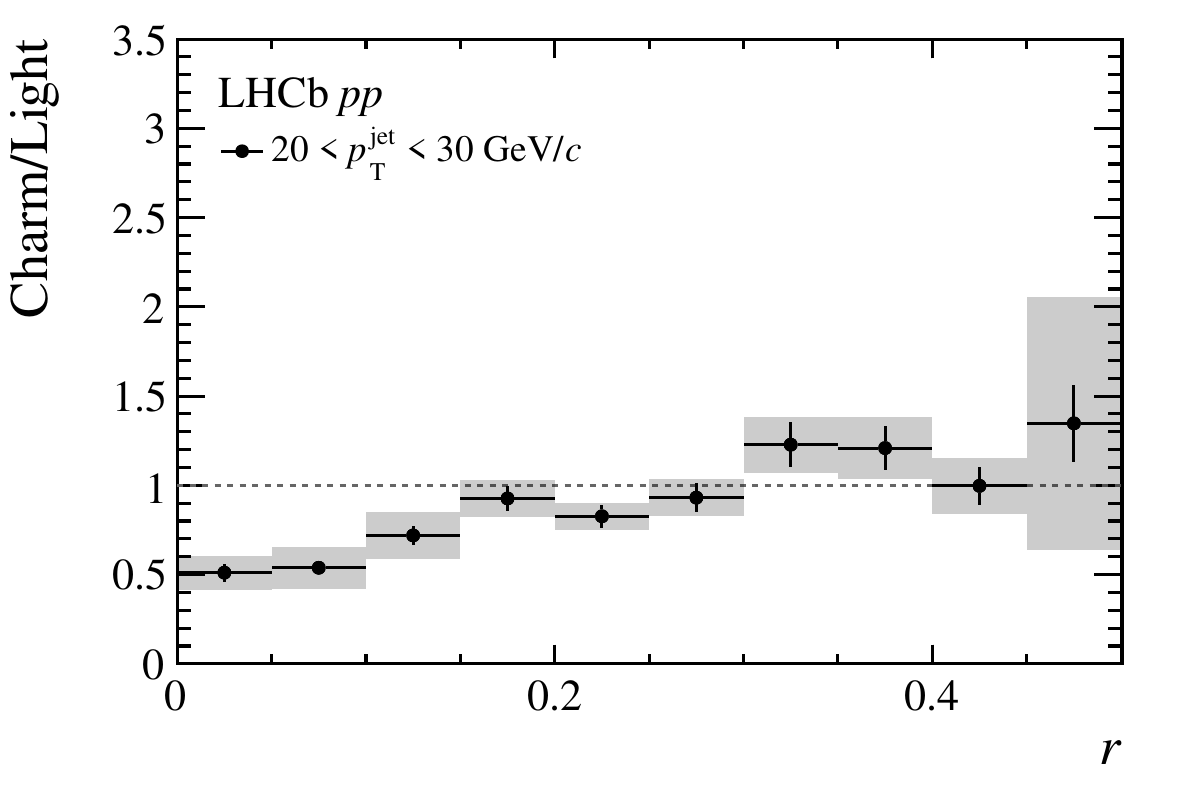}
        \includegraphics[width=0.42\linewidth]{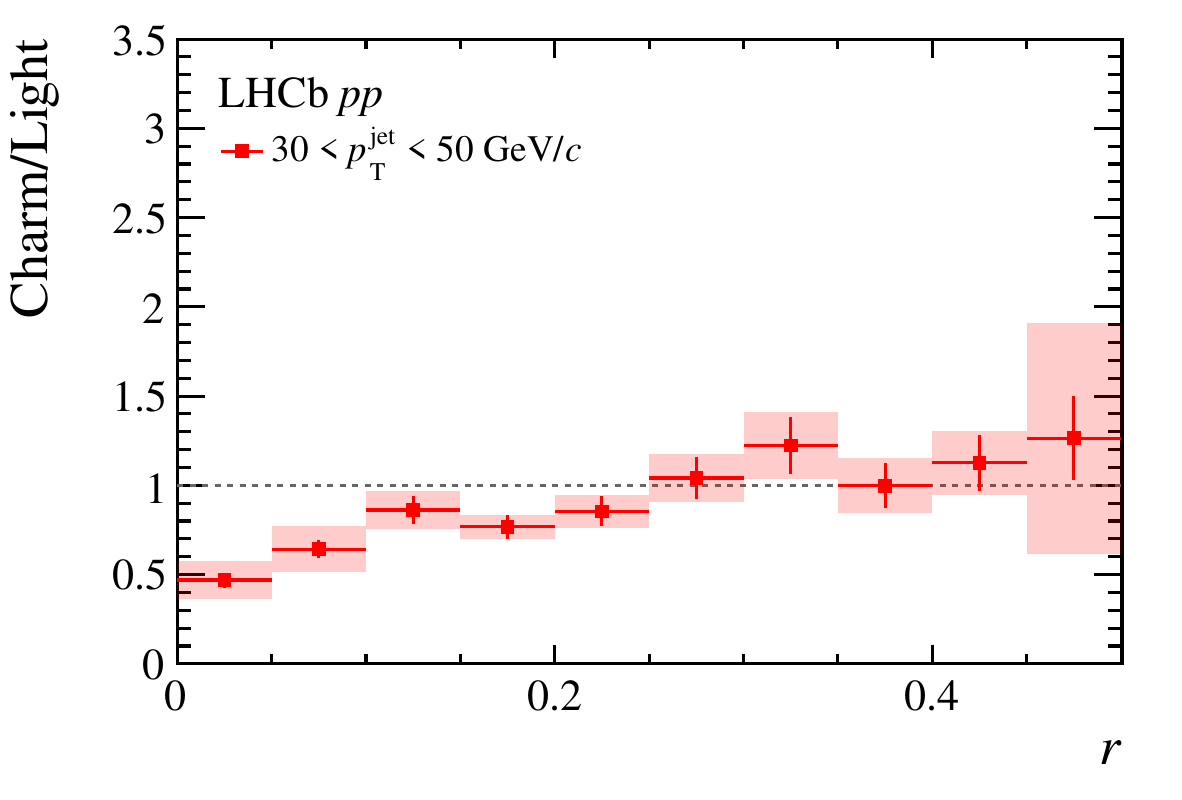}
        \includegraphics[width=0.42\linewidth]{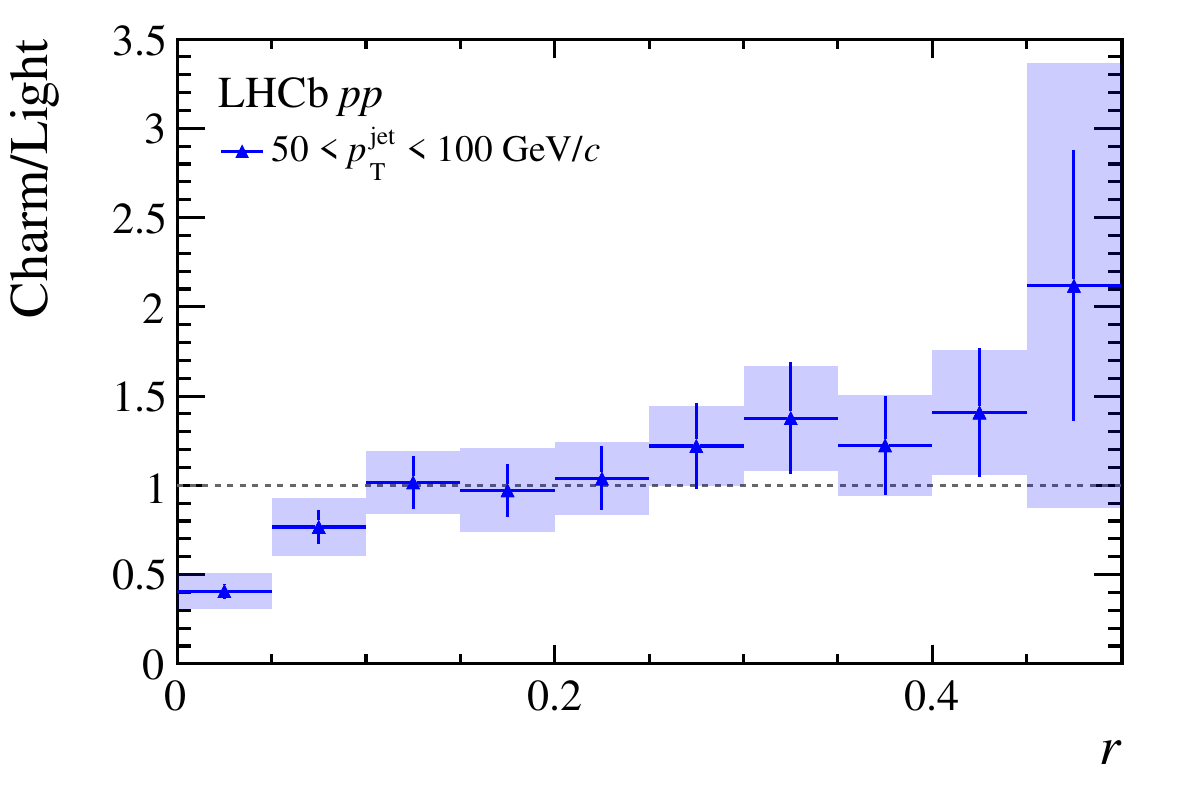}
    \end{center}
    \caption{Ratio of the measured \cquark-jet \rhad distributions at \mbox{$\sqs = 13\tev$} to the measured \rhad distributions in \Z-tagged jets at \mbox{$\sqs = 8\tev$}~\cite{LHCb-PAPER-2019-012}. Statistical uncertainties are indicated with bars and systematic uncertainties with shaded boxes.}
    \label{fig:CharmLightr}
\end{figure}

This section contains additional comparisons between the charged-hadron distributions measured in beauty and charm jets and those previously measured in \Z-tagged jets. Figures~\ref{fig:compZjets3050} and~\ref{fig:compZjets50100} show the \z and \jt distributions in the 30--50\gevc and 50--100\gevc \ptjet intervals, respectively. Figures \ref{fig:BeautyLightr} and \ref{fig:CharmLightr} show the ratios between the \rhad distributions measured in beauty and charm jets, respectively, to those previously measured in \Z-tagged jets at \mbox{$\sqs=8\tev$}.

\clearpage
\section{Distributions with a minimum \boldmath{\z} requirement}
\label{appendix:withzcut}
The very soft contribution at \z $\lesssim 0.01$ is difficult to address in theoretical calculations of the \jt and \rhad distributions. In this section, the \z, \jt and \rhad distributions computed with \z $> 0.01$ are presented. A \z $> 0.01$ selection is applied on the generator-level simulation when computing the purities, efficiencies and response matrices. No \z requirement is applied to the reconstructed simulation or data, in order to allow the true \z $> 0.01$ entries to be misreconstructed in nearby \z intervals, which is corrected for in the unfolding procedure. The data are then corrected with the set of purities, efficiencies and response matrices with the truth-level \z $> 0.01$ requirement implemented. It has been verified that this procedure recovers exactly the baseline \z distributions for \z $> 0.01$, which indicates that it successfully reproduces the baseline measurement while simply excluding data points below \z $= 0.01$. However, it must be stated that since this method relies on applying the \z requirement to the simulation, it inherently includes some model dependence as the relation between \z and \jt or \z and \rhad could vary depending on the choice of the used generator. Future two-dimensional measurements of the charged-hadron distributions in \z and \jt and in \z and \rhad would allow to remove intervals with \z $< 0.01$ without introducing a model dependence. 

\begin{table}[h!]
\begin{center}
\caption{Charged-hadron \z distribution in beauty jets, computed for \z $>$ 0.01. The quoted uncertainties are statistical and systematic. }
\resizebox{1\columnwidth}{!}{
\begin{tabular}{ c c c c }
\hline
\addlinespace
 \z & \multicolumn{3}{c}{\NormdNdz in beauty jets}\\
 & 20 $<$ \ptjet $<$ 30\gevc & 30 $<$ \ptjet $<$ 50\gevc & 50 $<$ \ptjet $<$ 100\gevc \\
 \hline
 0.010--0.015 & 20.7 $\pm$ 0.3 $\pm$ 2.8 & 67.5 $\pm$ 0.7 $\pm$ 9.0 & 151 $\pm$ 2 $\pm$ 16\\
 0.015--0.020 & 46.0 $\pm$ 0.4 $\pm$ 5.4 & 97.0 $\pm$ 0.8 $\pm$ 9.3 & 142 $\pm$ 2 $\pm$ 11\\
 0.020--0.030 & 65.9 $\pm$ 0.4 $\pm$ 5.3 & 95.7 $\pm$ 0.7 $\pm$ 6.3 & 115 $\pm$ 1 $\pm$ 7\\
 0.030--0.040 & 64.5 $\pm$ 0.4 $\pm$ 4.7 & 77.4 $\pm$ 0.6 $\pm$ 6.0 & 85.8 $\pm$ 1.0 $\pm$ 6.9\\
 0.040--0.050 & 55.3 $\pm$ 0.4 $\pm$ 4.8 & 62.6 $\pm$ 0.5 $\pm$ 4.9 & 66.5 $\pm$ 0.8 $\pm$ 5.0\\
 0.050--0.060 & 47.3 $\pm$ 0.3 $\pm$ 4.6 & 51.4 $\pm$ 0.4 $\pm$ 5.1 & 53.1 $\pm$ 0.7 $\pm$ 6.2\\
 0.060--0.080 & 37.5 $\pm$ 0.2 $\pm$ 4.7 & 39.5 $\pm$ 0.3 $\pm$ 4.7 & 40.5 $\pm$ 0.5 $\pm$ 4.8\\
 0.080--0.100 & 28.0 $\pm$ 0.2 $\pm$ 3.7 & 28.5 $\pm$ 0.2 $\pm$ 3.9 & 28.1 $\pm$ 0.3 $\pm$ 3.8\\
 0.100--0.125 & 20.3 $\pm$ 0.1 $\pm$ 2.9 & 19.8 $\pm$ 0.2 $\pm$ 2.7 & 19.1 $\pm$ 0.2 $\pm$ 3.1\\
 0.125--0.150 & 14.1 $\pm$ 0.1 $\pm$ 2.1 & 13.5 $\pm$ 0.1 $\pm$ 2.1 & 12.8 $\pm$ 0.2 $\pm$ 1.8\\
 0.150--0.200 & 8.38 $\pm$ 0.05 $\pm$ 1.29 & 8.10 $\pm$ 0.07 $\pm$ 1.07 & 7.58 $\pm$ 0.10 $\pm$ 1.22\\
 0.200--0.300 & 3.30 $\pm$ 0.02 $\pm$ 0.48 & 3.21 $\pm$ 0.03 $\pm$ 0.46 & 2.97 $\pm$ 0.04 $\pm$ 0.45\\
 0.300--0.400 & 0.988 $\pm$ 0.009 $\pm$ 0.121 & 0.992 $\pm$ 0.014 $\pm$ 0.107 & 0.921 $\pm$ 0.020 $\pm$ 0.145\\
 0.400--0.500 & 0.326 $\pm$ 0.005 $\pm$ 0.039 & 0.346 $\pm$ 0.008 $\pm$ 0.053& 0.334 $\pm$ 0.013 $\pm$ 0.062\\
 0.500--0.750 & 0.0679 $\pm$ 0.0015 $\pm$ 0.0156 & 0.0660 $\pm$ 0.0024 $\pm$ 0.0173& 0.0723 $\pm$ 0.0045 $\pm$ 0.0301\\
 \hline
\end{tabular}}
\end{center}
\label{tab:zcutbeautyzresults}
\end{table}

\begin{table}
\begin{center}
\caption{Charged-hadron \z distribution in charm jets, computed for \z $>$ 0.01. The quoted uncertainties are statistical and systematic.}
\resizebox{1\columnwidth}{!}{
\begin{tabular}{ c c c c }
\hline
\addlinespace
 \z & \multicolumn{3}{c}{\NormdNdz in charm jets}\\
 & 20 $<$ \ptjet $<$ 30\gevc & 30 $<$ \ptjet $<$ 50\gevc & 50 $<$ \ptjet $<$ 100\gevc \\
 \hline
 0.010--0.015 & 18.9 $\pm$ 0.3 $\pm$ 3.5 & 65.4 $\pm$ 1.1 $\pm$ 11.3 & 167 $\pm$ 4 $\pm$ 31\\
 0.015--0.020 & 38.6 $\pm$ 0.5 $\pm$ 6.4 & 87.9 $\pm$ 1.3 $\pm$ 14.0 & 149 $\pm$ 4 $\pm$ 26\\
 0.020--0.030 & 54.4 $\pm$ 0.5 $\pm$ 7.2 & 83.9 $\pm$ 1.0 $\pm$ 10.5 & 115 $\pm$ 3 $\pm$ 15\\
 0.030--0.040 & 54.7 $\pm$ 0.5 $\pm$ 4.9 & 68.9 $\pm$ 0.9 $\pm$ 6.9 & 88.2 $\pm$ 2.1 $\pm$ 12.2 \\
 0.040--0.050 & 47.2 $\pm$ 0.5 $\pm$ 3.5 & 55.6 $\pm$ 0.7 $\pm$ 4.2 & 69.9 $\pm$ 1.7 $\pm$ 9.8 \\
 0.050--0.060 & 39.9 $\pm$ 0.4 $\pm$ 2.6 & 45.0 $\pm$ 0.6 $\pm$ 3.0 & 54.3 $\pm$ 1.4 $\pm$ 7.2\\
 0.060--0.080 & 30.7 $\pm$ 0.3 $\pm$ 2.0 & 33.1 $\pm$ 0.4 $\pm$ 2.1 & 36.9 $\pm$ 0.9 $\pm$ 4.5\\
 0.080--0.100 & 22.5 $\pm$ 0.2 $\pm$ 1.7 & 23.3 $\pm$ 0.3 $\pm$ 1.7 & 24.7 $\pm$ 0.6 $\pm$ 3.6\\
 0.100--0.125 & 16.4 $\pm$ 0.2 $\pm$ 1.6 & 16.7 $\pm$ 0.2 $\pm$ 1.7 & 17.0 $\pm$ 0.4 $\pm$ 3.7\\
 0.125--0.150 & 11.8 $\pm$ 0.1 $\pm$ 1.8 & 11.4 $\pm$ 0.2 $\pm$ 1.7 & 11.2 $\pm$ 0.3 $\pm$ 2.6\\
 0.150--0.200 & 7.40 $\pm$ 0.07 $\pm$ 1.70 & 7.16 $\pm$ 0.09 $\pm$ 1.67 & 6.82 $\pm$ 0.17 $\pm$ 2.01\\
 0.200--0.300 & 3.34 $\pm$ 0.03 $\pm$ 1.12 & 3.18 $\pm$ 0.04 $\pm$ 1.17& 3.00 $\pm$ 0.07 $\pm$ 1.07\\
 0.300--0.400 & 1.23 $\pm$ 0.01 $\pm$ 0.47 & 1.19 $\pm$ 0.02 $\pm$ 0.47 & 1.09 $\pm$ 0.03 $\pm$ 0.38\\
 0.400--0.500 & 0.477 $\pm$ 0.007 $\pm$ 0.168 & 0.485 $\pm$ 0.011 $\pm$ 0.190 & 0.479 $\pm$ 0.022 $\pm$ 0.150\\
 0.500--0.750 & 0.0998 $\pm$ 0.0019 $\pm$ 0.0256 & 0.107 $\pm$ 0.003 $\pm$ 0.026 & \\
 \hline
\end{tabular}}
\end{center}
\label{tab:zcutcharmzresults}
\end{table}

\begin{table}
\begin{center}
\caption{Charged-hadron \jt distribution in beauty jets, computed for \z $>$ 0.01. The quoted uncertainties are statistical and systematic. }
\resizebox{1\columnwidth}{!}{
\begin{tabular}{ c c c c }
\hline
\addlinespace
 \jt [\gevc] & \multicolumn{3}{c}{\NormdNdjt in beauty jets}\\
 & 20 $<$ \ptjet $<$ 30\gevc & 30 $<$ \ptjet $<$ 50\gevc & 50 $<$ \ptjet $<$ 100\gevc \\
 \hline
 0.005--0.050 & 3.33 $\pm$ 0.03 $\pm$ 0.27& 3.47 $\pm$ 0.05 $\pm$ 0.45 & 3.23 $\pm$ 0.08 $\pm$ 0.42 \\
 0.050--0.100 & 7.48 $\pm$ 0.05 $\pm$ 0.54 & 7.87 $\pm$ 0.07 $\pm$ 0.70 & 7.45 $\pm$ 0.12 $\pm$ 0.58\\
 0.100--0.150 & 10.4 $\pm$ 0.1 $\pm$ 0.7 & 11.0 $\pm$ 0.1 $\pm$ 0.7 & 10.4 $\pm$ 0.1 $\pm$ 0.9 \\
 0.150--0.200 & 11.8 $\pm$ 0.1 $\pm$ 0.9 & 13.0 $\pm$ 0.1 $\pm$ 0.9 & 12.3 $\pm$ 0.2 $\pm$ 1.0\\
 0.200--0.250 & 11.6 $\pm$ 0.1 $\pm$ 0.9& 13.3 $\pm$ 0.1 $\pm$ 0.9 & 13.5 $\pm$ 0.2 $\pm$ 1.1\\
 0.250--0.300 & 10.8 $\pm$ 0.1 $\pm$ 1.0 & 12.5 $\pm$ 0.1 $\pm$ 1.0 & 13.3 $\pm$ 0.2 $\pm$ 1.0\\
 0.300--0.400 & 8.87 $\pm$ 0.05 $\pm$ 0.98 & 10.4 $\pm$ 0.1 $\pm$ 1.0 & 11.7 $\pm$ 0.1 $\pm$ 1.0\\
 0.400--0.500 & 6.34 $\pm$ 0.04 $\pm$ 0.74 & 7.55 $\pm$ 0.06 $\pm$ 0.75 & 8.80 $\pm$ 0.10 $\pm$ 0.83\\
 0.500--0.625 & 4.21 $\pm$ 0.03 $\pm$ 0.55 & 5.10 $\pm$ 0.04 $\pm$ 0.55 & 6.14 $\pm$ 0.07 $\pm$ 0.60\\
 0.625--0.750 & 2.66 $\pm$ 0.02 $\pm$ 0.36 & 3.35 $\pm$ 0.03 $\pm$ 0.41 & 4.15 $\pm$ 0.05 $\pm$ 0.33\\
 0.75--1.00 & 1.37 $\pm$ 0.01 $\pm$ 0.19 & 1.80 $\pm$ 0.01 $\pm$ 0.20 & 2.39 $\pm$ 0.03 $\pm$ 0.26\\
 1.00--1.25 & 0.568 $\pm$ 0.005 $\pm$ 0.080 & 0.821 $\pm$ 0.008 $\pm$ 0.094 & 1.15 $\pm$ 0.02 $\pm$ 0.16\\
 1.25--1.50 & 0.241 $\pm$ 0.003 $\pm$ 0.044 & 0.387 $\pm$ 0.005 $\pm$ 0.068 & 0.634 $\pm$ 0.011 $\pm$ 0.059\\
 1.50--2.00 & 0.0755 $\pm$ 0.0011 $\pm$ 0.0130 & 0.150 $\pm$ 0.002 $\pm$ 0.034 & 0.295 $\pm$ 0.006 $\pm$ 0.050\\
 2.00--3.00 & 0.00825 $\pm$ 0.00025 $\pm$ 0.00153 & 0.0260 $\pm$ 0.0007 $\pm$ 0.0103& 0.0769 $\pm$ 0.0024 $\pm$ 0.0188\\
 \hline
\end{tabular}}
\end{center}
\label{tab:zcutbeautyjTresults}
\end{table}

\begin{table}
\begin{center}
\caption{Charged-hadron \jt distribution in charm jets, computed for \z $>$ 0.01. The quoted uncertainties are statistical and systematic. }
\resizebox{1\columnwidth}{!}{
\begin{tabular}{ c c c c }
\hline
\addlinespace
 \jt [\gevc] & \multicolumn{3}{c}{\NormdNdjt in charm jets}\\
 & 20 $<$ \ptjet $<$ 30\gevc & 30 $<$ \ptjet $<$ 50\gevc & 50 $<$ \ptjet $<$ 100\gevc \\
 \hline
 0.005--0.050 & 3.39 $\pm$ 0.04 $\pm$ 0.29 & 3.26 $\pm$ 0.06 $\pm$ 0.49 & 3.29 $\pm$ 0.14 $\pm$ 0.62\\
 0.050--0.100 & 6.97 $\pm$ 0.07 $\pm$ 0.56 & 7.18 $\pm$ 0.10 $\pm$ 0.76 & 7.26 $\pm$ 0.22 $\pm$ 1.05 \\
 0.100--0.150 & 9.38 $\pm$ 0.09 $\pm$ 0.71 & 10.1 $\pm$ 0.1 $\pm$ 0.8 & 10.6 $\pm$ 0.3 $\pm$ 1.7 \\
 0.150--0.200 & 10.4 $\pm$ 0.1 $\pm$ 0.8 & 11.9 $\pm$ 0.1 $\pm$ 0.9 & 12.4 $\pm$ 0.3 $\pm$ 1.8 \\
 0.200--0.250 & 10.1 $\pm$ 0.1 $\pm$ 0.9 & 11.9 $\pm$ 0.1 $\pm$ 0.9 & 13.3 $\pm$ 0.3 $\pm$ 1.6\\
 0.250--0.300 & 9.25 $\pm$ 0.08 $\pm$ 0.91 & 11.0 $\pm$ 0.1 $\pm$ 0.9 & 12.9 $\pm$ 0.3 $\pm$ 1.5\\
 0.300--0.400 & 7.45 $\pm$ 0.06 $\pm$ 0.93 & 8.92 $\pm$ 0.10 $\pm$ 0.77 & 11.1 $\pm$ 0.2 $\pm$ 1.5 \\
 0.400--0.500 & 5.19 $\pm$ 0.04 $\pm$ 0.70 & 6.41 $\pm$ 0.08 $\pm$ 0.72 & 8.35 $\pm$ 0.19 $\pm$ 1.11 \\
 0.500--0.625 & 3.41 $\pm$ 0.03 $\pm$ 0.47 & 4.33 $\pm$ 0.05 $\pm$ 0.45 & 6.02 $\pm$ 0.14 $\pm$ 0.86\\
 0.625--0.750 & 2.09 $\pm$ 0.02 $\pm$ 0.26 & 2.79 $\pm$ 0.04 $\pm$ 0.25 & 3.99 $\pm$ 0.10 $\pm$ 0.54\\
 0.75--1.00 & 1.02 $\pm$ 0.01 $\pm$ 0.13 & 1.53 $\pm$ 0.02 $\pm$ 0.14 & 2.41 $\pm$ 0.06 $\pm$ 0.32\\
 1.00--1.25 & 0.385 $\pm$ 0.006 $\pm$ 0.032 & 0.709 $\pm$ 0.012 $\pm$ 0.066 & 1.25 $\pm$ 0.04 $\pm$ 0.17\\
 1.25--1.50 & 0.152 $\pm$ 0.004 $\pm$ 0.025 & 0.363 $\pm$ 0.009 $\pm$ 0.069 & 0.709 $\pm$ 0.028 $\pm$ 0.116\\
 1.50--2.00 & 0.0471 $\pm$ 0.0019 $\pm$ 0.0118 & 0.154 $\pm$ 0.005 $\pm$ 0.036 & 0.368 $\pm$ 0.016 $\pm$ 0.081\\
 2.00--3.00 & & 0.0308 $\pm$ 0.0020 $\pm$ 0.0106 & 0.113 $\pm$ 0.008 $\pm$ 0.025\\
 \hline
\end{tabular}}
\end{center}
\label{tab:zcutcharmjTresults}
\end{table}

\begin{table}
\begin{center}
\caption{Charged-hadron \rhad distribution in beauty jets, computed for \z $>$ 0.01. The quoted uncertainties are statistical and systematic. }
\begin{tabular}{ c c c c }
\hline
\addlinespace
 \rhad & \multicolumn{3}{c}{\NormdNdr in beauty jets}\\
 & 20 $<$ \ptjet $<$ 30\gevc & 30 $<$ \ptjet $<$ 50\gevc & 50 $<$ \ptjet $<$ 100\gevc \\
 \hline
 0.00--0.05 & 6.07 $\pm$ 0.04 $\pm$ 0.57 & 11.4 $\pm$ 0.1 $\pm$ 1.3 & 22.8 $\pm$ 0.2 $\pm$ 2.6\\
 0.05--0.10 & 14.4 $\pm$ 0.1 $\pm$ 1.4 & 23.6 $\pm$ 0.2 $\pm$ 2.6 & 34.6 $\pm$ 0.3 $\pm$ 4.2\\
 0.10--0.15 & 17.2 $\pm$ 0.1 $\pm$ 2.0 & 23.4 $\pm$ 0.2 $\pm$ 2.6 & 25.6 $\pm$ 0.3 $\pm$ 3.0\\
 0.15--0.20 & 16.6 $\pm$ 0.1 $\pm$ 2.2 & 19.3 $\pm$ 0.1 $\pm$ 2.3 & 18.3 $\pm$ 0.2 $\pm$ 2.5\\
 0.20--0.25 & 14.9 $\pm$ 0.1 $\pm$ 1.9 & 15.2 $\pm$ 0.1 $\pm$ 1.6 & 13.5 $\pm$ 0.2 $\pm$ 1.5\\
 0.25--0.30 & 12.8 $\pm$ 0.1 $\pm$ 1.3 & 12.1 $\pm$ 0.1 $\pm$ 1.1 & 10.4 $\pm$ 0.1 $\pm$ 0.7\\
 0.30--0.35 & 10.9 $\pm$ 0.1 $\pm$ 1.0 & 10.0 $\pm$ 0.1 $\pm$ 0.7 & 8.45 $\pm$ 0.12 $\pm$ 0.74\\
 0.35--0.40 & 9.32 $\pm$ 0.06 $\pm$ 0.65 & 8.47 $\pm$ 0.07 $\pm$ 0.55 & 6.98 $\pm$ 0.10 $\pm$ 0.62\\
 0.40--0.45 & 7.50 $\pm$ 0.05 $\pm$ 0.52 & 7.07 $\pm$ 0.07 $\pm$ 0.79 & 5.95 $\pm$ 0.09 $\pm$ 0.89\\
 0.45--0.50 & 4.44 $\pm$ 0.04 $\pm$ 1.67 & 4.53 $\pm$ 0.05 $\pm$ 1.85 & 4.02 $\pm$ 0.08 $\pm$ 1.73\\
 \hline
\end{tabular}
\end{center}
\label{tab:zcutbeautyrresults}
\end{table}

\begin{table}
\begin{center}
\caption{Charged-hadron \rhad distribution in charm jets, computed for \z $>$ 0.01. The quoted uncertainties are statistical and systematic. }
\begin{tabular}{ c c c c }
\hline
\addlinespace
 \rhad & \multicolumn{3}{c}{\NormdNdr in charm jets}\\
 & 20 $<$ \ptjet $<$ 30\gevc & 30 $<$ \ptjet $<$ 50\gevc & 50 $<$ \ptjet $< $ 100\gevc \\
 \hline
 0.00--0.05 & 7.17 $\pm$ 0.06 $\pm$ 1.26 & 12.6 $\pm$ 0.1 $\pm$ 2.7& 22.1 $\pm$ 0.5 $\pm$ 4.8\\
 0.05--0.10 & 14.4 $\pm$ 0.1 $\pm$ 3.0 & 20.4 $\pm$ 0.2 $\pm$ 3.8& 28.4 $\pm$ 0.6 $\pm$ 5.1\\
 0.10--0.15 & 14.8 $\pm$ 0.1 $\pm$ 2.6 & 18.4 $\pm$ 0.2 $\pm$ 2.0& 23.5 $\pm$ 0.5 $\pm$ 2.9\\
 0.15--0.20 & 13.3 $\pm$ 0.1 $\pm$ 1.3 & 15.4 $\pm$ 0.2 $\pm$ 1.1& 19.4 $\pm$ 0.5 $\pm$ 3.6\\
 0.20--0.25 & 11.5 $\pm$ 0.1 $\pm$ 0.8 & 12.8 $\pm$ 0.2 $\pm$ 1.2& 15.1 $\pm$ 0.4 $\pm$ 2.4\\
 0.25--0.30 & 10.0 $\pm$ 0.1 $\pm$ 0.9 & 10.7 $\pm$ 0.1 $\pm$ 1.2& 12.0 $\pm$ 0.3 $\pm$ 1.9\\
 0.30--0.35 & 8.76 $\pm$ 0.09 $\pm$ 0.91 & 9.24 $\pm$ 0.13 $\pm$ 1.33& 10.1 $\pm$ 0.3 $\pm$ 1.7\\
 0.35--0.40 & 7.67 $\pm$ 0.08 $\pm$ 0.93 & 7.92 $\pm$ 0.11 $\pm$ 1.15& 8.10 $\pm$ 0.25 $\pm$ 1.41\\
 0.40--0.45 & 6.43 $\pm$ 0.07 $\pm$ 0.83 & 6.80 $\pm$ 0.10 $\pm$ 1.01& 6.92 $\pm$ 0.22 $\pm$ 1.42\\
 0.45--0.50 & 3.86 $\pm$ 0.06 $\pm$ 2.01 & 4.56 $\pm$ 0.09 $\pm$ 2.32& 4.66 $\pm$ 0.19 $\pm$ 2.65\\
 \hline
\end{tabular}
\end{center}
\label{tab:zcutcharmrresults}
\end{table}

\clearpage

\addcontentsline{toc}{section}{References}
\bibliographystyle{LHCb}
\bibliography{main,standard,LHCb-PAPER,LHCb-CONF,LHCb-DP,LHCb-TDR}

@article{LHCb-DP-2018-001,
      author         = "Aaij, R. and others",
      title          = "{Selection and processing of calibration samples to measure the particle identification performance of the LHCb experiment in Run 2}",
      eprint         = "1803.00824",
      archivePrefix  = "arXiv",
      primaryClass   = "hep-ex",
      report         = "LHCb-DP-2018-001",
      year           = "2019",
      journal        = "Eur. Phys. J. Tech. Instr.",
      volume         = "6",
      pages          = "1",
      doi            = "10.1140/epjti/s40485-019-0050-z",
}

@article{LHCb-DP-2017-001,
      author         = "d'Argent, Ph. and others",
      title          = "{Improved performance of the LHCb Outer Tracker in LHC
                        Run 2}",
      year           = "2017",
      eprint         = "1708.00819",
      archivePrefix  = "arXiv",
      primaryClass   = "physics.ins-det",
      report         = "LHCb-DP-2017-001",
      journal        = "JINST",
      volume         = "12",
      pages          = "P11016",
      doi            = "10.1088/1748-0221/12/11/P11016",
}

@article{LHCb-DP-2014-002,
      author         = "Aaij, R. and others",
      title          = "{LHCb detector performance}",
      collaboration  = "LHCb collaboration",
      journal        = "Int. J. Mod. Phys.",
      volume         = "A30",
      pages          = "1530022",
      doi            = "10.1142/S0217751X15300227",
      year           = "2015",
      eprint         = "1412.6352",
      archivePrefix  = "arXiv",
      primaryClass   = "hep-ex",
      report         = "LHCB-DP-2014-002, CERN-PH-EP-2014-290",
}

@article{LHCb-DP-2014-001,
      author         = "Aaij, R. and others",
      title          = "{Performance of the LHCb Vertex Locator}",
      journal        = "JINST",
      volume         = "9",
      pages          = "P09007",
      doi            = "10.1088/1748-0221/9/09/P09007",
      year           = "2014",
      eprint         = "1405.7808",
      archivePrefix  = "arXiv",
      primaryClass   = "physics.ins-det",
      report         = "LHCB-DP-2014-001",
}

@article{LHCb-DP-2013-002,
      author         = "Aaij, R. and others",
      title          = "{Measurement of the track reconstruction efficiency at LHCb}",
      collaboration  = "LHCb collaboration",
      journal        = "JINST",
      volume         = "10",
      pages          = "P02007",
      doi            = "10.1088/1748-0221/10/02/P02007",
      year           = "2015",
      eprint         = "1408.1251",
      archivePrefix  = "arXiv",
      primaryClass   = "hep-ex",
      report         = "CERN-LHCB-DP-2013-002",
}

@article{LHCb-DP-2012-004,
      author         = "Aaij, R. and others",
      title          = "{The \lhcb trigger and its performance in 2011}",
      journal        = "JINST",
      volume         = "8",
      pages          = "P04022",
      doi            = "10.1088/1748-0221/8/04/P04022",
      year           = "2013",
      eprint         = "1211.3055",
      archivePrefix  = "arXiv",
      primaryClass   = "hep-ex",
      report         = "LHCb-DP-2012-004",
}

@article{LHCb-DP-2012-003,
      author         = "Adinolfi, M. and others",
      title          = "{Performance of the \lhcb RICH detector at the LHC}",
      journal        = "Eur. Phys. J.",
      volume         = "C73",
      pages          = "2431",
      doi            = "10.1140/epjc/s10052-013-2431-9",
      year           = "2013",
      eprint         = "1211.6759",
      archivePrefix  = "arXiv",
      primaryClass   = "physics.ins-det",
      report         = "LHCb-DP-2012-003",
}

@article{LHCb-DP-2012-002,
      author       = "Alves Jr., A A and others",
      title        = "{Performance of the LHCb muon system}",
      journal      = "JINST",
      volume       = "8",
      pages        = "P02022",
      doi          = "10.1088/1748-0221/8/02/P02022",
      year         = "2013",
      eprint       = "1211.1346",
      archivePrefix= "arXiv",
      primaryClass = "physics.ins-det",
      report       = "LHCb-DP-2012-002",
}

@article{LHCb-DP-2008-001,
      author         = "Alves~Jr., A. A. and others",
      title          = "{The \lhcb detector at the LHC}",
      collaboration  = "LHCb collaboration",
      journal        = "JINST",
      volume         = "3",
      pages          = "S08005",
      doi            = "10.1088/1748-0221/3/08/S08005",
      year           = "2008",
      number         = "LHCb-DP-2008-001",
}

@article{LHCb-PAPER-2022-013,
      author         = "Aaij, R. and others",
      title          = "{Multidifferential study of identified charged hadron distributions in $\Z$-tagged jets in proton-proton collisions at $\sqs=13$~\tev}",
      collaboration  = "LHCb collaboration",
      report         = "{LHCb-PAPER-2022-013, CERN-EP-2022-161}",
      eprint         = "2208.11691",
      archivePrefix  = "arXiv",
      primaryClass   = "hep-ex",
      year           = "2023",
      journal        = "Phys. Rev.",
      volume         = "D108",
      pages          = "L031103",
      doi            = "10.1103/PhysRevD.108.L031103",
}

@article{LHCb-PAPER-2020-018,
      author         = "Aaij, R. and others",
      title          = "{Measurement of differential $\bquark \bquarkbar$ and $\cquark \cquarkbar$ dijet cross-sections in the forward region of $pp$ collisions at $\sqrt{s} = 13$~\tev}",
      collaboration  = "LHCb collaboration",
      report         = "{LHCb-PAPER-2020-018, CERN-EP-2020-174}",
      eprint         = "2010.09437",
      archivePrefix  = "arXiv",
      primaryClass   = "hep-ex",
      year           = "2021",
      journal        = "JHEP",
      volume         = "02",
      pages          = "023",
      doi            = "10.1007/JHEP02(2021)023",
}

@article{LHCb-PAPER-2019-012,
      author         = "Aaij, R. and others",
      title          = "{Measurement of charged hadron production in \Z-tagged jets in proton-proton collisions at $\sqs=8$~\tev}",
      collaboration  = "LHCb collaboration",
      report         = "{LHCb-PAPER-2019-012 CERN-EP-2019-063}",
      eprint         = "1904.08878",
      archivePrefix  = "arXiv",
      primaryClass   = "hep-ex",
      year           = "2019",
      journal        = "Phys. Rev. Lett.",
      volume         = "123",
      pages          = "232001",
      doi            = "10.1103/PhysRevLett.123.232001",
}

@article{LHCb-PAPER-2016-011,
      author         = "Aaij, R. and others",
      title          = "{Measurement of forward \W and \Z boson production in 
                        association with jets in proton-proton collisions at \mbox{$\sqs=$8~\tev}}",
      collaboration  = "LHCb collaboration",
      year           = "2016",
      journal        = "JHEP",
      volume         = "05",
      pages          = "131",
      doi            = "10.1007/JHEP05(2016)131",
      report         = "{LHCb-PAPER-2016-011 CERN-EP-2016-092}",
      eprint         = "1605.00951",
      archivePrefix  = "arXiv",
      primaryClass   = "hep-ex",
}

@article{LHCb-PAPER-2015-016,
      author         = "Aaij, R. and others",
      title          = "{Identification of beauty and charm quark jets at LHCb}",
      collaboration  = "LHCb collaboration",
      year           = "2015",
      eprint         = "1504.07670",
      archivePrefix  = "arXiv",
      primaryClass   = "hep-ex",
      report         = "{CERN-PH-EP-2015-101 LHCb-PAPER-2015-016}",
      journal        = "JINST",
      volume         = "10",
      pages          = "P06013",
      doi            = "10.1088/1748-0221/10/06/P06013",
}

@article{LHCb-PAPER-2013-058,
      author         = "Aaij, R. and others",
      title          = "{Study of forward \Z+jet production in \proton\proton 
                         collisions at $\sqs = $7~\tev}",
      collaboration  = "LHCb collaboration",
      journal        = "JHEP",
      volume         = "01",
      pages          = "033",
      doi            = "10.1007/JHEP01(2014)033",
      year           = "2014",
      eprint         = "1310.8197",
      archivePrefix  = "arXiv",
      primaryClass   = "hep-ex",
      report         = "LHCb-PAPER-2013-058 CERN-PH-EP-2013-198",
}

@article{LHCb-PAPER-2025-010,
      author         = "Aaij, R. and others",
      title          = "{Measurement of the Lund plane for light- and beauty-quark jets}",
      collaboration  = "LHCb collaboration",
      report         = "{LHCb-PAPER-2025-010, CERN-EP-2025-093}",
      eprint         = "2505.23530",
      archivePrefix  = "arXiv",
      primaryClass   = "hep-ex",
      note           = "{Submitted to Phys. Rev. D}",
      year           = "2025",
}

@article{Metz:2016swz,
    author = "Metz, Andreas and Vossen, Anselm",
    title = "{Parton fragmentation functions}",
    eprint = "1607.02521",
    archivePrefix = "arXiv",
    primaryClass = "hep-ex",
    doi = "10.1016/j.ppnp.2016.08.003",
    journal = "Prog. Part. Nucl. Phys.",
    volume = "91",
    pages = "136--202",
    year = "2016"
}

@article{Kaufmann:2015hma,
    author = "Kaufmann, Tom and Mukherjee, Asmita and Vogelsang, Werner",
    title = "{Hadron fragmentation inside jets in hadronic collisions}",
    eprint = "1506.01415",
    archivePrefix = "arXiv",
    primaryClass = "hep-ph",
    doi = "10.1103/PhysRevD.92.054015",
    journal = "Phys. Rev.",
    volume = "D92",
    number = "5",
    pages = "054015",
    year = "2015",
    extraPrefix    = "Erratum",
    extraVolume    = "101",
    extraPages     = "079901",
    extraYear      = "2020",
    extraDoi       = "10.1103/PhysRevD.101.079901"
}

@article{Procura:2009vm,
    author = "Procura, Massimiliano and Stewart, Iain W.",
    title = "{Quark fragmentation within an identified jet}",
    eprint = "0911.4980",
    archivePrefix = "arXiv",
    primaryClass = "hep-ph",
    reportNumber = "MIT-CTP-4063, TUM-EFT-3-09",
    doi = "10.1103/PhysRevD.81.074009",
    journal = "Phys. Rev.",
    volume = "D81",
    pages = "074009",
    year = "2010",
    extraPrefix    = "Erratum",
    extraVolume    = "83",
    extraPages     = "039902",
    extraYear      = "2011",
    extraDoi       = "10.1103/PhysRevD.83.039902"
}

@article{Jain:2011xz,
    author = "Jain, Ambar and Procura, Massimiliano and Waalewijn, Wouter J.",
    title = "{Parton fragmentation within an identified jet at NNLL}",
    eprint = "1101.4953",
    archivePrefix = "arXiv",
    primaryClass = "hep-ph",
    reportNumber = "TUM-EFT-18-10",
    doi = "10.1007/JHEP05(2011)035",
    journal = "JHEP",
    volume = "05",
    pages = "035",
    year = "2011"
}

@article{Procura:2011aq,
    author = "Procura, Massimiliano and Waalewijn, Wouter J.",
    title = "{Fragmentation in jets: Cone and threshold effects}",
    eprint = "1111.6605",
    archivePrefix = "arXiv",
    primaryClass = "hep-ph",
    doi = "10.1103/PhysRevD.85.114041",
    journal = "Phys. Rev.",
    volume = "D85",
    pages = "114041",
    year = "2012"
}

@article{Kang:2016ehg,
    author = "Kang, Zhong-Bo and Ringer, Felix and Vitev, Ivan",
    title = "{Jet substructure using semi-inclusive jet functions in SCET}",
    eprint = "1606.07063",
    archivePrefix = "arXiv",
    primaryClass = "hep-ph",
    doi = "10.1007/JHEP11(2016)155",
    journal = "JHEP",
    volume = "11",
    pages = "155",
    year = "2016"
}

@article{Kang:2017glf,
    author = "Kang, Zhong-Bo and Liu, Xiaohui and Ringer, Felix and Xing, Hongxi",
    title = "{The transverse momentum distribution of hadrons within jets}",
    eprint = "1705.08443",
    archivePrefix = "arXiv",
    primaryClass = "hep-ph",
    doi = "10.1007/JHEP11(2017)068",
    journal = "JHEP",
    volume = "11",
    pages = "068",
    year = "2017"
}

@article{ATLAS:2021agf,
    author = "Aad, Georges and others",
    collaboration = "ATLAS",
    title = "{Measurement of $b$-quark fragmentation properties in jets using the decay $B^{\pm} \to J/\psi K^{\pm}$ in $pp$ collisions at $ \sqrt{s} $ = 13 TeV with the ATLAS detector}",
    eprint = "2108.11650",
    archivePrefix = "arXiv",
    primaryClass = "hep-ex",
    reportNumber = "CERN-EP-2021-123",
    doi = "10.1007/JHEP12(2021)131",
    journal = "JHEP",
    volume = "12",
    pages = "131",
    year = "2021"
}

@article{Cacciari:2008gp,
    author = "Cacciari, Matteo and Salam, Gavin P. and Soyez, Gregory",
    title = "{The anti-$k_t$ jet clustering algorithm}",
    eprint = "0802.1189",
    archivePrefix = "arXiv",
    primaryClass = "hep-ph",
    reportNumber = "LPTHE-07-03",
    doi = "10.1088/1126-6708/2008/04/063",
    journal = "JHEP",
    volume = "04",
    pages = "063",
    year = "2008"
}

@article{Cacciari:2011ma,
    author = "Cacciari, Matteo and Salam, Gavin P. and Soyez, Gregory",
    title = "{FastJet user manual}",
    eprint = "1111.6097",
    archivePrefix = "arXiv",
    primaryClass = "hep-ph",
    reportNumber = "CERN-PH-TH-2011-297",
    doi = "10.1140/epjc/s10052-012-1896-2",
    journal = "Eur. Phys. J.",
    volume = "C72",
    pages = "1896",
    year = "2012"
}

@article{ALICE:2019cbr,
    author = "Acharya, Shreyasi and others",
    collaboration = "ALICE",
    title = "{Measurement of the production of charm jets tagged with D$^{0}$ mesons in $pp$ collisions at $ \sqrt{\mathrm{s}}=7 $ TeV}",
    eprint = "1905.02510",
    archivePrefix = "arXiv",
    primaryClass = "nucl-ex",
    reportNumber = "CERN-EP-2019-081",
    doi = "10.1007/JHEP08(2019)133",
    journal = "JHEP",
    volume = "08",
    pages = "133",
    year = "2019"
}

@article{ALICE:2023jgm,
    author = "Acharya, Shreyasi and others",
    collaboration = "ALICE",
    title = "{Measurement of the fraction of jet longitudinal momentum carried by $\Lambda_c^+$ baryons in $pp$ collisions}",
    eprint = "2301.13798",
    archivePrefix = "arXiv",
    primaryClass = "nucl-ex",
    reportNumber = "CERN-EP-2023-005",
    doi = "10.1103/PhysRevD.109.072005",
    journal = "Phys. Rev.",
    volume = "D109",
    number = "7",
    pages = "072005",
    year = "2024"
}

@article{ALICE:2022mur,
    author = "Acharya, Shreyasi and others",
    collaboration = "ALICE",
    title = "{Measurement of the production of charm jets tagged with D$^{0}$ mesons in $pp$ collisions at $ \sqrt{s} $ = 5.02 and 13 TeV}",
    eprint = "2204.10167",
    archivePrefix = "arXiv",
    primaryClass = "nucl-ex",
    reportNumber = "CERN-EP-2022-070",
    doi = "10.1007/JHEP06(2023)133",
    journal = "JHEP",
    volume = "06",
    pages = "133",
    year = "2023"
}

@article{ATLAS:2011chi,
    author = "Aad, Georges and others",
    collaboration = "ATLAS",
    title = "{Measurement of $D^{*\pm}$ meson production in jets from pp collisions at $\sqrt{s}$ = 7 TeV with the ATLAS detector}",
    eprint = "1112.4432",
    archivePrefix = "arXiv",
    primaryClass = "hep-ex",
    reportNumber = "CERN-PH-EP-2011-180",
    doi = "10.1103/PhysRevD.85.052005",
    journal = "Phys. Rev.",
    volume = "D85",
    pages = "052005",
    year = "2012"
}

@article{ATLAS:2022miz,
    author = "Aad, Georges and others",
    collaboration = "ATLAS",
    title = "{Measurements of jet observables sensitive to $b$-quark fragmentation in $t\bar{t}$ events at the LHC with the ATLAS detector}",
    eprint = "2202.13901",
    archivePrefix = "arXiv",
    primaryClass = "hep-ex",
    reportNumber = "CERN-EP-2021-116",
    doi = "10.1103/PhysRevD.106.032008",
    journal = "Phys. Rev.",
    volume = "D106",
    number = "3",
    pages = "032008",
    year = "2022"
}

@lhcbreport{CMS:2024gds,
    collaboration = "CMS",
    title = "{Jet fragmentation function and groomed substructure of bottom quark jets in proton-proton collisions at 5.02 TeV}",
    number = "CMS-PAS-HIN-24-005",
    year = "2024"
}

@article{CMS:2020geg,
    author = "Sirunyan, Albert M and others",
    collaboration = "CMS",
    title = "{Measurement of b jet shapes in proton-proton collisions at $\sqrt{s} =$ 5.02 TeV}",
    eprint = "2005.14219",
    archivePrefix = "arXiv",
    primaryClass = "hep-ex",
    reportNumber = "CMS-HIN-18-020, CERN-EP-2020-060",
    doi = "10.1007/JHEP05(2021)054",
    journal = "JHEP",
    volume = "05",
    pages = "054",
    year = "2021"
}

@article{ATLAS:2011myc,
    author = "Aad, Georges and others",
    collaboration = "ATLAS",
    title = "{Measurement of the jet fragmentation function and transverse profile in proton-proton collisions at a center-of-mass energy of 7 TeV with the ATLAS detector}",
    eprint = "1109.5816",
    archivePrefix = "arXiv",
    primaryClass = "hep-ex",
    reportNumber = "CERN-PH-EP-2011-148",
    doi = "10.1140/epjc/s10052-011-1795-y",
    journal = "Eur. Phys. J.",
    volume = "C71",
    pages = "1795",
    year = "2011"
}

@article{DELPHI:2011aa,
    author = "Abdallah, J. and others",
    collaboration = "DELPHI",
    title = "{A study of the b-quark fragmentation function with the DELPHI detector at LEP I and an averaged distribution obtained at the Z pole}",
    eprint = "1102.4748",
    archivePrefix = "arXiv",
    primaryClass = "hep-ex",
    reportNumber = "CERN-PH-EP-2010-057",
    doi = "10.1140/epjc/s10052-011-1557-x",
    journal = "Eur. Phys. J.",
    volume = "C71",
    pages = "1557",
    year = "2011"
}

@article{OPAL:2002plk,
    author = "Abbiendi, G. and others",
    collaboration = "OPAL",
    title = "{Inclusive analysis of the b quark fragmentation function in Z decays at LEP}",
    eprint = "hep-ex/0210031",
    archivePrefix = "arXiv",
    reportNumber = "CERN-EP-2002-051",
    doi = "10.1140/epjc/s2003-01229-x",
    journal = "Eur. Phys. J.",
    volume = "C29",
    pages = "463--478",
    year = "2003"
}

@article{ALEPH:2001pfo,
    author = "Heister, A. and others",
    collaboration = "ALEPH",
    title = "{Study of the fragmentation of b quarks into B mesons at the Z peak}",
    eprint = "hep-ex/0106051",
    archivePrefix = "arXiv",
    reportNumber = "CERN-EP-2001-039",
    doi = "10.1016/S0370-2693(01)00690-6",
    journal = "Phys. Lett.",
    volume = "B512",
    pages = "30--48",
    year = "2001"
}

@article{SLD:2002poq,
    author = "Abe, Koya and others",
    collaboration = "SLD",
    title = "{Measurement of the b quark fragmentation function in $Z^0$ decays}",
    eprint = "hep-ex/0202031",
    archivePrefix = "arXiv",
    reportNumber = "SLAC-PUB-9087",
    doi = "10.1103/PhysRevD.65.092006",
    journal = "Phys. Rev.",
    volume = "D65",
    pages = "092006",
    year = "2002",
    extraPrefix = "Erratum", 
    extraVolume = "66", 
    extraPages = "079905", 
    extraYear = "2002",
    extraDoi = "10.1103/PhysRevD.66.079905"
}

@article{ALEPH:1999syy,
    author = "Barate, R. and others",
    collaboration = "ALEPH",
    title = "{Study of charm production in Z decays}",
    eprint = "hep-ex/9909032",
    archivePrefix = "arXiv",
    reportNumber = "CERN-EP-99-094",
    doi = "10.1007/s100520000421",
    journal = "Eur. Phys. J.",
    volume = "C16",
    pages = "597--611",
    year = "2000"
}

@article{OPAL:1994cct,
    author = "Akers, R. and others",
    collaboration = "OPAL",
    title = "{A measurement of the production of $D^{*\pm}$ mesons on the $Z^0$ resonance}",
    reportNumber = "CERN-PPE-94-217",
    doi = "10.1007/BF01564819",
    journal = "Z. Phys.",
    volume = "C67",
    pages = "27--44",
    year = "1995"
}

@article{ARGUS:1991vjh,
    author = "Albrecht, H. and others",
    collaboration = "ARGUS",
    title = "{Inclusive production of $D^0$, $D^+$ and $D^{*}(2010)^{+}$ mesons in B decays and nonresonant $e^+e^-$ annihilation at 10.6 GeV}",
    reportNumber = "DESY-91-023",
    doi = "10.1007/BF01559430",
    journal = "Z. Phys.",
    volume = "C52",
    pages = "353--360",
    year = "1991"
}

@article{ALICE:2020pga,
    author = "Acharya, Shreyasi and others",
    collaboration = "ALICE",
    title = "{Jet fragmentation transverse momentum distributions in pp and pPb collisions at $ \sqrt{s} $, $ \sqrt{s_{\mathrm{NN}}} $ = 5.02 TeV}",
    eprint = "2011.05904",
    archivePrefix = "arXiv",
    primaryClass = "nucl-ex",
    reportNumber = "CERN-EP-2020-220",
    doi = "10.1007/JHEP09(2021)211",
    journal = "JHEP",
    volume = "09",
    pages = "211",
    year = "2021"
}

@article{ALICE:2021aqk,
    author = "Acharya, S. and others",
    collaboration = "ALICE",
    title = "{Direct observation of the dead-cone effect in quantum chromodynamics}",
    eprint = "2106.05713",
    archivePrefix = "arXiv",
    primaryClass = "nucl-ex",
    reportNumber = "CERN-EP-2021-107",
    doi = "10.1038/s41586-022-04572-w",
    journal = "Nature",
    volume = "605",
    number = "7910",
    pages = "440--446",
    year = "2022",
    extraPrefix    = "Erratum",
    extraVolume    = "607",
    extraPages     = "E22",
    extraYear      = "2022",
    extraDoi       = "10.1038/s41586-022-05026-z",
}

@article{CTEQ:1993hwr,
    author = "Brock, Raymond and others",
    collaboration = "CTEQ",
    title = "{Handbook of perturbative QCD}",
    reportNumber = "FERMILAB-PUB-93-094, ANL-HEP-PR-95-29",
    doi = "10.1103/RevModPhys.67.157",
    journal = "Rev. Mod. Phys.",
    volume = "67",
    pages = "157--248",
    year = "1995"
}

@article{Bertone:2018ecm,
    author = "Bertone, V. and Hartland, N. P. and Nocera, E. R. and Rojo, J. and Rottoli, L.",
    collaboration = "NNPDF",
    title = "{Charged hadron fragmentation functions from collider data}",
    eprint = "1807.03310",
    archivePrefix = "arXiv",
    primaryClass = "hep-ph",
    reportNumber = "NIKHEF-2018-027, OUTP-18-01P",
    doi = "10.1140/epjc/s10052-018-6130-4",
    journal = "Eur. Phys. J.",
    volume = "C78",
    number = "8",
    pages = "651",
    year = "2018",
    extraPrefix = "Erratum", 
    extraVolume = "84", 
    extraPages = "155", 
    extraYear = "2024",
    extraDoi = "10.1140/epjc/s10052-024-12502-5"
}

@article{Soleymaninia:2018uiv,
    author = "Soleymaninia, Maryam and Goharipour, Muhammad and Khanpour, Hamzeh",
    title = "{First QCD analysis of charged hadron fragmentation functions and their uncertainties at next-to-next-to-leading order}",
    eprint = "1805.04847",
    archivePrefix = "arXiv",
    primaryClass = "hep-ph",
    doi = "10.1103/PhysRevD.98.074002",
    journal = "Phys. Rev.",
    volume = "D98",
    number = "7",
    pages = "074002",
    year = "2018"
}

@article{Soleymaninia:2020bsq,
    author = "Soleymaninia, Maryam and Goharipour, Muhammad and Khanpour, Hamzeh and Spiesberger, Hubert",
    title = "{Simultaneous extraction of fragmentation functions of light charged hadrons with mass corrections}",
    eprint = "2008.05342",
    archivePrefix = "arXiv",
    primaryClass = "hep-ph",
    doi = "10.1103/PhysRevD.103.054045",
    journal = "Phys. Rev.",
    volume = "D103",
    number = "5",
    pages = "054045",
    year = "2021"
}

@article{Soleymaninia:2019sjo,
    author = "Soleymaninia, Maryam and Goharipour, Muhammad and Khanpour, Hamzeh",
    title = "{Impact of unidentified light charged hadron data on the determination of pion fragmentation functions}",
    eprint = "1901.01120",
    archivePrefix = "arXiv",
    primaryClass = "hep-ph",
    doi = "10.1103/PhysRevD.99.034024",
    journal = "Phys. Rev.",
    volume = "D99",
    number = "3",
    pages = "034024",
    year = "2019"
}

@article{Gao:2024nkz,
    author = "Gao, Jun and Liu, ChongYang and Shen, XiaoMin and Xing, Hongxi and Zhao, Yuxiang",
    title = "{Simultaneous determination of fragmentation functions and test on momentum sum rule}",
    eprint = "2401.02781",
    archivePrefix = "arXiv",
    primaryClass = "hep-ph",
    doi = "10.1103/PhysRevLett.132.261903",
    journal = "Phys. Rev. Lett.",
    volume = "132",
    number = "26",
    pages = "261903",
    year = "2024"
}

@article{Gao:2024dbv,
    author = "Gao, Jun and Liu, ChongYang and Shen, XiaoMin and Xing, Hongxi and Zhao, Yuxiang",
    title = "{Global analysis of fragmentation functions to charged hadrons with high-precision data from the LHC}",
    eprint = "2407.04422",
    archivePrefix = "arXiv",
    primaryClass = "hep-ph",
    doi = "10.1103/PhysRevD.110.114019",
    journal = "Phys. Rev.",
    volume = "D110",
    number = "11",
    pages = "114019",
    year = "2024"
}

@article{Ellis:1992qq,
    author = "Ellis, Stephen D. and Kunszt, Zoltan and Soper, Davison E.",
    title = "{Jets at hadron colliders at order $\alpha_{s}^{3}$: A look inside}",
    eprint = "hep-ph/9208249",
    archivePrefix = "arXiv",
    reportNumber = "UW-PT-92-01, DOE-ER-40614-16",
    doi = "10.1103/PhysRevLett.69.3615",
    journal = "Phys. Rev. Lett.",
    volume = "69",
    pages = "3615--3618",
    year = "1992"
}

@article{Dokshitzer:1991fd,
    author = "Dokshitzer, Yuri L. and Khoze, Valery A. and Troian, S. I.",
    title = "{On specific QCD properties of heavy quark fragmentation (`dead cone')}",
    doi = "10.1088/0954-3899/17/10/023",
    journal = "J. Phys.",
    volume = "G17",
    pages = "1602--1604",
    year = "1991"
}

@article{ALICE:2025igw,
    author = "Acharya, Shreyasi and others",
    collaboration = "ALICE",
    title = "{Energy-energy correlators in charm-tagged jets in proton-proton collisions at $\sqrt{s} = 13$ TeV}",
    eprint = "2504.03431",
    archivePrefix = "arXiv",
    primaryClass = "hep-ex",
    reportNumber = "CERN-EP-2025-082",
    month = "4",
    year = "2025"
}

@article{ALICE:2023oww,
    author = "Acharya, Shreyasi and others",
    collaboration = "ALICE",
    title = "{Multiplicity dependence of charged-particle intra-jet properties in pp collisions at $\sqrt{s}$ = 13 TeV}",
    eprint = "2311.13322",
    archivePrefix = "arXiv",
    primaryClass = "hep-ex",
    reportNumber = "CERN-EP-2023-264",
    doi = "10.1140/epjc/s10052-024-13228-0",
    journal = "Eur. Phys. J.",
    volume = "C84",
    number = "10",
    pages = "1079",
    year = "2024"
}

@article{ALICE:2018ype,
    author = "Acharya, Shreyasi and others",
    collaboration = "ALICE",
    title = "{Charged jet cross section and fragmentation in proton-proton collisions at $\sqrt{s}$ = 7 TeV}",
    eprint = "1809.03232",
    archivePrefix = "arXiv",
    primaryClass = "nucl-ex",
    reportNumber = "CERN-EP-2018-235",
    doi = "10.1103/PhysRevD.99.012016",
    journal = "Phys. Rev.",
    volume = "D99",
    number = "1",
    pages = "012016",
    year = "2019"
}

@article{DAgostini:1994fjx,
    author = "D'Agostini, G.",
    title = "{A Multidimensional unfolding method based on Bayes' theorem}",
    reportNumber = "DESY-94-099",
    doi = "10.1016/0168-9002(95)00274-X",
    journal = "Nucl. Instrum. Meth. ",
    volume = "A362",
    pages = "487--498",
    year = "1995"
}

@cernreport{Battaglia:2004coa,
    author = "Battaglia, M. and Orava, R. and Salmi, L.",
    title = "{A study of depletion of fragmentation particles at small angles in b-jets with the DELPHI detector at LEP} ",
    number = "DELPHI-2004-037-CONF-712",
    url = "https://cds.cern.ch/record/989441",
    month = "7",
    year = "2004"
}

@article{ATLAS:2019rqw,
    author = "Aad, Georges and others",
    collaboration = "ATLAS",
    title = "{Properties of jet fragmentation using charged particles measured with the ATLAS detector in $pp$ collisions at $\sqrt{s}=13$ TeV}",
    eprint = "1906.09254",
    archivePrefix = "arXiv",
    primaryClass = "hep-ex",
    reportNumber = "CERN-EP-2019-090",
    doi = "10.1103/PhysRevD.100.052011",
    journal = "Phys. Rev.",
    volume = "D100",
    number = "5",
    pages = "052011",
    year = "2019"
}

@article{LHCb-PROC-2011-006,
  author="Clemencic, M and others",
  title="{The \lhcb simulation application, Gauss: Design, evolution and experience}",
  journal="J. Phys. Conf. Ser.",
  volume={331},
  pages={032023},
  doi={10.1088/1742-6596/331/3/032023},
  year={2011},
}

@article{LHCb-PROC-2010-056,
      author         = "Belyaev, I. and others",
      title          = "{Handling of the generation of primary events
                         in Gauss, the LHCb simulation framework}",
      journal="J. Phys. Conf. Ser.",
      volume={331},
      pages={032047},
      doi={10.1088/1742-6596/331/3/032047},
      year={2011},
}

@article{Sjostrand:2006za,
      author         = {Sj\"{o}strand, Torbj\"{o}rn and Mrenna, Stephen and Skands, Peter"},
      title          = "{PYTHIA 6.4 physics and manual}",
      journal        = "JHEP",
      volume         = "05",
      pages          = "026",
      doi            = "10.1088/1126-6708/2006/05/026",
      year           = "2006",
      eprint         = "hep-ph/0603175",
      archivePrefix  = "arXiv",
      primaryClass   = "hep-ph",
}

@article{Sjostrand:2007gs,
      author         = {Sj\"{o}strand, Torbj\"{o}rn and Mrenna, Stephen and
                        Skands, Peter"},
      title          = "{A brief introduction to PYTHIA 8.1}",
      journal        = "Comput. Phys. Commun.",
      volume         = "178",
      pages          = "852-867",
      doi            = "10.1016/j.cpc.2008.01.036",
      year           = "2008",
      eprint         = "0710.3820",
      archivePrefix  = "arXiv",
      primaryClass   = "hep-ph",
      reportNumber   = "CERN-LCGAPP-2007-04, LU-TP-07-28,
                        FERMILAB-PUB-07-512-CD-T",
}

@Article{Agostinelli:2002hh,
     author    = "Agostinelli, S. and others",
 collaboration = "Geant4 collaboration",
     title     = "{Geant4: A simulation toolkit}",
     journal   = "Nucl. Instrum. Meth.",
     volume    = "A506",
     year      = "2003",
     pages     = "250",
     doi       = "10.1016/S0168-9002(03)01368-8",
}

@article{Allison:2006ve,
      author         = "Allison, John and Amako, K. and Apostolakis, J. and
                        Araujo, H. and Dubois, P.A. and others",
 collaboration = "Geant4 collaboration",
      title          = "{Geant4 developments and applications}",
      journal        = "IEEE Trans.Nucl.Sci.",
      volume         = "53",
      pages          = "270",
      doi            = "10.1109/TNS.2006.869826",
      year           = "2006",
      reportNumber   = "SLAC-PUB-11870",
}

@Article{Lange:2001uf,
     author    = "Lange, D. J.",
     title     = "{The EvtGen particle decay simulation package}",
     journal   = "Nucl. Instrum. Meth.",
     volume    = "A462",
     year      = "2001",
     pages     = "152-155",
     doi       = "10.1016/S0168-9002(01)00089-4",
}

@article{davidson2015photos,
	author="Davidson, N. and Przedzinski, T. and Was, Z.",
	title = "{PHOTOS interface in C++: Technical and physics documentation}",
      	eprint={1011.0937},
      	archivePrefix={arXiv},
      	primaryClass={hep-ph},
	journal = {Comput. Phys. Commun.},
	volume = {199},
	pages = {86},
	year = {2016},
	doi = {https://doi.org/10.1016/j.cpc.2015.09.013},
}

@Article{AdaBoost,
    author = "Freund, Yoav and Schapire, Robert E.",
    title = "A decision-theoretic generalization of on-line learning and an application to boosting",
  journal = 	 "J. Comput. Syst. Sci.",
  volume         = "55",
  pages          = "119",
  year = 	 {1997},
  doi            = "10.1006/jcss.1997.1504",
}

@Misc{mciteplus,
  author = 	 {Shell, Michael},
  title = 	 {Mciteplus: Enhanced multicitations},
  howpublished = {\href{http://www.michaelshell.org/tex/mciteplus/} {http://www.michaelshell.org/tex/mciteplus/}},
}

@book{Breiman,
  author = 	 {Breiman, L. and Friedman, J. H. and Olshen,
                  R. A. and Stone, C. J.},
  title = 	 {Classification and regression trees},
  publisher = 	 {Wadsworth international group},
  year = 	 {1984},
  address = 	 {Belmont, California, USA},
}

@article{Stripping,
      author        = "Grieser, Nathan and Rodrigues, Eduardo and Sahoo, Niladri and Sheng, 
                       Shuqi and Skidmore, Nicole and Smith, Mark",
      title         = "{The LHCb stripping project: Sustainable legacy data processing for high-energy physics}",
      eprint        = "2509.05294",
      archivePrefix = "arXiv",
      primaryClass  = "hep-ex",
      year          = "2025"
}

\newpage
\centerline
{\large\bf LHCb collaboration}
\begin
{flushleft}
\small
R.~Aaij$^{38}$\lhcborcid{0000-0003-0533-1952},
A.S.W.~Abdelmotteleb$^{57}$\lhcborcid{0000-0001-7905-0542},
C.~Abellan~Beteta$^{51}$\lhcborcid{0009-0009-0869-6798},
F.~Abudin{\'e}n$^{57}$\lhcborcid{0000-0002-6737-3528},
T.~Ackernley$^{61}$\lhcborcid{0000-0002-5951-3498},
A. A. ~Adefisoye$^{69}$\lhcborcid{0000-0003-2448-1550},
B.~Adeva$^{47}$\lhcborcid{0000-0001-9756-3712},
M.~Adinolfi$^{55}$\lhcborcid{0000-0002-1326-1264},
P.~Adlarson$^{85}$\lhcborcid{0000-0001-6280-3851},
C.~Agapopoulou$^{14}$\lhcborcid{0000-0002-2368-0147}
% ...

{\footnotesize \it

% ...
$^{14}$Universit{\'e} Paris-Saclay, CNRS/IN2P3, IJCLab, Orsay, France\\
% ...
$^{38}$Nikhef National Institute for Subatomic Physics, Amsterdam, Netherlands\\
% ...
$^{47}$Instituto Galego de F{\'\i}sica de Altas Enerx{\'\i}as (IGFAE), Universidade de Santiago de Compostela, Santiago de Compostela, Spain\\
% ...
$^{51}$Physik-Institut, Universit{\"a}t Z{\"u}rich, Z{\"u}rich, Switzerland\\
% ...
$^{55}$H.H. Wills Physics Laboratory, University of Bristol, Bristol, United Kingdom\\
% ...
$^{57}$Department of Physics, University of Warwick, Coventry, United Kingdom\\
% ...
$^{60}$School of Physics and Astronomy, University of Glasgow, Glasgow, United Kingdom\\
$^{61}$Oliver Lodge Laboratory, University of Liverpool, Liverpool, United Kingdom\\
% ...
$^{69}$Syracuse University, Syracuse, NY, United States\\
% ...
$^{85}$Department of Physics and Astronomy, Uppsala University, Uppsala, Sweden, associated to $^{60}$\\
% ...
}
\end{flushleft}

\end{document}